\documentclass[a4paper,11pt]{article}
\pdfoutput=1 
\usepackage{jheppub} 
\usepackage{booktabs}
\usepackage{multirow}
\usepackage{amsmath}
\usepackage{slashed}
\usepackage{amstext,amssymb}
\usepackage{graphicx}
\usepackage{graphicx,wrapfig,lipsum}
\usepackage{xmpmulti}
\usepackage{animate}
\usepackage{epstopdf}
\usepackage{xcolor,colortbl}
\usepackage{multimedia}
\usepackage{hyperref}
\usepackage{url}
\usepackage{xspace}
\usepackage{color}
\usepackage{units}
\usepackage{tikz}
\usepackage{tikz-feynman}
\tikzfeynmanset{compat=1.1.0}
\usepackage{wrapfig}
\usetikzlibrary{snakes}
\usepackage{braket}
\usepackage{bigstrut}
\usepackage{subcaption}
\usepackage{float}
\usepackage{makecell}
\usepackage{bm}
\usepackage{bbm}
\usepackage[left]{lineno}

\newcommand{\mathsym}[1]{{}}

\newcommand{\baz}{\begin{array}{cc}}
\newcommand{\bad}{\begin{array}{ccc}}
\newcommand{\ba}{\begin{array}{c}}
\newcommand{\ea}{\end{array}}
\newcommand{\be}{\begin{equation}}
\newcommand{\ee}{\end{equation}}
\newcommand{\bea}{\begin{eqnarray}}
\newcommand{\eea}{\end{eqnarray}}

\newcommand{\bi}{\begin{itemize}}
\newcommand{\ei}{\end{itemize}}
\newcommand{\bmt}{\begin{pmatrix}}
\newcommand{\emt}{\end{pmatrix}}
\newcommand{\bt}{\begin{tabular}}
\newcommand{\et}{\end{tabular}}

\newcommand{\benu}{\begin{enumerate}}
\newcommand{\eenu}{\end{enumerate}}

\newcommand{\bav}{\begin{array}{cccc}}

\usepackage{wrapfig}
\definecolor{light-gray}{gray}{0.95}
\usepackage{tcolorbox}

\title{Probing $\bm{0\nu\beta\beta}$ and $\bm{\mu\to e\gamma}$ via Fully Determined Dirac Mass Terms in LRSM with Double Seesaw}
\author[a]{Pratik Adarsh}
\author[a]{, Rajrupa Banerjee}
\author[b]{, Purushottam Sahu}
\author[a,c]{, Utkarsh Patel}
\author[a,d]{, Sudhanwa Patra}

\affiliation[a]{Department of Physics, Indian Institute of Technology Bhilai, Kutelabhata, Durg 491002, India}
\affiliation[b]{Department of Physics, Indian Institute of Technology Bombay, Powai, Mumbai 400076, India}
\affiliation[c]{Saha Institute of Nuclear Physics, 1/AF Bidhan Nagar, Kolkata 700064, India}
\affiliation[d]{Institute of Physics, Bhubaneswar, Sachivalaya Marg, Sainik School Post, Bhubaneswar 751005, India}

\emailAdd{pratikad@iitbhilai.ac.in}
\emailAdd{rajrupab@iitbhilai.ac.in}
\emailAdd{purushottam.phys@gmail.com}
\emailAdd{utkarsh.patel@saha.ac.in}
\emailAdd{sudhanwa@iitbhilai.ac.in}
\abstract{Neutrinoless double beta decay ($0\nu\beta\beta$) and charged lepton flavor violation (cLFV) experiments provide promising avenues to probe new physics contributions from extended neutrino sectors in beyond Standard Model (BSM) scenarios. We consider a Left-Right Symmetric Model (LRSM) extended with three generations of sterile neutrinos to realize a double type-I seesaw mechanism for light neutrino mass generation. The double seesaw induces maximal lepton number violation in the right-handed sector and facilitates enhanced Majorana masses for right-handed neutrinos, thereby leading to their dominant contributions in both cLFV and $0\nu\beta\beta$ processes. We perform a comprehensive exploration of the parameter space for new-physics contributions to the cLFV decay $\mu \to e \gamma$ and to $0\nu\beta\beta$, considering two different textures for the Dirac mass matrices: (i) a \emph{symmetry-motivated} limit with $M_D \propto \mathbbm{1}$, and (ii) a texture \emph{fully determined} by the model framework. A detailed analysis of the common parameter regions accessible to current experiments like \textsf{KamLAND-Zen} and \textsf{LEGEND-200}, and upcoming experiments, such as \textsf{MEG-II} and \textsf{LEGEND-1000}, is presented, underscoring the phenomenological relevance of this framework. Our results aim to provide optimistic benchmarks for future searches targeting right-handed current-mediated neutrino interactions.}


\begin{document} 
\maketitle
\section{Introduction}\label{sec:intro}
In the precision era of particle physics, the observed neutrino oscillations \cite{SNO:2002tuh,Super-Kamiokande:2016yck,T2K:2019efw,DayaBay:2012fng,DoubleChooz:2011ymz}, and thereby the confirmation of non-zero neutrino masses and lepton flavor violation (LFV) in the neutral sector, have provided strong motivation for physics beyond the Standard Model (SM). Since both lepton flavor and lepton number are accidental symmetries of the SM, processes that violate these symmetries, namely charged lepton flavor violation (cLFV) \cite{Kuno:1999jp,Calibbi:2017uvl,Lindner:2016bgg,Ardu:2022sbt} and lepton number violation (LNV) \cite{Furry:1939qr,Schechter:1981bd,Rodejohann:2012xd,Dolinski:2019nrj,Deppisch:2012nb}, are well-motivated phenomena to explore.

The SM predicts massless neutrinos. Moreover, in a minimal extension of the SM that allows neutrinos to acquire masses through renormalizable Yukawa interactions with the Higgs boson, the predicted cLFV rates remain extremely suppressed due to the Glashow-Iliopoulos-Maiani (GIM) mechanism \cite{Glashow:1970gm}, placing them well beyond the reach of current and upcoming experiments \cite{Davidson:2022jai,Frau:2024rzt,Perrevoort:2024qtc,Palo:2025oyq}. The SM also offers no insight into whether neutrinos are Dirac or Majorana particles \cite{Majorana:1937vz,Majorana:translation}. If neutrinos are Majorana in nature, they can mediate the rare LNV process known as neutrinoless double beta ($0\nu\beta\beta$) decay \cite{Schechter:1981bd} via light active neutrino exchange. The predicted decay rates for this standard mechanism are typically small, with only a limited parameter space, particularly involving quasi-degenerate \cite{Bilenky:2001rz} light neutrinos, lying within the sensitivity reach of experiments. However, such a quasi-degenerate neutrino spectrum is increasingly disfavored by cosmological observations \cite{Planck:2018vyg,Lattanzi:2020iik}. Moreover, identifying the correct mass hierarchy through this channel, given the current constraints on the sum of light neutrino masses, would require a multi-ton scale detector that is not feasible in the near future. Therefore, any comparative observation of $0\nu\beta\beta$ decay in upcoming experiments would strongly point toward contributions beyond the Standard Model. Thus, any positive signal of cLFV or $0\nu\beta\beta$ decay would serve as compelling evidence for such scenarios. In this context, the search for cLFV and LNV at the intensity frontier constitutes a clear and unambiguous probe of new physics (NP).

The most economical way to obtain small neutrino masses along with enhanced, observable cLFV and $0\nu\beta\beta$ rates is by extending the SM to realize the type-I seesaw \cite{Minkowski:1977sc,Mohapatra:1979ia,Gell-Mann:1979vob} mechanism. However, in such scenarios, sizeable non-standard contributions require large active-sterile mixing, which in turn demands a degree of fine-tuning to keep the light neutrino masses and their observed splittings small. This drawback can be overcome by implementing the double seesaw mechanism \cite{Mohapatra:1986bd} in which high-mass scale sterile neutrinos induce Majorana mass to right-handed neutrinos (RHNs). The double seesaw explains the observed neutrino mass splitting without requiring unnatural fine-tuning of parameters. 

The realization of the double seesaw mechanism in a simple extension of the SM requires, in addition to the usual right-handed neutrinos $N_{R_i}$ ($i=1,2,3$), an extra set of left-handed gauge-singlet fermions $S_{L_i}$ ($i=1,2,3$) \cite{Lindner:2005pk}. The singlets $S_{L_i}$ carry large Majorana masses $M_S$, while symmetry considerations forbid a bare Majorana mass term for the RHNs. The RHNs couple via Yukawa interactions both to the SM lepton doublets and to the singlet fermions, giving rise to the Dirac mass matrices $M_D$ and $M_{RS}$, respectively, whereas any direct $\nu_L$-$S_L$ coupling is either absent or highly suppressed. In the hierarchical limit $M_S \gg M_{RS} \gg M_D$, the double (or cascade) seesaw is realized: in the first step, the heavy singlets generate a large Majorana mass for the RHNs via a type-I seesaw, and in the second step, these RHNs generate small Majorana masses for the light neutrinos. Such frameworks become even more compelling when embedded in a Left-Right Symmetric Model (LRSM) \cite{Mohapatra:1974gc,Pati:1974yy,Senjanovic:1975rk,Mohapatra:1980yp,Deshpande:1990ip}, which restores parity symmetry at high energies and naturally accommodates the seesaw structure since the RHNs are part of the neutral lepton doublet and their bare Majorana mass is automatically forbidden by $U(1)_{B-L}$ gauge invariance. Table~\ref{tab:LRSM_seesaw_summary} classifies the realization of different seesaw mechanisms in LRSM based on the scalar and fermion field content, along with the key conditions.   

\begin{table}
	\centering
	\scriptsize
	\renewcommand{\arraystretch}{1.3}
	\begin{tabular}{|c|c|c|c|c|}
		\hline
		\textbf{Scalar Sector} & \textbf{Extra Fermions} & \textbf{Seesaw Type} & \textbf{Key Conditions / Features} & \textbf{Ref.} \\
		\hline
		Scalar bidoublet +&\multirow{2}{*}{None}&\multirow{2}{*}{None}&Minimal LRSM setup; neutrinos &\multirow{2}{*}{\cite{Hati:2018tge}}\\
		Scalar doublets&&&are Dirac-type in nature&\\
		\hline
			Scalar bidoublet +&\multirow{2}{*}{None}&Type-I +&Minimal LRSM setup; neutrinos get&\multirow{2}{*}{\cite{Senjanovic:1975rk}}\\
		Scalar triplets&&Type-II&Majorana masses via triplet VEVs&\\
		\hline
		Scalar bidoublet +&Sterile neutrino&\multirow{2}{*}{Inverse Seesaw}&Small Majorana mass for sterile&\multirow{2}{*}{\cite{Mohapatra:1986bd,Brdar:2018sbk}}\\
		Scalar doublets&(per generation)&&state; negligible LH doublet VEV&\\
		\hline
			Scalar bidoublet +&Sterile neutrino&\multirow{2}{*}{Linear Seesaw}&Small Majorana mass for sterile&\multirow{2}{*}{\cite{Deppisch:2015cua,Malinsky:2005bi}}\\
		Scalar doublets&(per generation)&&state; non-zero LH doublet VEV&\\
		\hline
			Scalar bidoublet +&Sterile neutrino&\multirow{2}{*}{Double Seesaw}&Large Majorana mass for sterile&\multirow{2}{*}{\cite{Mohapatra:1986bd,Patra:2023ltl}}\\
		Scalar doublets&(per generation)&&state; negligible LH doublet VEV&\\
		\hline
	\end{tabular}
	\caption{Classification of different seesaw mechanisms in Left-Right Symmetric Models (LRSM) based on scalar sector and fermion content.}
	\label{tab:LRSM_seesaw_summary}
\end{table}

A TeV-scale LRSM offers rich phenomenology at both the energy frontier, through same-sign dilepton plus two jets signatures at the LHC \cite{Keung:1983uu,Ferrari:2000sp,Nemevsek:2011hz,Chakrabortty:2012pp,Aguilar-Saavedra:2012grq,Chen:2013foz,Gluza:2015goa,Dev:2015kca,Senjanovic:2010nq,Deppisch:2015qwa}, and the intensity frontier, via processes such as $0\nu\beta\beta$ decay and cLFV transitions like $\mu \to e \gamma$. This dual sensitivity provides a natural link between high-energy collider signals and low-energy precision observables. Numerous studies have examined new physics contributions to $0\nu\beta\beta$ decay \cite{Mohapatra:1981pm,Picciotto:1982qe,Hirsch:1996qw,SuperNEMO:2010wnd,Tello:2010am,Chakrabortty:2012mh,Nemevsek:2012iq,Patra:2012ur,Awasthi:2013ff,Barry:2013xxa,BhupalDev:2013ntw,Huang:2013kma,BhupalDev:2014qbx,Ge:2015yqa,Borah:2015ufa,Awasthi:2015ota,Bambhaniya:2015ipg,Gu:2015uek,Awasthi:2016kbk} and cLFV processes \cite{Das:2012ii,Tello:2010am,Barry:2013xxa,Awasthi:2015ota,Bambhaniya:2015ipg,Borah:2016iqd,Riazuddin:1981hz,Pal:1983bf,Mohapatra:1992uu,Cirigliano:2004mv,Bajc:2009ft,Lee:2013htl,Borah:2013lva,Bonilla:2016fqd} within TeV-scale LRSM frameworks featuring various seesaw realizations. In these scenarios, the heavy-neutrino masses and mixings, parametrized in terms of the Dirac mass matrix, play a pivotal role in determining the size of new physics effects in $0\nu\beta\beta$ and cLFV processes. Consequently, different treatments of the Dirac mass matrix in the literature lead to distinct predictions for the corresponding decay rates. For instance, references~\cite{Bambhaniya:2015ipg,Dash:2021pbx,Patra:2023ltl,Pritimita:2016fgr,Bonilla:2016fqd} adopt a Dirac mass matrix proportional to the identity matrix, motivated by certain flavor symmetries \cite{Hagedorn:2014wha,Altarelli:2004za}, to quantify the heavy-neutrino mixing for cLFV and $0\nu\beta\beta$ decay rate predictions. References~\cite{Pritimita:2016fgr,Bonilla:2016fqd} instead approximate the Dirac mass matrix by the up-type quark mass matrix, while references~\cite{Dev:2009aw,Awasthi:2013ff,Dash:2021pbx} consider an $SO(10)$ grand unified theory (GUT)-motivated structure.

In this paper, we perform a comprehensive analysis of new-physics contributions to the charged lepton flavor violation observable $\mu \to e \gamma$ and to neutrinoless double beta ($0\nu\beta\beta$) decay within the doublet LRSM employing a double seesaw mechanism (bidoublet plus doublets, no triplets), which is a structurally simpler setup than the extensively studied triplet-based LRSMs. To the best of our knowledge, this is the first cLFV study in the considered model setup. Unlike inverse or linear seesaw realizations in LRSM, which typically yield pseudo-Dirac heavy states and limited rates for LNV and cLFV, the double seesaw naturally accommodates large right-handed neutrino (RHN) Majorana masses in a TeV-scale LRSM, thereby opening a distinctive parameter space featuring sizable right-handed currents and potentially observable $0\nu\beta\beta$ and $\mu \to e \gamma$ signals.

We consider two benchmark choices for the Dirac mass matrix: (\textbf{Case I}) a symmetry-motivated limit with $M_D \propto \mathbbm{1}$, and (\textbf{Case II}) a \emph{model-determined} texture in which $M_D$ and $M_{RS}$ are fixed by LR symmetry ($\mathcal{C}$) and the screening condition (i.e., cancellation of Dirac structures in the light-neutrino mass matrix), combined with low-energy oscillation parameters, without arbitrary phases or free parameters~\cite{Patel:2023voj,Nemevsek:2012iq,Huang:2013kma}. The latter case yields direct, testable correlations between oscillation parameters and intensity-frontier observables, and remains compatible with leptogenesis~\cite{Patel:2023voj}. For Case~I, we follow~\cite{Patra:2023ltl}, which explores $0\nu\beta\beta$ in a sub-TeV RHN regime that enhances non-standard contributions; for Case~II, we follow~\cite{Patel:2023voj}, in which the Dirac mass matrix is fully determined, enabling definite predictions for both $0\nu\beta\beta$ and cLFV. Throughout, we work in parameter regions favored by $0\nu\beta\beta$ sensitivity, complemented by cLFV-relevant choices, and identify regions simultaneously compatible with $\mu \to e \gamma$ and $0\nu\beta\beta$.


The paper is structured as follows. In section~\ref{sec:framework_motivation}, we summarize the left–right symmetric model, including its particle content and interactions. The generation of Majorana masses for neutrinos (both light and heavy) via the double seesaw mechanism is presented in a dedicated subsection. The heavy gauge bosons associated with the $SU(2)_{R}$ gauge group, along with their masses and mixing, are also briefly described in the subsequent subsection. Section~\ref{DMT_NM} outlines the key ingredients for cLFV and neutrinoless double beta decay rate calculations: the Dirac mass terms and neutrino mixing for both Case~I and Case~II. In section~\ref{sec:general-cLFV-LNV}, the general contribution to the LNV ($0\nu\beta\beta$ decay) and cLFV ($\mu\to e\gamma$) processes is discussed for the considered model framework. Section~\ref{sec:pehnoCaseI} and section~\ref{sec:pehnoCaseII} discuss the predictions for BR($\mu\to e\gamma$) and $0\nu\beta\beta$ decay rates in Case~I and Case~II, respectively. Finally, in section~\ref{sec:concl}, we conclude by summarizing the key results and outlining prospects for future work.

\section{Model Framework and Motivation}\label{sec:framework_motivation}

The Standard Model (SM) accounts for the $(V - A)$ structure of weak interactions and parity violation through its gauge symmetry $\mathcal{G} \equiv SU(3)_C \times SU(2)_L \times U(1)_Y$, where right-handed fermions are singlets under $SU(2)_L$. However, the SM does not explain the fundamental origin of parity violation. A natural approach to address this shortcoming is to consider a theory that restores parity symmetry at higher energy scales. This motivates the Left-Right Symmetric Model (LRSM), an extension of the SM based on the gauge group
\begin{equation}\label{GLR}
	\mathcal{G}_{LR} \equiv SU(3)_C \times SU(2)_L \times SU(2)_R \times U(1)_{B-L},
\end{equation}
where $B - L$ denotes the baryon minus lepton number. In this framework, the electric charge is defined as
\begin{equation}
	Q = T_{3L} + T_{3R} + \frac{B - L}{2},
\end{equation}
with $T_{3L}$ and $T_{3R}$ being the third components of isospin corresponding to $SU(2)_L$ and $SU(2)_R$, respectively.

The fermion content of LRSM includes all SM fermions along with a right-handed neutrino $N_R$. The left- and right-handed fermion doublets, with their transformation properties under $SU(2)_L \times SU(2)_R \times U(1)_{B-L}$, are:
\begin{eqnarray}
	&&q_L = \begin{pmatrix} u_L \\ d_L \end{pmatrix} \sim [2,1,1/3], \quad 
	q_R = \begin{pmatrix} u_R \\ d_R \end{pmatrix} \sim [1,2,1/3], \nonumber \\
	&&\ell_L = \begin{pmatrix} \nu_L \\ e_L \end{pmatrix} \sim [2,1,-1], \quad
	\ell_R = \begin{pmatrix} N_R \\ e_R \end{pmatrix} \sim [1,2,-1]. \nonumber
\end{eqnarray}
For brevity, the $SU(3)_C$ quantum numbers are omitted.

The scalar sector of the LRSM consists of a Higgs bidoublet $\Phi$ and two Higgs doublets, $H_L$ and $H_R$. Their field content is:
\begin{equation*}
	\Phi = 
	\begin{pmatrix} 
		\phi_1^0 & \phi_2^+ \\
		\phi_1^- & \phi_2^0 
	\end{pmatrix} \sim [2,2,0], \quad
	H_L = 
	\begin{pmatrix} 
		h_L^+ \\
		h_L^0 
	\end{pmatrix} \sim [2,1,1], \quad
	H_R = 
	\begin{pmatrix} 
		h_R^+ \\
		h_R^0 
	\end{pmatrix} \sim [1,2,1].
\end{equation*}

The pattern of spontaneous symmetry breaking (SSB) proceeds as follows:
\begin{center}
	\textbf{\underline{SSB of LRSM}}
\end{center}
\begin{equation*}
	SU(2)_L \times \underbrace{SU(2)_R \times U(1)_{B-L}}
\end{equation*}
\begin{equation*}
	\hspace{4cm} \Big\downarrow \braket{H_R^0}
\end{equation*}
\begin{equation*}
	\hspace{1cm} \underbrace{SU(2)_L \times U(1)_Y}
\end{equation*}
\begin{equation*}
	\hspace{5.2cm} \Big\downarrow \braket{\phi_{1}^0,\, \phi_{2}^0} \subset \Phi
\end{equation*}
\begin{equation*}
	\hspace{2cm} U(1)_{em}
\end{equation*}

At the LR breaking scale, the neutral component $h_R^0$ of $H_R$ acquires a vacuum expectation value (VEV), $\langle H_R^0 \rangle \equiv v_R$, breaking $SU(2)_R \times U(1)_{B-L}$ to $U(1)_Y$. This generates masses for the right-handed gauge bosons $W_R^\pm$ and $Z'$, which depend on $v_R$. Although $H_L$ does not participate in symmetry breaking, it is included to maintain left-right symmetry.

Electroweak symmetry breaking, $SU(2)_L \times U(1)_Y \rightarrow U(1)_{em}$, is achieved when the neutral components $\phi_1^0$ and $\phi_2^0$ of $\Phi$ acquire VEVs: $\langle \phi_1^0 \rangle \equiv v_1$ and $\langle \phi_2^0 \rangle \equiv v_2$, with the SM Higgs VEV $v = \sqrt{v_1^2 + v_2^2} \simeq 246$ GeV.

The Yukawa Lagrangian involving quarks and leptons is given by:
\begin{equation}
	-\mathcal{L}_{\text{Yuk}} \supset 
	\overline{q_L} \left[Y_1 \Phi + Y_2 \widetilde{\Phi} \right] q_R 
	+ \overline{\ell_L} \left[Y_3 \Phi + Y_4 \widetilde{\Phi} \right] \ell_R + \text{h.c.},
\end{equation}
where $\widetilde{\Phi} = \sigma_2 \Phi^* \sigma_2$ and $\sigma_2$ is the second Pauli matrix. Upon electroweak symmetry breaking, the VEV of the bidoublet becomes:
\begin{equation}
	\langle \Phi \rangle = 
	\begin{pmatrix}
		v_1 & 0 \\
		0 & v_2 
	\end{pmatrix},
\end{equation}
resulting in the following mass matrices:
\begin{eqnarray}
	&& M_u = Y_1 v_1 + Y_2 v_2, \nonumber \\
	&& M_d = Y_1 v_2 + Y_2 v_1, \nonumber \\
	&& M_e = Y_3 v_2 + Y_4 v_1, \nonumber \\
	&& M_D = Y_3 v_1 + Y_4 v_2.
\end{eqnarray}
Here, $M_u$, $M_d$, $M_e$, and $M_D$ denote the up-type quark, down-type quark, charged lepton, and Dirac neutrino mass matrices, respectively.

Assuming $v_2 \ll v_1$ and $|Y_3| \ll |Y_4|$, one obtains suppressed Dirac neutrino masses. Under these conditions, the mass matrices simplify to:
\begin{eqnarray}\label{VEVconditionResult}
	M_e &\approx& Y_4 v_1, \\
	M_D &\approx& v_1 \left( Y_3 + \frac{M_e v_2}{v_1^2} \right) \approx v Y_3 \equiv v Y_D,
	\label{MDY3}
\end{eqnarray}
where $v \approx v_1$.

The minimal LRSM described here is based on the gauge group $SU(2)_L \times SU(2)_R \times U(1)_{B-L}$, with corresponding gauge couplings $g_L$, $g_R$, and $g_{BL}$. To ensure $g_L = g_R$, a discrete left-right (LR) symmetry must be imposed. Two common choices for this symmetry are: (i) generalized parity $\mathcal{P}$ and (ii) charge conjugation $\mathcal{C}$~\cite{Maiezza:2010ic}. Under the discrete left-right symmetry transformations, the fields transform as follows:
\begin{equation}
	\mathcal{P}:\left\{
	\begin{aligned}
		& \ell_L \leftrightarrow \ell_R,\quad q_L \leftrightarrow q_R, \\
		& \Phi \leftrightarrow \Phi^{\dagger},\quad H_L \leftrightarrow H_R,\quad \tilde{\Phi} \leftrightarrow \tilde{\Phi}^{\dagger}
	\end{aligned}
	\right.
	\quad
	\mathcal{C}:\left\{
	\begin{aligned}
		& \ell_L \leftrightarrow \ell_R^c,\quad q_L \leftrightarrow q_R^c, \\
		& \Phi \leftrightarrow \Phi^{T},\quad H_L \leftrightarrow H_R^*,\quad \tilde{\Phi} \leftrightarrow \tilde{\Phi}^{T}
	\end{aligned}
	\right.
\end{equation}

Imposing either parity ($\mathcal{P}$) or charge conjugation ($\mathcal{C}$) as the discrete left-right symmetry constrains the Yukawa couplings and mass matrices of the theory. In our framework, this leads to a predictive structure for the Dirac neutrino mass matrix and the coupling between right-handed neutrinos and left-handed sterile states. These features play a key role in the generation of light neutrino masses through a double seesaw mechanism. We utilize this setup to study low-energy observables such as neutrinoless double beta decay ($0\nu\beta\beta$) and charged lepton flavor violating (cLFV) processes, which serve as sensitive probes of the underlying new physics embedded in the left-right symmetric theory.

\subsection{Neutrino masses via Double Seesaw Mechanism}\label{subsec:DS_NeuMass}
In our left-right symmetric framework, we incorporate additional gauge-singlet fermions $S_L$, one for each generation, to realize a double seesaw mechanism for neutrino mass generation, following \cite{Patra:2023ltl}. The neutral fermion sector consists of the standard left-handed neutrinos $\nu_L$, their right-handed counterparts $N_R$, and the sterile singlets $S_L$. The relevant neutral fermion mass terms in the flavor basis of $(\nu_L, N_R^c, S_L)$ can be written as:
\begin{equation}\label{massLag}
	\mathcal{L}_{\text{mass}} \supset \frac{1}{2} 
	\begin{pmatrix}
		\overline{\nu_L} & \overline{N_R^c} & \overline{S_L}
	\end{pmatrix}
	\begin{pmatrix}
		0 & M_D & 0 \\
		M_D^T & 0 & M_{RS} \\
		0 & M_{RS}^T & M_S
	\end{pmatrix}
	\begin{pmatrix}
		\nu_L^c \\
		N_R \\
		S_L^c
	\end{pmatrix} + \text{h.c.}
\end{equation}

Here, $M_D$ is the Dirac neutrino mass matrix generated via the Yukawa couplings to the bidoublet scalar $\Phi$, $M_{RS}$ connects the right-handed neutrinos $N_R$ with the singlet fermions $S_L$, and $M_S$ is the Majorana mass matrix for $S_L$. We work in the regime where $M_S \gg M_{RS} \gg M_D$, which enables a two-step seesaw suppression of the light neutrino masses.

Crucially, the vacuum expectation value (VEV) of the left-handed scalar doublet is taken to be zero, i.e., $\langle H_L \rangle = 0$, in order to forbid a Dirac mass term between $\nu_L$ and $S_L$. This preserves the desired block structure of the neutral fermion mass matrix and ensures that the light neutrino masses arise entirely from the double seesaw mechanism.

The heavy sterile fermions $S_L$ are first integrated out, inducing an effective Majorana mass for the right-handed neutrinos:
\begin{equation}
	M_N \simeq - M_{RS} M_S^{-1} M_{RS}^T.
\end{equation}
Subsequently, the light neutrino masses are generated via a type-I seesaw involving $M_N$:
\begin{equation}
	m_\nu \simeq - M_D M_N^{-1} M_D^T.
\end{equation}

Collecting all neutrino mass matrices, we have:
\begin{eqnarray} \label{MajMasses}
	&& m_\nu \cong M_D ({M^T_{RS}})^{-1}  M_S M^{-1}_{RS} M^T_D \quad \text{(mass matrix of light neutrinos)} ,\nonumber \\
	&& m_N \equiv M_R \cong -M_{RS} M^{-1}_S M^T_{RS} \quad \text{(mass matrix of RH neutrinos)} ,\nonumber \\
	&& m_S \cong M_S \quad \text{(mass matrix of sterile neutrinos)} \, .
\end{eqnarray}

The structure and scale of the mass matrices play a crucial role in shaping low-energy observables such as neutrinoless double beta decay and charged lepton flavor violation, which are sensitive to both the heavy neutrino mixings and the properties of the heavy neutral states. At the same time, the double seesaw mechanism naturally accounts for the smallness of light neutrino masses while keeping the intermediate-scale states $N_R$ and $S_L$ phenomenologically relevant. These features provide testable predictions for current and future experiments, thereby offering valuable insight into the origin of neutrino mass within the left-right symmetric framework.

\subsection{Gauge boson masses and mixing}\label{subsec:gauge_boson_masses}

The extended gauge group of the LRSM (eq.~\ref{GLR}) introduces additional gauge bosons associated with the $SU(2)_R$ and $U(1)_{B-L}$ sectors: the right-handed charged gauge bosons $W_R^\pm$ and an extra neutral gauge boson $Z'$. These new gauge bosons are characteristic features of the LRSM, and their masses are determined by the scale of right-handed symmetry breaking, denoted by $v_R$, which is typically much larger than the electroweak scale. In this subsection, we briefly present the analytical expressions for the masses of $W_R$ and $Z'$, as well as the $W_L$–$W_R$ mixing angle ($\xi$).

The physical mass of the right-handed charged gauge boson is given by
\begin{equation}\label{massWR}
	M_{W_R} \simeq \frac{1}{2} g_R v_R.
\end{equation} 

The mass of the additional neutral gauge boson $Z'$ is approximately
\begin{equation}\label{massZ'}
	M_{Z'} \simeq \frac{\sqrt{g_{BL}^2 + g_R^2}}{g_R} M_{W_R} \simeq 1.2 M_{W_R}.
\end{equation}

Finally, the mixing angle between $W_L$ and $W_R$ is given by
\begin{equation}\label{mixingWLWR}
	\begin{aligned}
		\tan 2\xi &\approx 8 \frac{g_L}{g_R} \frac{v_1 v_2}{v_R^2}, \\
		\Rightarrow \quad \xi &\simeq \frac{v_1 v_2 g_R^2}{M_{W_R}^2},
	\end{aligned}
\end{equation}
where $v_R$ is the vacuum expectation value (VEV) of the neutral component of the right-handed Higgs doublet $H_R$, i.e., $\langle H^0_R \rangle \equiv v_R$. Here, we take $g_L = g_R \approx 0.632$, and $v_1$, $v_2$ denote the VEVs of the neutral components of the bidoublet $\Phi$.

Recent experimental searches by both \textsf{ATLAS} and \textsf{CMS} collaborations have placed stringent constraints on the masses of $W_R$, $Z'$, and the heavy Majorana neutrinos in the context of the LRSM. The \textsf{ATLAS} collaboration~\cite{ATLAS:2023cjo} reports that for Majorana $N_R$ with $m_{N_R} < 1$~TeV, the lower bound on the right-handed gauge boson mass reaches
\begin{equation}
	m_{W_R} \gtrsim 6.4~\text{TeV}.
\end{equation}
Additionally, for $m_{W_R} = 4.8$~TeV, Majorana neutrinos are excluded up to
\begin{equation}
	m_{N_R} < 
	\begin{cases}
		3.5~\text{TeV} & \text{(electron channel)} \\
		3.6~\text{TeV} & \text{(muon channel)}.
	\end{cases}
\end{equation}
Furthermore, in the theoretical framework of a left-right symmetric model, the \textsf{CMS} Collaboration~\cite{CMS:2023ooo} excludes $Z'$ bosons with masses up to approximately 4.42 TeV, assuming dominant decays into pairs of heavy Majorana neutrinos. These represent the most stringent exclusion limits to date on the $m_{W_R}$-$m_{N_R}$ and $M_{Z'}$ parameter spaces under the assumptions of the LRSM.

In summary, we have outlined the analytical expressions for the gauge boson masses and reviewed the relevant experimental bounds. In the forthcoming phenomenological analysis, we will adopt benchmark values consistent with these constraints. Before that, we proceed to examine the Dirac mass terms and neutrino mixing structure in the next section.

\section{Dirac Mass Terms and Neutrino Mixing}\label{DMT_NM}
 The new physics contributions to charged lepton flavor violation (cLFV), such as the radiative decay $\mu \to e \gamma$, and neutrinoless double beta decay ($0\nu\beta\beta$) arise from the masses and mixings of the additional neutral fermions present in the considered model framework. These mixings can be further expressed in terms of the Dirac Yukawa matrices or, equivalently, the Dirac mass terms. 
 
 The Dirac mass terms $M_D$ and $M_{RS}$, as well as the Majorana mass term $M_S$, in eq.~(\ref{massLag}) arise from the following interaction Lagrangian:
 \begin{eqnarray}\label{LRDSM}
 	\begin{aligned}
 		\mathcal{L}_{\text{int}} &= -\mathcal{L}_{M_D} - \mathcal{L}_{M_{RS}} - \mathcal{L}_{M_S} \\
 		&= -\sum_{\alpha, \beta} \overline{\nu_{\alpha L}} [M_D]_{\alpha \beta} N_{\beta R}
 		-\sum_{\alpha, \beta} \overline{S_{\alpha L}} [M_{RS}]_{\alpha \beta} N_{\beta R}
 		- \frac{1}{2} \sum_{\alpha, \beta} \overline{S^c_{\alpha R}} [M_S]_{\alpha \beta} S_{\beta L}
 		+ \text{h.c.},
 	\end{aligned}
 \end{eqnarray}
 where the sums run over the three generations $\alpha, \beta = 1,2,3$. Various textures for the Dirac mass matrix have been explored in the literature for studies of $0\nu\beta\beta$ and cLFV. Among these, the orthogonal parametrization (also known as the Casas-Ibarra parametrization \cite{Casas:2001sr}) is widely used in scenarios where the neutrino masses arise via the Type-I seesaw mechanism. In this approach, the Dirac mass matrix is written as $M_D = i\, U_{\nu} \hat{m}_\nu^{1/2} R^T \hat{m}_N^{1/2} U_N^T$, where $R$ is a complex orthogonal matrix satisfying $RR^T = R^T R = 1$, and $\hat{m}_\nu = \mathrm{diag}(m_1, m_2, m_3)$, $\hat{m}_N = \mathrm{diag}(m_{N_1}, m_{N_2}, m_{N_3})$ are diagonal matrices of light and heavy neutrino masses, respectively. Here, $U_\nu$ denotes the PMNS matrix, and $U_N$ is the matrix that diagonalizes the right-handed neutrino mass matrix. This parametrization is particularly useful for scanning the full allowed parameter space \cite{Ibarra:2011xn,Deppisch:2005zm}. Another widely used texture for the Dirac mass matrix assumes it to be proportional to the identity matrix \cite{Bambhaniya:2015ipg,Dash:2021pbx,Patra:2023ltl,Pritimita:2016fgr}, which can, in principle, be motivated by certain flavor symmetries \cite{Hagedorn:2014wha,Altarelli:2004za}. Additionally, the Dirac neutrino mass matrix is often taken to be equal or proportional to the up-type quark mass matrix, as motivated by grand unified theories (GUTs) \cite{Joshipura:2001ya,Joshipura:1999is,Pritimita:2016fgr,Bonilla:2016fqd}. The Dirac mass matrix structure can also be motivated within $SO(10)$ GUT frameworks \cite{Dash:2021pbx,Awasthi:2013ff,Dev:2009aw}.
 
 Other parametrizations of the Dirac mass matrix, explored in the context of leptogenesis but also applicable to cLFV and $0\nu\beta\beta$ studies, can be found in Refs.~\cite{Pascoli:2003uh,Branco:2011zb,Joshipura:2001ui,Rahat:2020mio}.
 
In the following subsections, we will explore the cases relevant to our framework where the Dirac mass matrix is taken proportional to the identity matrix, as considered in earlier work \cite{Patra:2023ltl}, and where the Dirac mass matrix is derived from the model framework, as done in \cite{Patel:2023voj}. We will also discuss the corresponding neutrino mixing matrices for each case, as they are essential for cLFV and $0\nu\beta\beta$ studies.

\subsection{Case I: Symmetry-Motivated Dirac Mass Structure}\label{subsec:MDIdentity}
Following ref.~\cite{Patra:2023ltl}, we begin by discussing the symmetry-motivated case where the Dirac mass matrices $M_D$ and $M_{RS}$ (as mentioned in eq.~(\ref{LRDSM})) are considered proportional to the identity matrix:
\begin{equation}\label{MDMRS_I}
	M_D = k_d I, \hspace{1cm} M_{RS} = k_{rs} I.
\end{equation}
Here $k_d$ and $k_{rs}$ are real constants with $|k_d|<|k_{rs}|$. The simultaneous proportionality of the Dirac mass matrices $M_D$ and $M_{RS}$ to the identity matrix implies that $M_D \propto M_{RS}$, with the proportionality constant equal to $k_d/k_{rs}$. This feature, often referred to as the \textit{screening} or \textit{cancellation of Dirac structures}, can naturally arise in the presence of underlying flavor symmetries that enforce degenerate or diagonal Yukawa textures at leading order. In particular, discrete non-Abelian symmetries such as $A_4$~\cite{Altarelli:2005yx,Altarelli:2006kg}, $S_4$~\cite{Hagedorn:2006ug,Lam:2008rs}, $T'$~\cite{Feruglio:2007uu}, and $\Delta(27)$~\cite{Bhattacharyya:2012pi} can yield Dirac or Yukawa mass matrices proportional to the identity matrix through specific triplet-triplet contractions and vacuum alignments of the associated flavon fields. Symmetry-motivated realizations of proportional Dirac structures have also been discussed in the context of the double seesaw mechanism in refs.~\cite{Lindner:2005pk,Ludl:2015tha,Smirnov:2018luj}.

With the structure of Dirac mass matrices considered in eq.~(\ref{MDMRS_I}), we simplify the neutrino mass matrices from double seesaw relations mentioned in eq.~(\ref{MajMasses}) to get the relations between light neutrino Majorana mass matrix ($m_\nu$), right-handed neutrino Majorana mass matrix ($m_N$) and sterile neutrino Majorana mass matrix ($m_S$):
\begin{equation}\label{MnuMnMs}
		m_\nu =\frac{k_d^2}{k^2_{rs}} m_S, \hspace{1cm}
		m_N=-k_d^2 m_\nu^{-1}.
\end{equation}
Considering the choice of basis in which charged lepton mass matrix is diagonal, we diagonalize the light neutrino Majorana mass matrix with the Pontecorvo-Maki-Nakagawa-Sakata (PMNS) mixing matrix $U_{\text{PMNS}}\equiv U_\nu$ \cite{Pontecorvo:1957qd,Maki:1962mu,Pontecorvo:1967fh} as
\begin{equation}\label{MnuCap}
	\hat{m}_\nu\equiv \text{diag}(m_1,m_2,m_3)=U_\nu^\dagger m_\nu U_\nu^*.
\end{equation}
It is evident from eq.~(\ref{MnuMnMs}) that light and sterile neutrino Majorana mass matrices have the same structure due to the cancellation of the Dirac structures. Therefore, if $U_S$ is the unitary mixing matrix which diagonalizes the sterile Majorana mass matrix ($	\hat{m}_S\equiv \text{diag}(m_{S_1},m_{S_2},m_{S_3})=U_S^\dagger m_S U_S^*$), then it can be equal to the PMNS mixing matrix which diagonalize the light neutrino Majorana mass matrix:
\begin{equation}\label{UsUnu}
	U_S=U_\nu
\end{equation}
Given that the eigenvalues of $m_\nu$ are positive, the presence of a negative factor in the relation $m_N = -k_d^2 m_\nu^{-1}$ implies that the unitary matrix $U_N$ which diagonalizes $m_N$  ($\hat{m}_N\equiv \text{diag}(m_{N_1},m_{N_2},m_{N_3})=U_N^\dagger m_N U_N^*$) must satisfy:
\begin{equation}\label{UNUnu}
	U_N=i U_\nu^*
\end{equation}
Thus, for the considered scenario, the physical Majorana masses of light ($m_i$), right-handed ($m_{N_i}$) and sterile ($m_{S_i}$) neutrinos are related as follows:
\begin{equation}\label{phyMnu}
	m_i= \frac{k_d^2}{m_{N_i}}=\frac{k_d^2}{k_{rs}^2}m_{S_i}, \hspace{1cm} i=1,2,3.
\end{equation}
Throughout our analysis, we will use the standard parametrization of the PMNS matrix \cite{ParticleDataGroup:2018ovx}:
\begin{eqnarray}\label{UPMNS}
	\begin{aligned}
		U_{\rm {PMNS}}&\equiv U_{\nu}\\
		&=	\begin{pmatrix} c_{13}c_{12}&c_{13}s_{12}&s_{13}e^{-i\delta}\\
			-c_{23}s_{12}-c_{12}s_{13}s_{23}e^{i\delta}&c_{12}c_{23}-s_{12}s_{13}s_{23}e^{i\delta}&s_{23}c_{13}\\
			s_{12}s_{23}-c_{12}c_{23}s_{13}e^{i\delta}&-c_{12}s_{23}-s_{12}s_{13}c_{23}e^{i\delta}&c_{13}c_{23}
		\end{pmatrix}\underbrace{\begin{pmatrix}
				1 &0&0\\
				0 &e^{i\alpha/2}&0\\
				0 &0&e^{i\beta/2}
		\end{pmatrix}}_{\text{Majorana phase matrix}}
	\end{aligned}
\end{eqnarray}
where $\delta$ is the Dirac CP-violating phase ($0 \leq \delta \leq 2\pi$), and $\alpha, \beta$ are the Majorana CP-violating phases ($0 \leq \alpha, \beta \leq 2\pi$). The parameters $c_{ij} = \cos\theta_{ij}$ and $s_{ij} = \sin\theta_{ij}$ denote the mixing angles. The experimentally measured values of the neutrino oscillation parameters for the active neutrino mass spectra with normal ordering (NO) and inverted ordering (IO) are presented in the table~\ref{tab:exp_nu}.

\begin{table}
	\centering
\begin{tabular}{|c|cccc|}
	\hline Parameter & best-fit $\pm 1 \sigma$ & best-fit $\pm 1 \sigma$ & $3 \sigma$ range & $3 \sigma$ range \\
	& NO & IO & NO & IO\\
	\hline 
	&&&&\\[-8pt]
	$\Delta m_{21}\left[10^{-5} \mathrm{eV}^2\right]$  & $7.49^{+0.19}_{-0.19}$ & $7.49^{+0.19}_{-0.19}$ & $6.92-8.05$ & $6.92-8.05$ \\[5pt]
	$\left|\Delta m_{31}^2\right|\left[10^{-3} \mathrm{eV}^2\right]$ & $2.513^{+0.021}_{-0.019}$ & $-2.484^{+0.020}_{-0.020}$ & $2.451-2.578$ & $-(2.547 -2.421)$ \\[5pt]
	$\sin ^2 \theta_{12} / 10^{-1}$ & $3.08^{+0.12}_{-0.11}$ & $3.08^{+0.12}_{-0.11}$ & $2.75-3.45$ & $2.75-3.45$ \\[5pt]
	$\theta_{12} /^{\circ}$ & $33.68^{+0.73}_{-0.70}$ & $33.68^{+0.73}_{-0.70}$ & $31.63-35.95$ & $31.63-35.95$ \\[5pt]
	$\sin ^2 \theta_{23} / 10^{-1}$ & $4.70^{+0.17}_{-0.13}$ & $5.50^{+0.12}_{-0.15}$ & $4.35-5.85$ & $4.40-5.84$ \\[5pt]
	$\theta_{23} /^{\circ}$ & $43.3^{+1.0}_{-0.8}$ & $47.9^{+0.7}_{-0.8}$ & $41.3-49.9$ & $41.5-49.8$ \\[5pt]
	$\sin ^2 \theta_{13} / 10^{-2}$ & $2.215_{-0.056}^{+0.058}$ & $2.231_{-0.056}^{+0.056}$ & $2.030-2.388$ & $2.060-2.409$ \\[5pt]
	$\theta_{13} /^{\circ}$ & $8.56_{-0.11}^{+0.11}$ & $8.59_{-0.11}^{+0.11}$ & $8.19-8.89$ & $8.25-8.93$ \\[5pt]
	$\delta /^{\circ}$ & $212_{-41}^{+26}$ & $274_{-25}^{+22}$ & $124-364$ & $201-335$ \\[5pt]
	\hline
\end{tabular}
	\caption{The current updated estimates of experimental values of
	neutrino oscillation parameters for global best-fits and for $1\sigma$ to $3\sigma$ ranges taken from NuFITv-6.0~\cite{Esteban:2024eli}.}
	\label{tab:exp_nu}
\end{table}

The physical light neutrino masses ($m_i$) can be expressed in terms of the lightest neutrino mass $m_{1(3)}$ for normal ordering (NO) and inverted ordering (IO) mass spectrum of neutrinos, respectively. In the case of NO mass spectrum of light neutrinos ($m_1<m_2<m_3$), we have:
	\begin{equation}\label{NOlightneutrino}
	\begin{aligned}
		& m_1=\text { lightest neutrino mass }\\
		& m_2=\sqrt{m_1^2+\Delta m_{\mathrm{sol}}^2}, \\
		& m_3=\sqrt{m_1^2+\Delta m_{\mathrm{atm}}^2}.
	\end{aligned}
\end{equation}
Following eq.~(\ref{phyMnu}), the relations between physical masses of light and sterile neutrinos and between right-handed and light neutrinos become:
	\begin{eqnarray}\label{mimSi_NO}
	m_{S_1} = \frac{m_{1}}{m_{3}} m_{S_3}\,, \quad 
	m_{S_2} = \frac{m_{2}}{m_{3}} m_{S_3}\,, \quad  m_{S_1} <  m_{S_2} < m_{S_3}\,
\end{eqnarray}
and
	\begin{eqnarray}\label{mimNi_NO}
	m_{N_2} = \frac{m_{1}}{m_{2}} m_{N_1}\,, \quad 
	m_{N_3} = \frac{m_{1}}{m_{3}} m_{N_1}\,, \quad  m_{N_3} <  m_{N_2} < m_{N_1}\,
\end{eqnarray}
respectively in the NO mass spectrum of light neutrinos. Similarly, for IO mass spectrum of light neutrinos ($m_3<m_1<m_2$), we have:
\begin{equation}\label{IOlightneutrino}
	\begin{aligned}
		& m_3=\text { lightest neutrino mass }\\
		& m_1=\sqrt{m_3^2+\Delta m_{\mathrm{atm}}^2},\\
		& m_2=\sqrt{m_3^2+\Delta m_{\mathrm{sol}}^2+\Delta m_{\mathrm{atm}}^2}.
	\end{aligned}
\end{equation}
From eq.~(\ref{phyMnu}), the relations between physical masses of light and sterile neutrinos and between right-handed and light neutrinos become:
	\begin{eqnarray}\label{mimSi_IO}
	m_{S_1} = \frac{m_{1}}{m_{2}} m_{S_2}\,, \quad
	m_{S_3} = \frac{m_{3}}{m_{2}} m_{S_2}\,, \quad  m_{S_3} <  m_{S_1} < m_{S_2}\,
	\label{eqn:MSe2} 
\end{eqnarray}
and
\begin{eqnarray}\label{mimNi_IO}
	m_{N_1} = \frac{m_{3}}{m_{1}} m_{N_3}\,, \quad 
	m_{N_2} = \frac{m_{3}}{m_{2}} m_{N_3}\,, \quad  m_{N_2} <  m_{N_1} < m_{N_3}\,
\end{eqnarray}
respectively in the IO mass spectrum of light neutrinos. For both orderings, we have $\Delta m_{\mathrm{sol}}^2=\Delta m_{21}^2$ and $\Delta m_{\mathrm{atm}}^2=\left|\Delta m_{31}^2\right|$ \cite{deSalas:2020pgw}.

From eq.~(\ref{massLag}) we have $9\times9$ neutral fermion mass matrix in the flavor basis of $(\nu_L, N_R^c, S_L)$:
\begin{equation}\label{Mlrdsm}
	\mathcal{M}_{\text{LRDSM}}= 	\begin{bmatrix}
		0 & M_D & 0 \\
		M_D^T & 0 & M_{RS} \\
		0 & M_{RS}^T & M_S
	\end{bmatrix}
\end{equation}
The diagonalization of $\mathcal{M}_{\text{LRDSM}}$ is done by a generalized unitary transformation to get physical masses of neutral fermions as:
\begin{equation}
	V^\dagger \mathcal{M}_{\text{LRDSM}} V^* = \hat{\mathcal{M}}_{\text{LRDSM}}=\text{diag}(m_i,m_{N_i},m_{S_i}),\hspace{0.5 cm} i,j,k=1,2,3.
\end{equation}
It also leads to the following relation between neutral fermions' weak eigenstate and mass eigenstate:
\begin{equation}
	\ket{\Psi}_{\text{flavor}}=V \ket{\Psi}_{\text{mass}}
\end{equation}
where, $\ket{\Psi}_{\text{flavor}}=(\nu_{\alpha L}\,N^c_{\beta R}\,S_{\gamma L})^T$, $\ket{\Psi}_{\text{mass}}=(\nu_{i}\,N^c_{j}\,S_{k})^T$ and 
\begin{equation}\label{NeutrinoMixMat}
	V=\begin{bmatrix}
		V^{\nu\nu}_{\alpha i} & V^{\nu N}_{\alpha j} & V^{\nu S}_{\alpha k} \\
		V^{N\nu}_{\beta i} & V^{NN}_{\beta j} & V^{NS}_{\beta k} \\
		V^{S\nu}_{\gamma i} & 	V^{SN}_{\gamma j} & 	V^{SS}_{\gamma k} 
	\end{bmatrix}\text{  (Neutrino Mixing Matrix)}.
\end{equation}
Here, the indices $\alpha,\beta,\gamma$ run over three generations of light, right-handed and sterile neutrinos, respectively, whereas the indices $i,j,k$ run over corresponding mass states.

The neutrino mixing matrix defined in eq.~(\ref{NeutrinoMixMat}) is a $9\times9$ unitary matrix, resulted by matrix multiplication of three other $9\times9$ unitary matrices: two unitary transformation matrices which block diagonalizes the $\mathcal{M}_{\text{LRDSM}}$ (eq.~(\ref{Mlrdsm})) and a unitary matrix which further diagonalize the block diagonalized neutral fermion mass matrix to $9\times9$ diagonal matrix, consisting mass eigenvalues of light, right-handed and sterile neutrinos as diagonal elements. The derivation of unitary transformation matrices follows a general diagonalization procedure as discussed in \cite{Grimus:2000vj}. For phenomenological relevance, we here directly present the analytical form of the neutrino mixing matrix, defined in eq.~(\ref{NeutrinoMixMat}) (for detailed derivation, one can refer to appendix A of \cite{Patra:2023ltl}):
\begin{equation}\label{VmixForm}
	V=\begin{bmatrix}
		(1-\frac{1}{2}XX^\dagger)U_\nu &\quad XU_N & 0\\
		-(1-\frac{1}{2}YY^\dagger)X^\dagger U_\nu &\quad (1-\frac{1}{2}YY^\dagger)(1-\frac{1}{2}XX^\dagger)U_N &YU_S\\
		Y^\dagger X^\dagger U_\nu & \quad-Y^\dagger U_N & (1-\frac{1}{2}YY^\dagger)U_S
	\end{bmatrix},
\end{equation}
where  $X^\dagger=-{M_{RS}^{-1}}^TM_SM_{RS}^{-1}M_D^T$, $\,Y=M_{RS}M_S^{-1}$.

For our phenomenological analysis of cLFV and $0\nu\beta\beta$ in LRSM with the double seesaw mechanism, the neutrino mixing submatrices relevant to us in eq.~(\ref{NeutrinoMixMat}) are \cite{Patra:2023ltl}:

\textbf{Self-mixing of right-handed neutrinos:}
\begin{equation}
	V^{NN} = \left(1 - \frac{1}{2} YY^\dagger \right) \left(1 - \frac{1}{2} XX^\dagger \right) U_N \approx U_N,
\end{equation}

\textbf{Mixing between right-handed and sterile neutrinos:}
\begin{equation}
	V^{NS} = Y U_S = M_{RS} M_S^{-1} U_S.
\end{equation}

In the current scenario, where the Dirac mass matrices are proportional to the identity matrix (eq.~ (\ref{MDMRS_I})), we obtain the following from eqs. (\ref{MnuCap}), (\ref{UNUnu}), and (\ref{UsUnu}):

\begin{eqnarray}
	V^{NN} &=& i U_{\nu}^*,\\
	V^{NS} &=& k_{rs} U_\nu^* \, \text{diag}(1/m_{S_1}, 1/m_{S_2}, 1/m_{S_3}).
\end{eqnarray}

It is worthwhile to note that in ref. \cite{Patra:2023ltl}, the phenomenological discussion was carried out assuming discrete LR symmetry as Parity. However, since the Yukawa matrices are considered real and diagonal, this choice does not affect the results in the current context. In the next subsection, where we discuss the case in which the Dirac mass terms are derived, the choice of LR symmetry plays a significant role.

Before proceeding to \textbf{Case II}, we summarize the relevant results obtained for \textbf{Case I}, where the Dirac mass matrices are proportional to the identity matrix, in Table~\ref{tab:case1} for the reader's convenience.

\begin{table}
	\centering
	\renewcommand{\arraystretch}{1.4}
	\begin{tabular}{|c|c|c|}
		\hline
		\multicolumn{3}{|c|}{\textbf{Case I}} \\
		\hline
		\textbf{Dirac masses} & $M_D =k_d I$ & $M_{RS}=k_{rs} I$ \\
		\hline
		\textbf{$U_N$, $U_S$ relation} & $U_N = i U_\nu^*$ & $U_S = U_\nu$ \\
		\hline
		\multicolumn{1}{|c|}{$V^{NN}$} & \multicolumn{2}{c|}{\centering $i U_\nu^*$} \\
		\hline
			\multicolumn{1}{|c|}{$V^{NS}$} & \multicolumn{2}{c|}{\centering $k_{rs} U_\nu^* \, \text{diag}(1/m_{S_1}, 1/m_{S_2}, 1/m_{S_3})$} \\
		\hline
		& \textbf{NO} & \textbf{IO} \\
		\hline
		\textbf{$m_i$-$m_{S_i}$ relation} & 
		$m_{S_1} = \frac{m_1}{m_3} m_{S_3}$; \quad $m_{S_2} = \frac{m_2}{m_3} m_{S_3}$ & 
		$m_{S_1} = \frac{m_1}{m_2} m_{S_2}$; \quad $m_{S_3} = \frac{m_3}{m_2} m_{S_2}$ \\
		\hline
			\textbf{$m_i$-$m_{N_i}$ relation} & 
		$m_{N_2} = \frac{m_{1}}{m_{2}} m_{N_1}$; \quad $m_{N_3} = \frac{m_{1}}{m_{3}} m_{N_1}$ & 
		$m_{N_1} = \frac{m_{3}}{m_{1}} m_{N_3}$; \quad $m_{N_2} = \frac{m_{3}}{m_{2}} m_{N_3}$ \\
		\hline
	\end{tabular}
	\caption{Summary of relevant relations in Case I where Dirac mass matrices are proportional to identity. Here $k_{rs}=\sqrt{m_{N_i}m_{S_i}}$, $i=1,2,3$.}
	\label{tab:case1}
\end{table}

\subsection{Case II: Model-Determined Dirac Mass Structure}\label{subsec:derivedMD}

Here we adopt the approach of ref. \cite{Patel:2023voj}, which derives $M_D$ and $M_{RS}$ (see eq.~(\ref{LRDSM})) from the underlying model dynamics. We briefly discuss this scenario in reference to the preceding analysis, highlighting the differences that arise when the Dirac masses are derived and more general.

We start with the screening (cancellation) of Dirac structures, that is, $M_D$ being proportional to $M_{RS}$ so that light and sterile neutrino mass matrices have the same structure:
\begin{equation}\label{MDMRSprop}
	M_D =\frac{1}{k}M_{RS}^T.
\end{equation}
Here, $1/k$ is the proportionality constant, which is real. Plugging eq.~(\ref{MDMRSprop}) into the relation for light neutrino mass matrix in eq.~(\ref{MajMasses}) and simplifying, we get the relation between light neutrino and sterile neutrino masses as $m_S= k^2 m_\nu$. Thus, the PMNS matrix ($U_{\text{PMNS}}\equiv U_\nu$) again diagonalizes both light and sterile neutrino mass matrices:
\begin{equation}\label{USUnuDMD}
	U_S=U_\nu.
\end{equation}
The diagonalization of $m_\nu$ and $m_S$ happens in the same way as mentioned in eq.~(\ref{MnuCap}). The physical mass relation between sterile and light neutrinos becomes:
\begin{equation}\label{phyMnuMs}
	m_{S_i}=k^2 m_i, \quad i=1,2,3.
\end{equation}
Similar to case I, we again work in the basis where the charged lepton mass matrix is diagonal, and the right-handed neutrino Majorana mass matrix ($m_N$) is diagonalized by a unitary matrix $U_N$ as $\hat{m}_N=U_N^\dagger m_N U_N^*$.

Interestingly, unlike the previous case, a direct connection between the unitary matrices, $U_\nu$ and $U_N$, that diagonalizes light and right-handed Majorana mass matrices, respectively, is not straightforward. As Dirac mass matrices ($M_D$ or $M_{RS}$) are not proportional to identity, there is no direct correspondence between $m_N$ and $m_S$ (see eq.~(\ref{MajMasses})) or between $m_N$ and $m_\nu$ (see eq.~(\ref{phyMnuMs})). However, a connection can be established with an appropriate choice of discrete LR symmetry in the model. Noting that $m_N$ can be diagonalized by $U_N$ as $\hat{m}_N=U_N^\dagger m_N U_N^*$ and $m_N=-M_{RS} M^{-1}_S M^T_{RS}$ from eq.~(\ref{MajMasses}), we have:
\begin{equation}
	\hat{m}_N=-U_N^\dagger M_{RS} M^{-1}_S M^T_{RS} U_N^*.
\end{equation}
Inverting the relation $\hat{m}_S=U_\nu^\dagger M_S U_\nu^*$ to express $M^{-1}_S$, and substituting into the expression for $\hat{m}_N$, we obtain:
\begin{equation}\label{mnCapDmD}
	\begin{aligned}
		\hat{m}_N&=-U_N^\dagger M_{RS}U_\nu^*{\hat{m}_S}^{-1} U_\nu^\dagger M^T_{RS} U_N^*\\
		&=-U_N^\dagger M_{RS}U_\nu^*{\hat{m}_S}^{-1}(U_N^\dagger M_{RS}U_\nu^*)^T.
	\end{aligned}
\end{equation}
The left-hand side of eq.~(\ref{mnCapDmD}) is diagonal; therefore, the right-hand side must also be diagonal for consistency. Given that \( \hat{m}_S^{-1} \) is diagonal by construction, one possible implication is that \( U_N^\dagger M_{RS} U_\nu^* \) is diagonal\footnote{There might be other solutions, but these would complicate the structure of both $U_N$ and $M_{RS}$.}:
\begin{equation}\label{diagMRS}
	U_N^\dagger M_{RS} U_\nu^* = \hat{M}_{RS}.
\end{equation}

Equation (\ref{diagMRS}) implies that the Dirac mass matrix \( M_{RS} \) is diagonalized by a biunitary transformation. If \( M_{RS} \) is complex symmetric, then this diagonalization requires (see \cite{Giunti:2007ry}, Section 6.7.1):
\begin{equation}\label{UnUnuDmD}
	U_N = U_\nu.
\end{equation}

We choose charge conjugation (\( \mathcal{C} \)) as the discrete LR symmetry to make the Dirac matrices \( M_D \) and \( M_{RS} \) symmetric (see refs.~\cite{Patel:2023voj,Nemevsek:2012iq}) which leads to a simple relation between \( U_N \) and \( U_\nu \). From eq.~(\ref{MajMasses}), we can rewrite the right-handed Majorana mass ($m_N$) with the consideration that $M_{RS}$ is symmetric as:
\begin{equation}\label{mNsymm}
	m_N=-M_{RS} M^{-1}_S M_{RS}.
\end{equation}
By rearranging terms in eq.~(\ref{mNsymm}) and using eqs.~(\ref{MDMRSprop}), (\ref{phyMnuMs}), and (\ref{UnUnuDmD}), we obtain the analytical expressions for the derived Dirac mass matrices \( M_{RS} \) and \( M_D \) as:
\begin{equation}\label{DiracMassesDmD}
	\begin{aligned}
		M_{RS} &= i k\, U_\nu \sqrt{\hat{m}_\nu \hat{m}_N}\, U_\nu^T,\\
		M_D &= i\, U_\nu \sqrt{\hat{m}_\nu \hat{m}_N}\, U_\nu^T.
	\end{aligned}
\end{equation}
Here, \( k = \sqrt{m_{S_i}/m_i} \) with \( i = 1,2,3 \) (see eq.~(\ref{phyMnuMs})). For a detailed derivation of the Dirac masses in eq.~(\ref{DiracMassesDmD}), see Appendix B of Ref.~\cite{Patel:2023voj}. We emphasize that the Dirac masses in eq.~(\ref{DiracMassesDmD}) are free from arbitrariness, similar to the previous case where they were considered proportional to the identity matrix. However, they are now more general, emerging from the underlying model dynamics, and are directly dependent on the light neutrino masses and mixing parameters. This dependence allows for a more predictive and constrained phenomenological analysis, particularly for processes such as neutrinoless double beta decay (\(0\nu\beta\beta\)) and charged lepton flavor violation (cLFV).

For this case (Case II) as well, we adopt the standard parametrization of the PMNS mixing matrix as defined in eq.~(\ref{UPMNS}). The light neutrino mass spectra for the normal ordering (NO) and inverted ordering (IO) scenarios are taken as defined earlier in eq.~(\ref{NOlightneutrino}) and eq.~(\ref{IOlightneutrino}), respectively. Since the relation between the light and sterile neutrino Majorana masses remains similar in both cases (see eq.~(\ref{phyMnuMs}) and eq.~(\ref{phyMnu})), the corresponding physical mass relations also remain unchanged for the NO and IO scenarios, as given in eq.~(\ref{mimSi_NO}) and eq.~(\ref{mimSi_IO}).

In contrast to the previous case, where the physical masses of the right-handed neutrinos were related to those of the light neutrinos (see eq.~(\ref{mimNi_NO}) and eq.~(\ref{mimNi_IO})), we now have this scenario in which the right-handed neutrino masses are treated as input parameters. As a result, their mass spectrum is independent of the NO and IO structure of the light neutrino masses.

The analytical form of the neutrino mixing matrix, as presented in eq.~(\ref{VmixForm}), remains unchanged in this case. However, the relevant neutrino mixing submatrices—specifically, \( V^{NN} \) and \( V^{NS} \)—now have different structures, given by:

\textbf{Self-mixing of right-handed neutrinos:}
\begin{equation}
	V^{NN} \approx U_N = U_\nu,
\end{equation}

\textbf{Mixing between right-handed and sterile neutrinos:}
\begin{equation}
	V^{NS} = M_{RS} M_S^{-1} U_S = \frac{i}{k} U_\nu \sqrt{\hat{m}_\nu^{-1} \hat{m}_N}.
\end{equation}

Since the present scenario does not have any specific hierarchy among the right-handed neutrino masses, we remain open to all six possible hierarchical arrangements. These are categorized and listed in Table~\ref{tab:RHN_hierarchy_cases} for clarity.

\begin{table}
	\centering
	\begin{tabular}{ll}
		\toprule
		\textbf{Case} &\hspace{1cm} \textbf{Hierarchy of RHN Masses} \\
		\midrule
		Case $m_{N_A}$  &\hspace{2.5cm} $m_{N_1} < m_{N_2} < m_{N_3}$ \\
		\hline
		Case $m_{N_B}$ &\hspace{2.5cm} $m_{N_2} < m_{N_1} < m_{N_3}$ \\
		\hline
		Case $m_{N_C}$ & \hspace{2.5cm}$m_{N_1} < m_{N_3} < m_{N_2}$ \\
		\hline
		Case $m_{N_D}$& \hspace{2.5cm} $m_{N_3} < m_{N_1} < m_{N_2}$\\
		\hline
		Case $m_{N_E}$ & \hspace{2.5cm}$m_{N_2} < m_{N_3} < m_{N_1}$ \\
		\hline
		Case $m_{N_F}$ &\hspace{2.5cm} $m_{N_3} < m_{N_2} < m_{N_1}$ \\
		\hline
		Case $m_{N_G}$ &\hspace{2.5cm} $m_{N_1} = m_{N_2} = m_{N_3}$ \\
		\bottomrule
	\end{tabular}
	\caption{Possible hierarchy patterns for the right-handed neutrino masses.}
	\label{tab:RHN_hierarchy_cases}
\end{table}

Before proceeding to the next section, we summarize the relevant results obtained for \textbf{Case II} in Table~\ref{tab:case2}.

\begin{table}
	\centering
	\renewcommand{\arraystretch}{1.4}
	\begin{tabular}{|c|c|c|}
		\hline
		\multicolumn{3}{|c|}{\textbf{Case II}} \\
		\hline
		\textbf{Dirac masses} & $M_D =i\, U_\nu \sqrt{\hat{m}_\nu \hat{m}_N}\, U_\nu^T$ & $M_{RS}=i k\, U_\nu \sqrt{\hat{m}_\nu \hat{m}_N}\, U_\nu^T$ \\
		\hline
		\textbf{$U_N$, $U_S$ relation} & $U_N =  U_\nu$ & $U_S = U_\nu$ \\
		\hline
		\multicolumn{1}{|c|}{$V^{NN}$} & \multicolumn{2}{c|}{\centering $U_\nu$} \\
		\hline
		\multicolumn{1}{|c|}{$V^{NS}$} & \multicolumn{2}{c|}{\centering $\frac{i}{k}U_\nu\sqrt{{\hat{m}_\nu}^{-1} \hat{m}_N}$} \\
		\hline
		& \textbf{NO} & \textbf{IO} \\
		\hline
		\textbf{$m_i$-$m_{S_i}$ relation} & 
		$m_{S_1} = \frac{m_1}{m_3} m_{S_3}$; \quad $m_{S_2} = \frac{m_2}{m_3} m_{S_3}$ & 
		$m_{S_1} = \frac{m_1}{m_2} m_{S_2}$; \quad $m_{S_3} = \frac{m_3}{m_2} m_{S_2}$ \\
		\hline
	\end{tabular}
	\caption{Summary of relevant relations in Case II where Dirac mass matrices are derived. Here $k = \sqrt{m_{S_i}/m_i}$, $i=1,2,3$.}
	\label{tab:case2}
\end{table}

\section{Model Contributions to LNV and cLFV Decay Processes}
\label{sec:general-cLFV-LNV}

Lepton number violation (LNV) and charged lepton flavor violation (cLFV) serve as powerful low-energy probes of new physics that can reveal the underlying nature of neutrinos. 
Within the framework of the LRSM, both types of processes receive several new contributions due to the presence of heavy Majorana neutrinos and right-handed gauge interactions, thereby providing a direct link between neutrino mass generation and rare decay observables. 
In particular, the coexistence of light, heavy, and sterile Majorana states in the double seesaw realization offers a rich structure where LNV and cLFV processes become inherently correlated through the same set of Yukawa couplings and mass parameters.

In this section, we present the \textit{model-dependent} contributions to two key processes: 
the neutrinoless double beta ($0\nu\beta\beta$) decay, which violates total lepton number by two units, and the radiative decay $\mu\!\to\!e\gamma$, which changes lepton flavor by one unit. 
While the former proceeds dominantly via right-handed currents at tree level, the latter is induced at the one-loop level through diagrams involving the same heavy and sterile neutrinos that participate in the double seesaw mechanism. 
Their simultaneous consideration thus provides a complementary understanding of lepton-number and lepton-flavor violation within a single theoretical setup.

The following subsections present the analytic expressions for these processes in the LRSM with double seesaw. These results are common to all subsequent phenomenological analyses and constitute the theoretical foundation for Sections~5 and~6, where they are evaluated respectively for the symmetry-motivated and model-determined Dirac mass structures.

\subsection{Contribution to $0\nu\beta\beta$}
Neutrinoless double beta decay (\(0\nu\beta\beta\)) is a promising TeV-scale process that can probe the Majorana nature of neutrinos and provide crucial insights into lepton number violation (LNV) beyond the Standard Model (BSM). In the minimal scenario, this decay proceeds via the exchange of light active Majorana neutrinos and is commonly referred to as the standard mass mechanism. However, various BSM frameworks predict additional LNV contributions to the \(0\nu\beta\beta\) amplitude, mediated by heavy Majorana particle exchange, which may dominate over or interfere with the standard mechanism.

Recent experimental progress in \(0\nu\beta\beta\) searches has significantly improved the lower bounds on the half-life (\(T^{0\nu}_{1/2}\)) of this process. A combined analysis of data from \textsf{GERDA}~\cite{GERDA:2020xhi}, \textsf{KamLAND-Zen-400+800}~\cite{KamLAND-Zen:2022tow}, and the early phase of the \textsf{LEGEND-200} experiment~\cite{LEGEND:2021bnm} has established a new observed limit of \(T^{0\nu}_{1/2} > 1.9\times 10^{26}\) years at 90\% C.L., with an anticipated sensitivity reaching up to \(2.3\times 10^{26}\) years for \({}^{76}\text{Ge}\). These results show no indication of a positive signal, thereby placing stringent constraints on both standard and non-standard mechanisms.
In what follows, we briefly present the standard and new physics contributions to the $0\nu\beta\beta$ decay amplitude and rate arising in our model. For a more detailed discussion, one may refer to~\cite{Patra:2023ltl}. The dominant contributions include:
\begin{itemize}
	\item The standard contribution mediated by the exchange of light active Majorana neutrinos \(\nu_i\),
	\item The non-standard contributions from heavy right-handed Majorana neutrinos \(N_i\) (\(i = 1, 2, 3\)) and sterile Majorana neutrinos \(S_i\) (\(i = 1, 2, 3\)).
\end{itemize}

\begin{figure}
	\centering
	\hspace{-0.2 in}\includegraphics[width=0.95\textwidth]{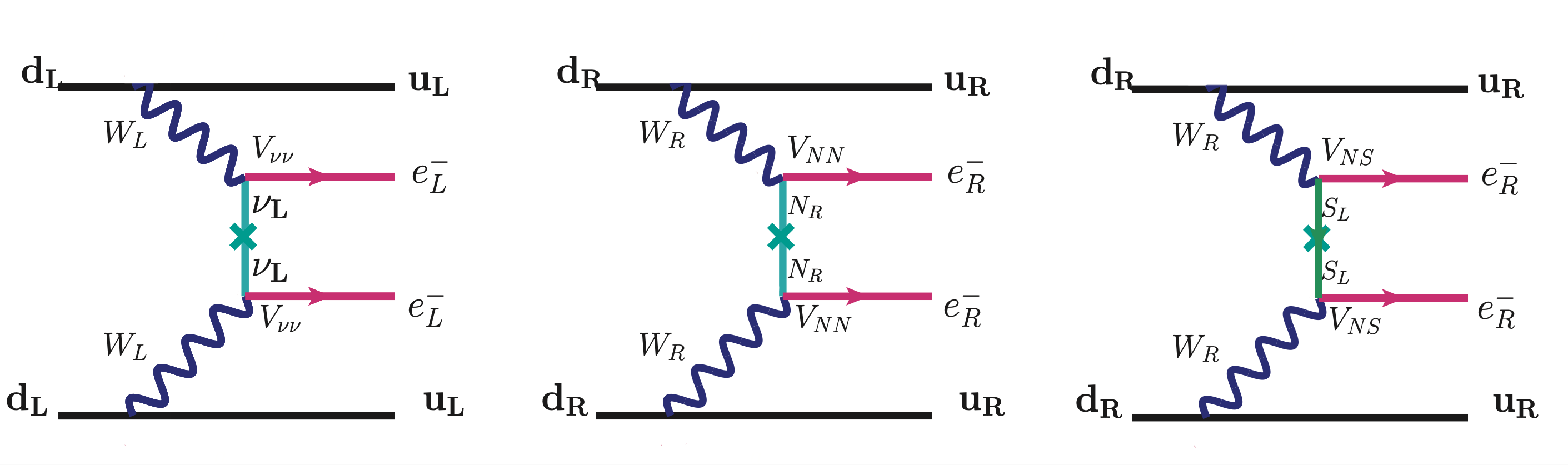}
	\caption{Representative Feynman diagrams contributing to neutrinoless double beta decay: light neutrino exchange (left), heavy right-handed neutrino exchange (middle), and sterile neutrino exchange (right).}
	\label{fig:0nubbFeynDiag}
\end{figure}

The Feynman diagrams corresponding to these dominant contributions to the $0\nu\beta\beta$ decay amplitude are shown in figure~\ref{fig:0nubbFeynDiag}.

The inverse half-life for these processes can be expressed in terms of the generalized effective Majorana mass (GEMM) as:
\begin{equation}\label{invThalf}
	[T^{0\nu}_{1/2}]^{-1}=G_{01}^{0\nu}\Big|\frac{\mathcal{M}_\nu^{0\nu}}{m_e}\Big|^2 |m_{\beta\beta,L,R}^{\text{eff}}|^2,
\end{equation}
where $G_{01}^{0\nu}$ is the phase-space factor, $\mathcal{M}_\nu^{0\nu}$ denotes the nuclear matrix elements (NME) for active neutrino exchange, $m_e$ is the electron mass, and a mildly quenched value of the axial coupling constant ($g_A = 1.00$) is used. The parameter $m_{\beta\beta,L,R}^{\text{eff}}$ appearing in eq.~(\ref{invThalf}) encapsulates the lepton-number violating effects and can be written as an individual sum of amplitudes associated with the exchange of different types of neutrino states:
\begin{equation}\label{Individualmeffective}
	|m_{\beta\beta,L,R}^{\mbox{eff}}|\equiv m^{\nu+N+S}_{ee} = \sqrt{\big( |m_{\beta\beta,L}^{\nu}|^{2} + |m_{\beta\beta,R}^{N}|^{2} + |m_{\beta\beta,R}^{S}|^{2}\big)}.
\end{equation}

However, a coherent sum of amplitudes associated with the exchange of heavy neutrino states $N_{1,2,3}$ and $S_{1,2,3}$ is required to properly account for interference between their contributions. In such a case, the GEMM parameter becomes:
\begin{equation}\label{Interferencemeffective}
	|m_{\beta\beta,L,R}^{\mbox{eff}}|\equiv m^{\nu+|N+S|}_{ee} = \sqrt{\big( |m_{\beta\beta,L}^{\nu}|^{2} + |m_{\beta\beta,R}^{N} + m_{\beta\beta,R}^{S}|^{2}\big)}.
\end{equation}

Here, the effective Majorana mass (EMM) terms for the standard (light neutrino mediated), RHN, and sterile neutrino channels are respectively expressed as:
\begin{equation}\label{meffStandard}
	|m^{\nu}_{\beta\beta,L}|  =\Big|\sum_{i=1}^{3}U^{2}_{ei}m_{i}\Big|=|m_{1}c_{12}^{2}c^{2}_{13}+m_{2}c_{13}^{2}s_{12}^{2}e^{i\alpha}+m_{3}s_{13}^{2}e^{i(\beta-2\delta)}|,
\end{equation}
\begin{equation}\label{meffRHN}
	m_{\beta\beta,R}^N=C_N \sum_{j}\frac{{[V_{ej}^{NN}]}^2}{m_{N_j}},
\end{equation}
\begin{equation}\label{meffSterile}
	m_{\beta\beta,R}^S=C_N \sum_{l}\frac{{[V_{el}^{NS}]}^2}{m_{S_l}}.
\end{equation}

The parameters appearing in the EMM terms in eqs.~(\ref{meffStandard})–(\ref{meffSterile}) are $m_1$, $m_2$, $m_3$, denoting the masses of light Majorana neutrinos $\nu_{1,2,3}$, and $V^{NN}$, $V^{NS}$ are the heavy neutrino mixing matrices as summarized in table~\ref{tab:case1} and table~\ref{tab:case2} for the two Dirac mass matrix structures considered in our work. The dimensionless proportionality factor $C_N$ is given by:
\begin{equation}\label{CN}
	C_N = m_e m_p \, \frac{\mathcal{M}^{0\nu}_N}{\mathcal{M}^{0\nu}_\nu}\frac{M_{W_L}^4}{M_{W_R}^4},
\end{equation}
where $m_e$, $m_p$, $M_{W_L}$, and $M_{W_R}$ are the masses of the electron, proton, $W_L$ boson, and $W_R$ boson respectively. $\mathcal{M}^{0\nu}_\nu$ and $\mathcal{M}^{0\nu}_N$ denote the NMEs corresponding to light and heavy neutrino exchange. The minimum and maximum values of the ratio $\mathcal{M}^{0\nu}_N/\mathcal{M}^{0\nu}_\nu$ for $^{76}$Ge and $^{136}$Xe are summarized in table~\ref{tab:NMEratio}. The values for these two isotopes do not vary significantly; hence, we will restrict ourselves to $^{76}$Ge as the reference in our numerical analysis.

\begin{table}
	\centering
	\renewcommand{\arraystretch}{1.3}
	\begin{minipage}[t]{0.48\textwidth}
		\centering
		\begin{tabular}{|c|c|c|}
			\hline
			\multicolumn{3}{|c|}{\textbf{NME Ratios and $G_{01}^{0\nu}$ for $^{76}$Ge}} \\
			\hline
			\textbf{Method} & $\mathcal{M}^{0\nu}_\nu$ & $\mathcal{M}^{0\nu}_N/\mathcal{M}^{0\nu}_\nu$ \\
			\hline
			IBM-2~\cite{Barea:2013bz} & 4.68 & 22.2 (min.) \\
			\hline
			QRPA-Jy~\cite{Hyvarinen:2015bda} & 5.26 & 76.3 (max.) \\
			\hline
			$G_{01}^{0\nu}$ [$10^{-14}$ yrs$^{-1}$]~\cite{Horoi:2017gmj} & \multicolumn{2}{c|}{0.22} \\
			\hline
		\end{tabular}
	\end{minipage}
	\hfill
	\begin{minipage}[t]{0.48\textwidth}
		\centering
		\begin{tabular}{|c|c|c|}
			\hline
			\multicolumn{3}{|c|}{\textbf{NME Ratios and $G_{01}^{0\nu}$ for $^{136}$Xe}} \\
			\hline
			\textbf{Method} & $\mathcal{M}^{0\nu}_\nu$ & $\mathcal{M}^{0\nu}_N/\mathcal{M}^{0\nu}_\nu$ \\
			\hline
			IBM-2~\cite{Barea:2013bz} & 3.05 & 23.8 (min.) \\
			\hline
			QRPA-Tu~\cite{Simkovic:2013qiy} & 2.18 & 69.7 (max.) \\
			\hline
			$G_{01}^{0\nu}$ [$10^{-14}$ yrs$^{-1}$] & \multicolumn{2}{c|}{1.5} \\
			\hline
		\end{tabular}
	\end{minipage}
	\caption{Minimum and maximum values of the ratio of nuclear matrix elements corresponding to light and heavy neutrino exchange for $^{76}$Ge (left) and $^{136}$Xe (right) using different methods. The last row shows the value of the phase-space factor. For values obtained using other methods and for different isotopes, one may refer to~\cite{Patra:2023ltl,Horoi:2017gmj}.}
	\label{tab:NMEratio}
\end{table}

The above expressions~(\ref{meffStandard})-(\ref{CN}) summarize the primary sources of lepton number violation in our framework, where both the right-handed and sterile neutrino sectors contribute significantly to the decay amplitude. Since these heavy Majorana states also participate in radiative transitions among charged leptons, the same couplings and mixings responsible for $0\nu\beta\beta$ decay can induce observable charged lepton flavor violating processes. In the following subsection, we discuss the general expression for the branching ratio of the $\mu \!\to\! e\gamma$ process within this theoretical setup.\\

\textbf{\underline{Impact of $M_{W_R}$ on $0\nu\beta\beta$ and collider reach}}\\
From eq.~(\ref{CN}), the heavy-channel EMMs carry $C_N\propto (M_{W_L}/M_{W_R})^4$. When the RH/sterile exchange dominates, the inverse half-life scales as
\[
[T^{0\nu}_{1/2}]^{-1}\ \propto\ |m_{\beta\beta,R}^{N,S}|^2\ \propto\ \left(\frac{M_{W_L}}{M_{W_R}}\right)^{\!8}\!,
\]
up to NME factors and mixing/mass weights. Numerically, increasing $M_{W_R}$ from $7\to 8$~TeV suppresses the heavy-channel rate by $(7/8)^8\simeq 0.34$, while $7\to 10$~TeV gives $(7/10)^8\simeq 5.8\times 10^{-2}$. Conversely, a modest downward shift (e.g.\ $7\to 6.5$~TeV) enhances the heavy-channel contribution by $\sim (7/6.5)^8\!\simeq\!1.7$ and simultaneously improves direct $W_R$ production prospects at the LHC/HL-LHC \cite{Keung:1983uu,Nemevsek:2011hz,Deppisch:2015qwa}. We therefore adopt $M_{W_R}=7$~TeV as a conservative baseline tailored to precision/intensity-frontier sensitivity, noting that collider-oriented benchmarks with slightly smaller $M_{W_R}$ can be chosen to place points within current or near-future LHC reach without altering our qualitative conclusions.

\subsection{Contribution to $\mu\to e\gamma$}
The Standard Model (SM) preserves lepton flavor because neutrinos are massless in the absence of right-handed neutrinos. However, the discovery of neutrino oscillations has confirmed that lepton flavor violation (LFV) is a physical reality in the neutral lepton sector. This naturally suggests that similar effects could also manifest in the charged sector, i.e., charged lepton flavor violation (cLFV), since lepton flavor is only an accidental symmetry in the SM.

With the discovery of neutrino oscillations and the consequent realization that neutrinos must have mass, the Standard Model extended minimally to include massive neutrinos does allow for cLFV processes such as $\mu \to e\gamma$, mediated by loop diagrams involving virtual $W_L$ bosons and neutrinos. However, these transitions are GIM suppressed (BR($\mu\to e \gamma$)$\sim 10^{-54}$) and far beyond the sensitivity of any current or near-future experiment. Therefore, any experimental observation of $\mu \to e\gamma$ would constitute a definitive signal of new physics beyond the Standard Model. Apart from the $\mu \to e\gamma$ process, there are various other cLFV processes such as $\mu \to 3e$, $\mu$–$e$ conversion in nuclei, $\tau \to \mu\gamma$, $\tau \to e\gamma$, and $\tau \to 3\mu$. So far, no positive signal has been observed for any of these processes, thus placing stringent upper limits on their branching ratios as summarized in Table~\ref{tab4}. We, however, limit our cLFV phenomenological discussion to the $\mu \to e\gamma$ process in this work. Prior to the case-specific analyses presented in the subsequent sections, we outline the general contributions to the $\mu \!\to\! e\gamma$ process and the corresponding expression for its branching ratio within this framework.

\begin{table}
	\centering
	\begin{tabular}{|c|c|}
		\hline
		\textbf{cLFV Process} & \textbf{Branching Ratio Bound (90\% C.L.) [Experiment]} \\
		\hline
		$\text{BR}(\mu \to e\gamma)$ & $< 4.2 \times 10^{-13}$ [MEG~\cite{MEG:2016leq}] \\
		$\text{BR}(\mu \to e\gamma)$ (future) & $\sim 6 \times 10^{-14}$ [MEG II~\cite{MEGII:2018kmf}] \\
		$\text{BR}(\mu \to 3e)$ & $< 1.0 \times 10^{-12}$ [SINDRUM~\cite{SINDRUM:1987nra}] \\
		$\text{BR}(\mu \to 3e)$ (future) & $\sim 10^{-16}$ [Mu3e~\cite{Mu3e:2020gyw}] \\
		$\mu$–$e$ conversion in Au & $< 7.0 \times 10^{-13}$ [SINDRUM II~\cite{SINDRUMII:2006dvw}] \\
		$\mu$–$e$ conversion (future, Al) & $\sim 10^{-17}$ [Mu2e~\cite{Mu2e:2014fns}, COMET~\cite{COMET:2018auw}] \\
		$\text{BR}(\tau \to \mu\gamma)$ & $< 4.4 \times 10^{-8}$ [BaBar~\cite{BaBar:2009hkt}] \\
		$\text{BR}(\tau \to e\gamma)$ & $< 3.3 \times 10^{-8}$ [BaBar~\cite{BaBar:2009hkt}] \\
		$\text{BR}(\tau \to 3\mu)$ & $< 2.1 \times 10^{-8}$ [Belle~\cite{Hayasaka:2010np}] \\
		\hline
	\end{tabular}
	\caption{Current experimental bounds and future sensitivities for various charged lepton flavor violating (cLFV) processes.}
	\label{tab4}
\end{table}

\begin{figure}
	\centering
	\begin{tikzpicture}
		\begin{feynman}
			\vertex (a) at (-1.2,0) {\(\mu_R\)};
			\vertex (b) at (2,0);
			\vertex (c) at (5.2,0) {\(e_R\)};
			
			\node at (2,-0.35) {\({N}_{R_i}\)};
			
			\diagram* {
				(a) -- [plain,thick] (b) -- [plain,thick] (c),
			};

			\draw[thick, decorate, decoration={snake, amplitude=1mm, segment length=2mm}] 
			(0,0) arc[start angle=180, end angle=0, radius=1.8cm] 
			node[pos=0.25, above left=2pt] {\(W_R\)}
			node[pos=0.75, above right=2pt] {\(W_R\)};

			\draw[photon,thick] (2,1.95) -- (2.5,3) node[above] {\(\Large\gamma\)};
			
			\node at (0,-0.35) {\scriptsize\(V_{\mu i}^{NN}\)};
			\node at (4,-0.35) {\scriptsize\({V_{e i}^{NN}}^*\)};
		\end{feynman}
		
		\begin{scope}[xshift=7.5cm]
			\begin{feynman}
				\vertex (a) at (-1.2,0) {\(\mu_R\)};
				\vertex (b) at (2,0);
				\vertex (c) at (5.2,0) {\(e_R\)};
				
				\node at (2,-0.35) {\({S}_{L_i}\)};
				
				\diagram* {
					(a) -- [plain,thick] (b) -- [plain,thick] (c),
				};

				\draw[thick, decorate, decoration={snake, amplitude=1mm, segment length=2mm}] 
				(0,0) arc[start angle=180, end angle=0, radius=1.8cm] 
				node[pos=0.25, above left=2pt] {\(W_R\)}
				node[pos=0.75, above right=2pt] {\(W_R\)};

				\draw[photon,thick] (2,1.95) -- (2.5,3) node[above] {\(\gamma\)};
				
				\node at (0,-0.35) {\scriptsize\(V_{\mu i}^{NS}\)};
				\node at (4,-0.35) {\scriptsize\({V_{e i}^{NS}}^*\)};
			\end{feynman}
		\end{scope}
	\end{tikzpicture}
	\caption{One-loop diagrams contributing to the charged lepton flavor violating process \(\mu \rightarrow e\gamma\) in the LRSM with double seesaw. The left diagram involves the exchange of heavy right-handed neutrinos \(N_{R_i}\), while the right diagram involves sterile neutrinos \(S_{L_i}\). The amplitudes are governed by the neutrino mixing matrices \(V^{NN}\) and \(V^{NS}\), which enter through the vertices and control the strength of lepton flavor violation in each diagram, leading to sizeable cLFV contributions.}
	\label{fig:LFVloop}
\end{figure}
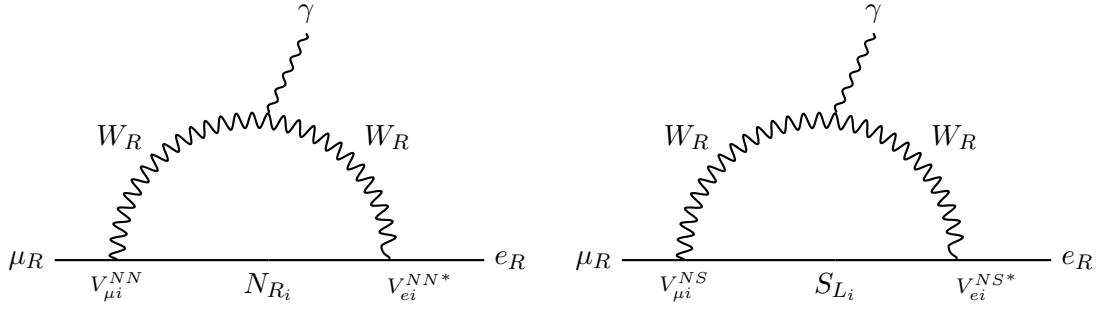

In the framework of the LRSM with double seesaw that we have considered, the only non-vanishing neutrino mixing matrices (see eq.~\ref{NeutrinoMixMat}) that provide new contributions to the $\mu \to e\gamma$ process are $V^{NN}$ and $V^{NS}$~\cite{Patra:2023ltl}. These mixing matrices contribute via right-handed currents, and the relevant one-loop Feynman diagrams mediated by RHNs ($N_{R_i}$) and sterile neutrinos ($S_{L_i}$) are shown in Figure~\ref{fig:LFVloop}. The analytic expression for the branching ratio corresponding to these diagrams is given by~\cite{Cirigliano:2004mv,Barry:2013xxa,Bambhaniya:2015ipg} 
\begin{equation}\label{BRmu2eg}
	\text{BR}(\mu \to e\gamma) = \frac{3\alpha_{em}}{2\pi}(|G_N + G_S|^2),
\end{equation}
where $\alpha_{em} \equiv \frac{e^2}{4\pi}$ is the fine-structure constant, and $G_N$ and $G_S$ are the form factors defined as
\begin{eqnarray}
	G_N &=& \sum_{i=1}^{3} \left[ V_{\mu i}^{NN} {V_{e i}^{NN}}^* |\xi|^2 G^\gamma(x_i) + V_{\mu i}^{NN} {V_{e i}^{NN}}^* \frac{m^2_{W_L}}{m^2_{W_R}} G^\gamma(y_i) \right], \label{GNFF} \\
	G_S &=& \sum_{i=1}^{3} \left[ V_{\mu i}^{NS} {V_{e i}^{NS}}^* |\xi|^2 G^\gamma(x_i) + V_{\mu i}^{NS} {V_{e i}^{NS}}^* \frac{m^2_{W_L}}{m^2_{W_R}} G^\gamma(y_i) \right], \label{GSFF}
\end{eqnarray}
with $x_i = (M_i/m_{W_L})^2$, $y_i = (M_i/m_{W_R})^2$, where $M_i = M_{N_i}, M_{S_i}$, and $\xi = \tan^{-1}(2v_1 v_2/v_R^2)/2$. The loop function $G^\gamma(x)$ is given by
\begin{equation}\label{loopFunc}
	G^\gamma(x) = -\frac{2x^3 + 5x^2 - x}{4(1 - x)^3} - \frac{3x^3}{2(1 - x)^4} \ln x.
\end{equation}

\textbf{Dependence on $M_{W_R}$}\\
From eqs.~(\ref{GNFF})–(\ref{GSFF}) and (\ref{BRmu2eg}), the dominant scaling arises through $G_{N,S}\propto (m_{W_L}^2/M_{W_R}^2)$, so that
\[
\text{BR}(\mu\to e\gamma)\;\propto\;\left(\frac{m_{W_L}^2}{M_{W_R}^2}\right)^{\!2}\times \left|\sum_i \cdots\, G^\gamma\!\left(\frac{M_i^2}{M_{W_R}^2}\right)\right|^{\!2}.
\]
Thus, for fixed heavy-fermion masses, the branching ratio decreases approximately as \(M_{W_R}^{-4}\), with an additional mild suppression from the loop function as \(y_i=m_i^2/M_{W_R}^2\) becomes smaller at larger \(M_{W_R}\). Numerically, increasing \(M_{W_R}\) from \(7\) to \(8\)~TeV reduces \(\text{BR}\) by a factor \((7/8)^4\simeq 0.59\), while raising it to \(10\)~TeV yields \((7/10)^4\simeq 0.24\). Conversely, lowering \(M_{W_R}\) enhances \(\text{BR}\) accordingly, subject to collider limits and the RHN-$W_R$ mass hierarchy discussed in subsection~\ref{subsec:gauge_boson_masses}.

\subsection{Contribution to other cLFV processes}

\underline{$\mu\to 3e$}\\

For the LRSM framework considered here—with a scalar sector comprising only a bidoublet ($\Phi$) and doublets ($H_{L,R}$), and no scalar triplets ($\Delta_{L,R}$)—the process $\mu\to 3e$ is absent at tree level. It can, however, proceed via one-loop diagrams with heavy and light neutrinos. These contributions are suppressed by loop factors and by the small light–heavy neutrino mixing relative to any putative tree-level effect. In our setup, the dominant mixings are $V^{NN}$ (RHN self-mixing) and $V^{NS}$ (RH–sterile mixing) (see subsection~\ref{DMT_NM}), so the relevant one-loop $\mu\to 3e$ amplitudes arise primarily from box diagrams with right-handed currents. We omit the lengthy branching-ratio expressions and refer the reader to \cite{Barry:2013xxa,Ilakovac:2012sh} for analytic formulas.\\

\noindent\underline{$\mu\to e$ conversion}\\

Within our setup, $\mu\to e$ conversion in nuclei receives contributions from the same loop and box topologies that contribute to $\mu\to 3e$ \cite{Barry:2013xxa}. Its analysis additionally involves hadronic physics and depends on both dipole and scalar/vector operators. A careful evaluation requires nuclear modeling and may introduce parameters beyond those considered for $\mu\to e\gamma$. We note that $\mu\to e$ conversion can be a sensitive probe, and a detailed study in our setup is an important direction for future work.

In this paper, we focus on the single-flavor transition $\mu\to e\gamma$ and do not analyze $\tau$-LFV modes. The emphasis on the muon channel is motivated by both experimental and theoretical considerations: upcoming experiments such as \textsf{MEG II} provide the highest sensitivity to $\mu\to e\gamma$, muon beams offer high event rates and a suitable lifetime, and LRSM-like frameworks often yield similar flavor structures across lepton sectors, making $\mu\to e\gamma$ a prototypical probe of charged-lepton flavor violation. Given the comparatively limited experimental reach for $\tau$ channels, we prioritize the muon channel in this work.

With the analytical expressions for $0\nu\beta\beta$ and for $\mathrm{BR}(\mu\to e\gamma)$ in hand, we now turn to the phenomenology using two Dirac-mass textures in the sections that follow.

\section{Phenomenology with Symmetry-Motivated Dirac Mass Structure (Case I)}\label{sec:pehnoCaseI}

In this section, we study the phenomenological implications of the symmetry-motivated Dirac mass structure introduced in subsection~\ref{subsec:MDIdentity}, where the Dirac neutrino mass matrix is proportional to the identity matrix, as suggested by an underlying flavor symmetry.

While the assumption of Dirac mass matrices proportional to the identity is motivated by certain flavor-symmetry arguments and provides a minimal and transparent framework for analysis, it represents a strong idealization. More general textures, including off-diagonal elements, complex phases, or texture zeros~\cite{Borgohain:2020csn}, can introduce additional sources of flavor mixing and CP violation. Such structures can be conveniently parametrized using the Casas--Ibarra parametrization~\cite{Casas:2001sr}, which captures the most general form of the Dirac mass matrix consistent with low-energy neutrino data. Relaxing the proportional-to-identity assumption may modify detailed predictions for observables such as the branching ratios of cLFV processes and the $0\nu\beta\beta$ decay rates, chiefly through changes in phase dependences and mass-ordering sensitivities. Presenting our results in this simplified limit allows us to highlight parameter correlations in a clear and tractable manner, without the complications of additional free parameters. A detailed study of the phenomenology arising from general Dirac mass textures, including the effects of new flavor and CP structures, is left for future work.

We analyze both lepton-flavor-violating processes, such as the radiative decay $\mu \to e \gamma$, and the lepton-number-violating process, namely neutrinoless double beta decay ($0\nu\beta\beta$), within this framework. We aim to identify regions of parameter space where both types of observables can lie within the sensitivity reach of upcoming experiments and to explore the complementarity between these two classes of probes.

In what follows the subsection~5.1 presents the predictions for $\mu \to e \gamma$, while subsection~5.2 discusses $0\nu\beta\beta$ within the cLFV-relevant parameter space and summarizes the combined allowed region.

\subsection{Charged lepton flavor violation: $\mu\to e \gamma$ predictions}\label{subsec:cLFVcase1}
We first consider Case I, as discussed in subsection~\ref{subsec:MDIdentity}, to compute the branching ratio \(\text{BR}(\mu \to e\gamma)\). This decay proceeds via the one-loop diagram shown in figure~\ref{fig:LFVloop}, and the relevant expressions are given in eqs.~(\ref{BRmu2eg})–(\ref{loopFunc}). The form factors and loop function defined in eqs.~(\ref{GNFF}),~(\ref{GSFF}), and~(\ref{loopFunc}) enter into the expression for the branching ratio. It is evident from these expressions that \(\text{BR}(\mu \to e\gamma)\) depends on the mass of the right-handed charged gauge boson \(M_{W_R}\), as well as on the mixing between the left- and right-handed gauge bosons \(\xi\) (defined in eq.~(\ref{mixingWLWR})).\footnotetext[2]{In the double seesaw model under discussion, we require \(m_{N_{1(3)}} \ll m_{S_{3(2)}}\) for the NO (IO) case. To ensure this hierarchy, we fix \(m_{S_{3(2)}} = 10\, m_{N_{1(3)}}\) \cite{Patra:2023ltl}.}

Before presenting the numerical predictions, we examined the compatibility of the benchmark value $M_{W_R}=4.8$~TeV, used in collider limits, with the right-handed neutrino mass hierarchy in Case~I. We find that this benchmark is not viable within our framework, since the heaviest RHN mass exceeds $M_{W_R}$, violating the hierarchy assumed in the ATLAS analysis~\cite{ATLAS:2023cjo} (see subsection~\ref{subsec:gauge_boson_masses}). The detailed derivation and corresponding RHN mass spectrum are provided in Appendix~\ref{app:WRbenchmark}. Consequently, we adopt a more conservative benchmark, $M_{W_R}=7$~TeV, which remains consistent with collider constraints and forms the basis for our subsequent cLFV analysis. 

To systematically investigate the impact of this setup on cLFV observables, we summarize the key benchmark values for the relevant parameters in Table~\ref{tab:benchmarks_caseI}. This includes the gauge boson mass \(M_{W_R}\), the corresponding left-right mixing angle \(\xi\) (see eq.~(\ref{mixingWLWR}) along with the condition on VEV: $v_2\ll v_1$, mentioned above eq.~(\ref{VEVconditionResult})), Dirac CP violating phase $\delta$, and the chosen range of the heaviest RHN mass and lightest active neutrino mass.

\begin{table}
	\centering
	\small 	
	\renewcommand{\arraystretch}{1.3}
	\begin{tabular}{|c|c|c|}
		\hline
		\multicolumn{3}{|c|}{\textbf{cLFV Benchmark Parameters for Case I}} \\
		\hline
		\textbf{Parameter} & \textbf{Normal Ordering (NO)} & \textbf{Inverted Ordering (IO)}\\
		\hline
		\(M_{W_R}\) & \multicolumn{2}{c|}{7 TeV (fixed)} \\
		\hline
		\(\delta\) (Dirac CP phase) & \multicolumn{2}{c|}{0 (fixed)} \\
		\hline
		\(\xi\) ($W_L-W_R$ mixing)& \multicolumn{2}{c|}{\(10^{-6}\) (fixed)} \\
		\hline
		Heaviest RHN mass & \(m_{N_1} = 100\text{--}6000\) GeV & \(m_{N_3} = 100\text{--}4000\) GeV \\
		\hline
		Lightest active neutrino mass & \(m_1 = 10^{-5}\text{--}1\) eV & \(m_3 = 10^{-5}\text{--}1\) eV \\
		\hline
		Heaviest sterile neutrino mass & \(m_{S_3} = 10\, m_{N_1}\) & \(m_{S_2} = 10\, m_{N_3}\) \\
		\hline
	\end{tabular}
	\caption{Benchmark ranges for the analysis of \(\mu \to e\gamma\) in Case I for both Normal and Inverted mass ordering. The RH gauge boson mass \(M_{W_R}\), Dirac CP phase \(\delta\), and left-right mixing angle \(\xi\) are fixed across both scenarios. The variation is done in the heaviest RHN and the lightest active neutrino mass. The heaviest sterile neutrino mass is fixed to be ten times the heaviest RHN mass in each case\protect\footnotemark[2].}
	\label{tab:benchmarks_caseI}
\end{table}

\begin{figure}
	\centering
	\begin{minipage}[b]{0.48\textwidth}
		\centering
		\includegraphics[width=\textwidth]{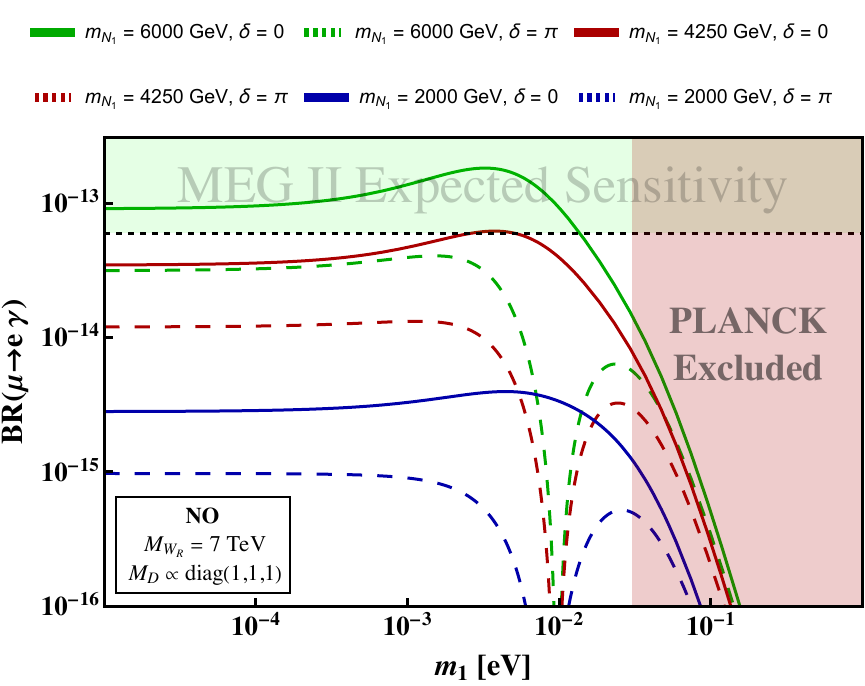}
	\end{minipage}
	\hfill
	\begin{minipage}[b]{0.48\textwidth}
		\centering
		\includegraphics[width=\textwidth]{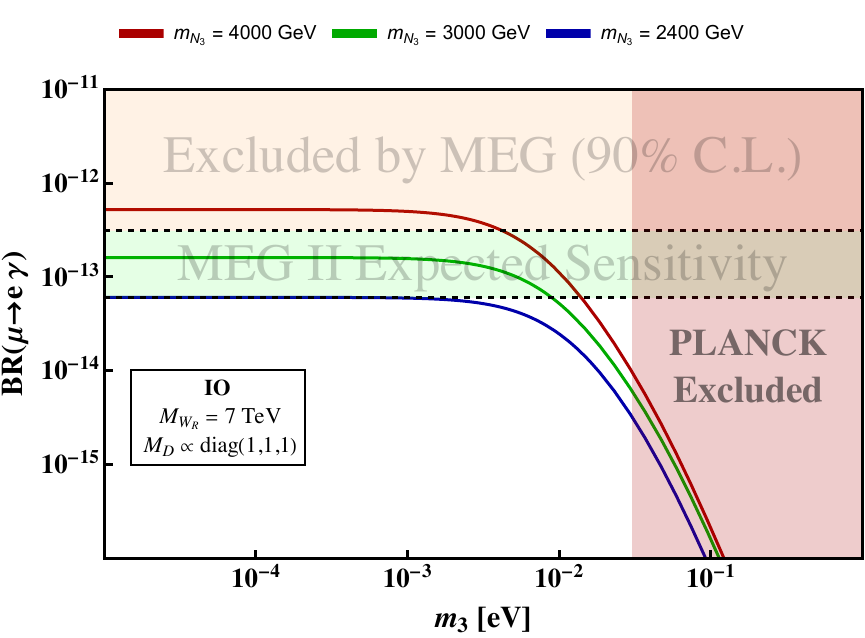}
	\end{minipage}
	\caption{Variation of \(\text{BR}(\mu \to e\gamma)\) with the lightest active neutrino mass for both normal ordering (left, \(m_{N_1}\) fixed) and inverted ordering (right, \(m_{N_3}\) fixed) scenarios. In the NO case, achieving branching ratios within the MEG II sensitivity range requires large \(m_{N_1}\) and \(\delta = 0\), while in the IO case, the dependence on \(\delta\) is negligible, and large values of \(m_{N_3}\) are already excluded by existing bounds.}
	\label{fig:BRvsLightestMass}
\end{figure}

In order to motivate our choice of benchmark parameters for the charged lepton flavor violating process \(\mu \to e\gamma\), we present a detailed numerical analysis of the branching ratio as a function of the lightest active neutrino mass for fixed values of the heaviest right-handed neutrino mass, \(m_{N_1}\) (for NO) and \(m_{N_3}\) (for IO). Figure~\ref{fig:BRvsLightestMass} illustrates the behavior of \(\text{BR}(\mu \to e\gamma)\) for both normal and inverted ordering scenarios. For visual clarity, in both panels of figure~\ref{fig:BRvsLightestMass}, we indicate the \textsf{PLANCK}-excluded region as a vertical shaded band, corresponding to the upper bound on the lightest neutrino mass derived from the cosmological constraint \(\sum m_\nu \lesssim 0.12\,\text{eV}\) \cite{Planck:2018vyg}. Specifically, we adopt a \textit{conservative upper bound} of \(m_{\text{lightest}} = 0.03\,\text{eV}\) for both NO and IO, although the exact bound for IO corresponds to a smaller value \(m_3^{\text{max}} \approx 0.016\,\text{eV}\). This choice ensures a unified presentation without significantly altering the exclusion region, as the numerical difference is marginal. Throughout this work, we follow this conservative assumption for the IO case.

In the normal ordering case, shown in the left panel of figure~\ref{fig:BRvsLightestMass}, the branching ratio remains above the \textsf{MEG II} sensitivity threshold for most of the parameter space unless both \(m_{N_1}\) is sufficiently large (e.g., above 4250 GeV) and the Dirac CP phase \(\delta\) is set to zero. This observation is supported by an analytical expression for the branching ratio derived for a representative lightest neutrino mass value around \(m_1 = 10^{-2}\) eV, where the plot indicates maximal sensitivity to the Dirac CP phase—exhibiting enhancement or cancellation for \(\delta = 0\) or \(\pi\), respectively. While the full analytical expression as a function of the lightest neutrino mass is too lengthy and impractical to present here, its behavior can be illustrated by the following simplified form obtained for \(m_1 = 10^{-2}\) eV and \(m_{N_1} = 5\) TeV:
\begin{equation}
	\text{BR}(\mu \to e\gamma) \approx 3.02 \times 10^{-14} + 3.01 \times 10^{-14} \cos\delta
\end{equation}
This consistently demonstrates that the maximal value of the branching ratio is achieved for \(\delta = 0\), across the scanned range of \(m_1\) and \(m_{N_1}\), thereby justifying this choice for our benchmark.

The inverted ordering case, shown in the right panel of figure~\ref{fig:BRvsLightestMass}, exhibits a very mild numerical dependence on \(\delta\), as the terms containing \(\delta\) are significantly suppressed compared to the dominant \(\delta\)-independent contributions in the branching ratio expression. As a result, the branching ratio remains largely unaffected by the choice of \(\delta\). However, for \(m_{N_3} \gtrsim 4\) TeV, the branching ratio becomes excessively large, entering regions already ruled out by existing experimental bounds. This necessitates restricting the heaviest RHN mass to lie below 4 TeV in the IO scenario. Given the suppressed numerical dependence on \(\delta\), we fix \(\delta = 0\) for the IO case and consider a scan over the lightest neutrino mass and heaviest RHN mass within experimentally viable ranges, as summarized in table~\ref{tab:benchmarks_caseI}.

It is worthwhile to note that the sizeable new-physics contribution to $\text{BR}(\mu\to e\gamma)$ in our framework arises from loop diagrams involving heavy fermions---namely, the right-handed neutrinos (RHNs) $N_{R_i}$ and sterile neutrinos $S_{L_i}$---as shown in figure~\ref{fig:LFVloop}. The approximate expression for the branching ratio in Case I can then be written as:
\begin{equation}
	\text{BR}(\mu\to e\gamma)= \underbrace{\boxed{\frac{3\alpha_{em}}{2\pi}}}_{\textcolor{blue}{\approx 3\times 10^{-3}}}\,\left|G_N + G_S\right|^2,
	\label{BRdefOOM}
\end{equation}
where the electromagnetic fine-structure constant appears as an overall prefactor, and the new physics enters via the form factors $G_N$ and $G_S$. These are given by
\begin{equation}\label{GNsumTerm}
	G_N = \underbrace{\boxed{\frac{m^2_{W_L}}{m^2_{W_R}}}}_{\textcolor{blue}{\approx 1.32\times10^{-4}}}
	\underbrace{\sum_{i=1}^{3} V_{\mu i}^{NN}\,{V_{e i}^{NN}}^*\, G^\gamma(y_i)}_{\text{$G_N$ Sum Term}},
\end{equation}
\begin{equation}\label{GSsumTerm}
	G_S = \underbrace{\boxed{\frac{m^2_{W_L}}{m^2_{W_R}}}}_{\textcolor{blue}{\approx 1.32\times10^{-4}}}
	\underbrace{\sum_{i=1}^{3} V_{\mu i}^{NS}\,{V_{e i}^{NS}}^*\, G^\gamma(y_i)}_{\text{$G_S$ Sum Term}},
\end{equation}
where $G^\gamma(y_i)$ is the loop function with argument $y_i = m_i^2 / M_{W_R}^2$, and the indices run over the three generations of heavy RHNs or sterile neutrinos. The prefactor captures the suppression due to the heavy $W_R$ mass, while the sums encapsulate the flavor mixing and loop dynamics. The detailed numerical decomposition for the benchmarks in Table~\ref{tab:benchmarks_caseI} is collected in Appendix~\ref{app:tables_caseI} (Tables~\ref{tab:RHNOoM_caseI} and~\ref{tab:SLOoM_caseI}).

\begin{figure}
	\centering
	\begin{subfigure}[b]{0.49\textwidth}
		\includegraphics[width=\linewidth]{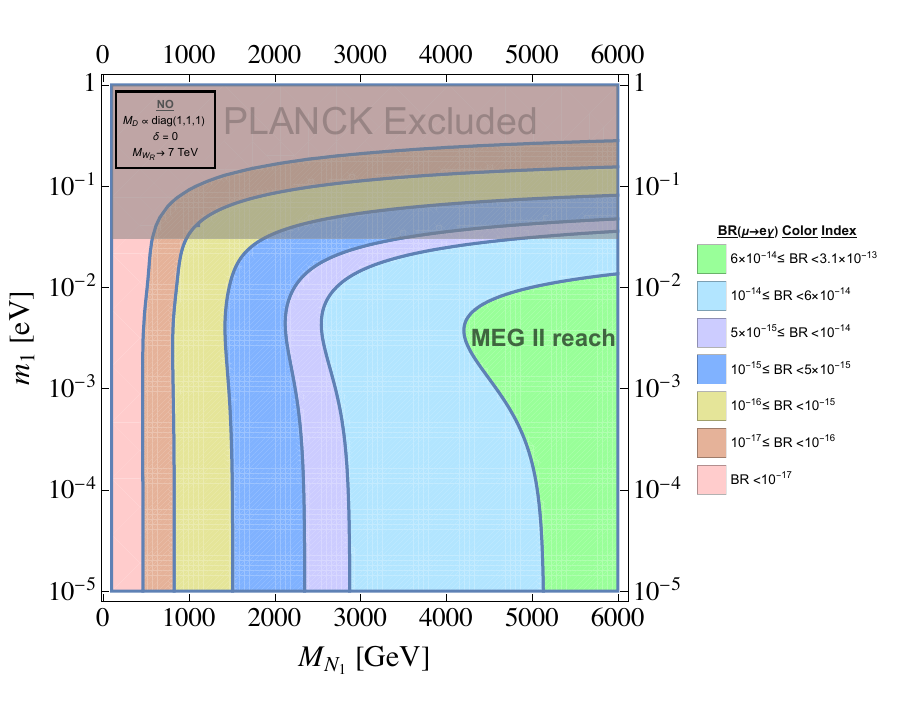}
		\caption{}
		\label{fig:DMD_RP_NO} 
	\end{subfigure}
	\begin{subfigure}[b]{0.49\textwidth}
		\includegraphics[width=\linewidth]{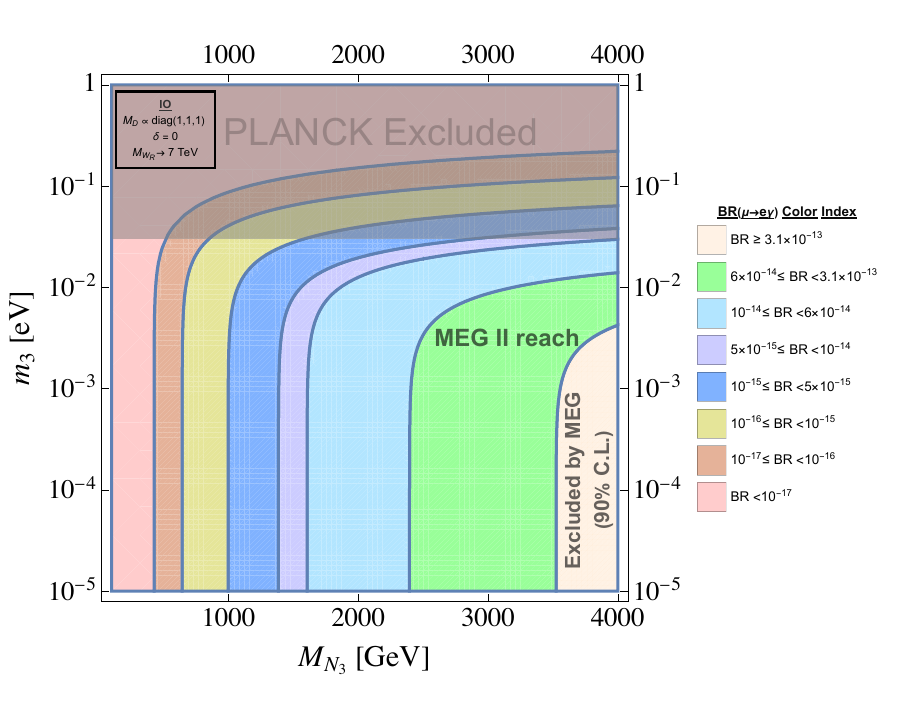}
		\caption{}
		\label{fig:DMD_RP_IO}
	\end{subfigure}
	\caption{Region plots of BR\((\mu \to e\gamma)\) in the plane of heaviest RHN mass and lightest active neutrino mass for Case I. The left and right panels correspond to the normal ordering (NO) and inverted ordering (IO) of the active neutrino mass spectrum, respectively. The plots highlight regions within the reach of \textsf{MEG II} (light green), currently excluded by \textsf{MEG} (faint orange), and ruled out by the \textsf{PLANCK} constraint on the sum of neutrino masses (translucent grey band). Various branching ratio ranges are indicated using the color bar on the right.}
	\label{fig:DMD_RP_NO_IO}
\end{figure}

We conclude our discussion of Case I by presenting region plots of \(\text{BR}(\mu \to e\gamma)\) as a function of the heaviest RHN mass and the lightest active neutrino mass in both NO and IO mass spectra, as shown in Figure~\ref{fig:DMD_RP_NO_IO}. The dependence of BR\((\mu \to e\gamma)\) in the \(m_{N_{1(3)}}\)-\(m_{1(3)}\) plane, as depicted in Figure~\ref{fig:DMD_RP_NO} and Figure~\ref{fig:DMD_RP_IO} for the NO and IO mass spectra respectively, corresponds to the benchmark points summarized in Table~\ref{tab:benchmarks_caseI}.

Assuming \(M_{W_R} = 7~\text{TeV}\), Figure~\ref{fig:DMD_RP_NO} for the NO scenario shows that BR\((\mu \to e\gamma)\) enters the \textsf{MEG II} sensitivity region only for large heaviest RHN masses, specifically \(m_{N_1} \gtrsim 4210~\text{GeV}\). In contrast, Figure~\ref{fig:DMD_RP_IO} for the IO case indicates that for \(m_{N_3} > 3530~\text{GeV}\), BR\((\mu \to e\gamma)\) lies in the excluded region of the \textsf{MEG} experiment for a wide range of the lightest neutrino mass \(m_3\).

In both Figures~\ref{fig:DMD_RP_NO} and \ref{fig:DMD_RP_IO}, the region accessible by \textsf{MEG II} is shown in light green, while the area excluded by the \textsf{PLANCK} constraint on the sum of active neutrino masses is indicated by a translucent grey band. Other ranges of the branching ratio are represented using different colours, with the corresponding colour indices displayed on the right-hand side of each plot. Additionally, in Figure~\ref{fig:DMD_RP_IO}, the region excluded by the current \textsf{MEG} limit is marked with a faint orange shade. The favourable ranges of the heaviest RHN mass and the lightest active neutrino mass for which BR\((\mu \to e\gamma)\) falls within the \textsf{MEG II} sensitivity region, as inferred from these region plots, are summarised in Table~\ref{tab:FavPara_caseI}.

\begin{table}
	\centering
	\renewcommand{\arraystretch}{0.93}
	\begin{tabular}{|c|c|c|}
		\hline
		\multicolumn{3}{|c|}{\textbf{cLFV Favourable Parameter Space for Case I}} \\
		\hline
		$m_\nu$ \textbf{Ordering} &\textbf{NO}&\textbf{IO}\\
		\hline
		\textbf{Heaviest RHN}&\multirow{2}{*}{$m_{N_1}\gtrsim4210$}&\multirow{2}{*}{$2400\lesssim m_{N_3}\lesssim 3530$}\\
		\textbf{mass (GeV)}&&\\
		\hline
		\textbf{Lightest active}&\multirow{2}{*}{$10^{-5}\lesssim m_{1}\lesssim1.26\times10^{-2}$}&\multirow{2}{*}{$10^{-5}\lesssim m_{1}\lesssim1.58\times10^{-2}$}\\
		\textbf{$\nu$ mass (eV)}&&\\
		\hline
	\end{tabular}
	\caption{Favourable parameter space for BR\((\mu \to e\gamma)\) to enter the \textsf{MEG II} sensitivity zone in Case I. The table presents the required ranges of the heaviest RHN mass and the lightest active neutrino mass for both normal (NO) and inverted (IO) ordering of active neutrinos. The lightest neutrino mass has a full range from \(10^{-5}\)~eV up to the \textsf{PLANCK}-excluded limit. The parameters \(M_{W_R} = 7\)~TeV, Dirac CP phase \(\delta = 0\), and left-right mixing angle \(\xi = 10^{-6}\) are fixed throughout.}
	\label{tab:FavPara_caseI}
\end{table}

\subsection{$0\nu\beta\beta$ predictions within the cLFV-relevant region}\label{subsec:0nubb_caseI_common}

For Case I, the effective Majorana mass (EMM) terms for the RHN and sterile neutrino channels, as defined in eq.~(\ref{meffRHN}) and eq.~(\ref{meffSterile}) respectively, take the form determined by the neutrino mixing matrices and the masses of the intermediate and lightest states of RHN and sterile neutrinos as summarized in table~\ref{tab:case1}. Here, we only quote the resulting expressions of EMM for the RHN and sterile neutrino channels and note that the expression for the EMM of the standard channel ($|m^{\nu}_{\beta\beta,L}|$) remains unchanged as expressed in eq.~(\ref{meffStandard}). A more involved discussion on this is presented in~\cite{Patra:2023ltl}. The expressions of EMM for the RHN and sterile neutrino channels are:

\begin{equation}\label{meffCaseI}
	\begin{aligned}
		|m_{\beta\beta,R}^N|_{\textbf{NO}} &= \frac{C_N}{m_{N_1} m_1} |m^{\nu}_{\beta\beta,L}|, \\
		|m_{\beta\beta,R}^N|_{\textbf{IO}} &= \frac{C_N}{m_{N_3} m_3} |m^{\nu}_{\beta\beta,L}|, \\
		|m_{\beta\beta,R}^S|_{\textbf{NO}} &= \bigg| \frac{C_N m_{N_1} m_1 m_3^2}{m_{S_3}^2 m_1^3} \left[ |U_{e1}|^2 + |U_{e2}|^2 e^{i\alpha} \frac{m_1^3}{m_2^3} + |U_{e3}|^2 e^{i(\beta-2\delta)} \frac{m_1^3}{m_3^3} \right] \bigg|, \\
		|m_{\beta\beta,R}^S|_{\textbf{IO}} &= \bigg| \frac{C_N m_{N_3} m_3 m_2^2}{m_{S_2}^2 m_3^3} \left[ |U_{e1}|^2 \frac{m_3^3}{m_1^3} + |U_{e2}|^2 e^{i\alpha} \frac{m_3^3}{m_2^3} + |U_{e3}|^2 e^{i(\beta-2\delta)} \right] \bigg|.
	\end{aligned}
\end{equation}
Here, $C_N$ is the dimensionless proportionality factor as defined in eq.~(\ref{CN}), and $\delta$, $\alpha$, $\beta$ are the Dirac CP phase and Majorana phases, respectively. $U_{ej}\ (j=1,2,3)$ are the first row elements of the $U_{\text{PMNS}}$ matrix, while $m_{N_{1(3)}}$ and $m_{S_{3(2)}}$ denote the heaviest RHN and sterile neutrino mass in the NO and IO cases, respectively. With the expressions of EMM for the RHN and sterile neutrino channels in eq.~(\ref{meffCaseI}), we now investigate the implications for neutrinoless double beta decay when the masses of the right-handed neutrinos lie within the charged lepton flavor violation (cLFV) relevant region, as determined by the projected sensitivity of the \textsf{MEG II} experiment~\cite{MEGII:2023ltw}. From figure~\ref{fig:DMD_RP_NO}, it is evident that for the NO mass spectrum of active neutrinos in case I, the mass of the heaviest right-handed neutrino is constrained within the range $m_{N_1} \in [4250, 6000]$~GeV, consistent with the \textsf{MEG II} sensitivity. In contrast, for the inverted ordering scenario (figure~\ref{fig:DMD_RP_IO}), values of the heaviest right-handed neutrino mass $m_{N_3} \gtrsim 3500$~GeV are excluded by \textsf{MEG} for a wide range of lightest active neutrino masses.

\begin{figure}
	\centering
	\begin{minipage}[b]{0.495\textwidth}
		\centering
		\includegraphics[width=\textwidth]{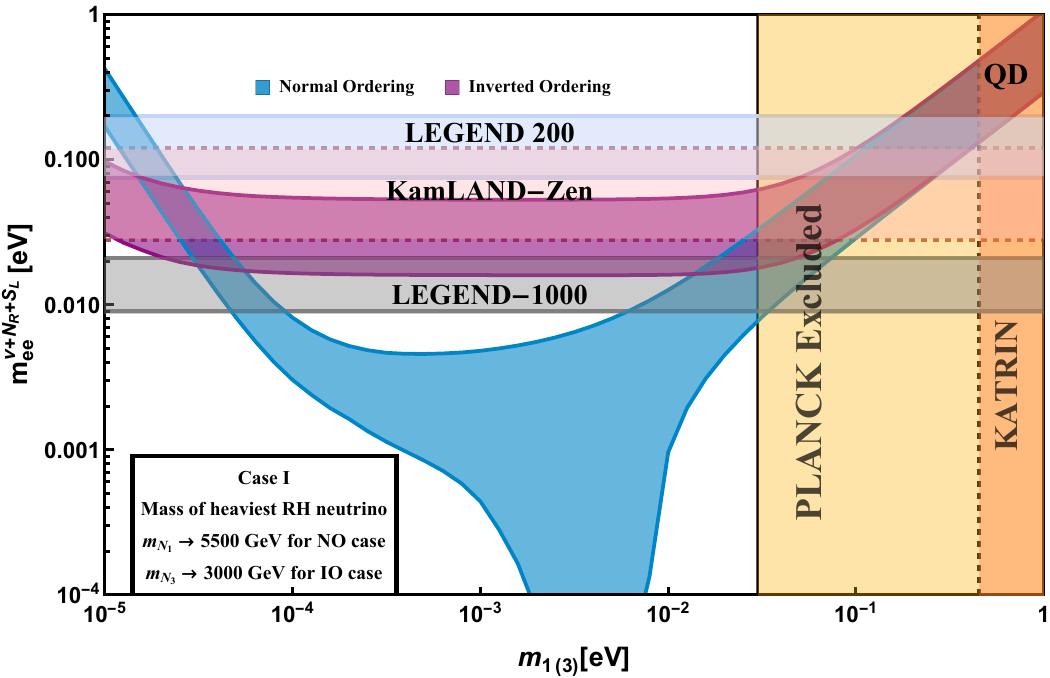}
	\end{minipage}
	\begin{minipage}[b]{0.495\textwidth}
		\centering
		\includegraphics[width=\textwidth]{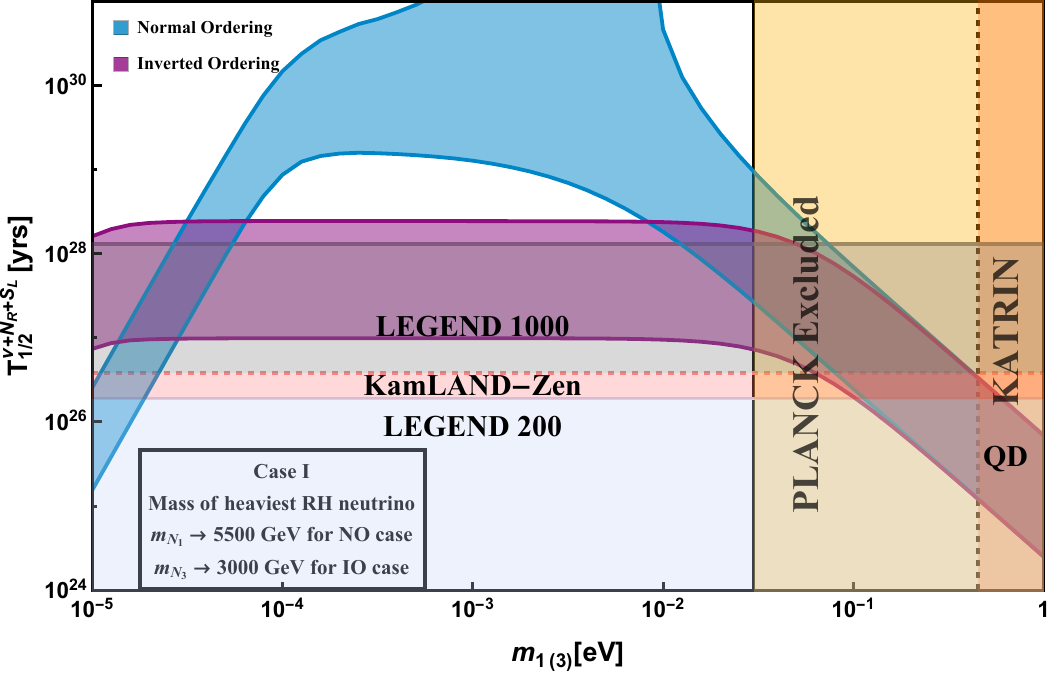}
	\end{minipage}
	\caption{Variation of the generalized effective Majorana mass $|m_{\beta\beta}^{\nu+N+S}|$ (left) and the corresponding $0\nu\beta\beta$ decay half-life $T_{1/2}^{\nu+N+S}$ (right) as functions of the lightest active neutrino mass, considering contributions from light active neutrinos, right-handed neutrinos, and sterile neutrinos for Dirac mass matrix in Case~I. The cyan and magenta bands represent the NO and IO spectra, respectively, with their overlap indicating the quasi-degenerate region. The light-pink, light-blue, and grey horizontal bands denote current and projected experimental limits from \textsf{KamLAND-Zen}, \textsf{LEGEND-200}, and \textsf{LEGEND-1000}, respectively. Vertical orange and yellow bands indicate bounds from \textsf{KATRIN} and \textsf{PLANCK} on the absolute neutrino mass scale.}
	\label{fig:onubbCaseI}
\end{figure}

Taking these bounds into account, we choose representative values $m_{N_1} = 5500$~GeV and $m_{N_3} = 3000$~GeV for the heaviest RHN mass in the NO and IO scenarios respectively, to define a viable parameter space for $0\nu\beta\beta$ decay within the cLFV ($\mu\to e\gamma$) relevant parameter space in case I. For the representative values of the heaviest RHN masses and with the analytical expressions of EMM for new physics contributions in eq.~(\ref{meffCaseI}), we plot the GEMM and the half-life of $0\nu\beta\beta$ decay as a function of the lightest active neutrino mass $m_{1(3)}$ in the NO and IO mass spectra, shown in figure~\ref{fig:onubbCaseI}.
The left and right panels of figure~\ref{fig:onubbCaseI} show the variation of GEMM and the variation of the $0\nu\beta\beta$ decay half-life, respectively, incorporating contributions from the standard mechanism (light active neutrinos), right-handed neutrinos, and sterile neutrinos. In both panels, the cyan and magenta bands represent the NO and IO mass spectra of active neutrinos, respectively. These bands are obtained by varying the oscillation parameters within their $3\sigma$ allowed ranges, as provided by the global fit data from NuFITv-6.0~\cite{Esteban:2024eli}. The Majorana phases $\alpha$, $\beta$, and the Dirac CP-violating phase $\delta$ are scanned uniformly over the interval $[0, 2\pi]$, the NME ratio is varied as: $22.2\lesssim\mathcal{M}^{0\nu}_N/\mathcal{M}^{0\nu}_\nu\lesssim76.3$, as discussed in table~\ref{tab:NMEratio}.  The light-blue and light-pink horizontal bands in figure~\ref{fig:onubbCaseI} correspond to the current experimental bounds on the effective Majorana mass from \textsf{LEGEND 200} (75--200 meV~\cite{LEGEND:2025jwu}) and \textsf{KamLAND-Zen} (29--122 meV~\cite{KamLAND-Zen:2024eml}), respectively. The light-grey band corresponds to the future sensitivity of \textsf{LEGEND 1000} (9--21 meV~\cite{LEGEND:2021bnm}). The vertical orange band denotes the direct limit on absolute active neutrino mass ($m_\nu < 0.45$ eV at 90\% CL) from \textsf{KATRIN}~\cite{KATRIN:2024cdt}, while the yellow vertical band represents the \textsf{PLANCK}-excluded region. The overlap of the NO and IO bands indicates the region of the quasi-degenerate (QD) spectrum.

In comparison with the standard mass mechanism, figure~\ref{fig:onubbCaseI} reveals a substantial enhancement in the effective Majorana mass arising from non-standard contributions for the Dirac mass matrix ($m_D$) in case I. As shown in the left panel of figure~\ref{fig:onubbCaseI}, for lightest neutrino masses $m_{1(3)} \lesssim 10^{-3}$~eV, the generalized effective Majorana mass $|m_{\beta\beta}^{\nu+N+S}|$ exhibits a steep dependence on $m_{1(3)}$, with non-standard contributions dominating over the standard light neutrino exchange. This behavior persists until $m_{1(3)} \gtrsim 3.85 \times 10^{-4}$~eV, beyond which the standard contribution becomes dominant. For $m_{1(3)} \gtrsim 5.75 \times 10^{-3}$~eV, the new physics effects become strongly suppressed, resulting in $|m_{\beta\beta}^{\nu+N+S}| \simeq |m_{\beta\beta}^{\nu}|$.

The non-standard contributions are particularly significant at low values of the lightest neutrino mass, e.g., around $m_{1(3)} \approx 2 \times 10^{-5}$~eV for the benchmark scenario with CP phases $\alpha = \beta = 0$. However, for values below $m_{1(3)} \lesssim 1.05 \times 10^{-5}$~eV, the total effective mass $|m_{\beta\beta}^{\nu+N+S}|$ enters regions excluded by current experimental bounds from \textsf{LEGEND-200} and \textsf{KamLAND-Zen}, as well as the projected bounds from \textsf{LEGEND-1000}. For the inverted ordering (IO) case, this exclusion is reached at slightly lower values of the lightest mass, specifically for $m_{1(3)} \lesssim 10^{-5}$~eV.

For the normal ordering (NO) scenario with $\alpha = \beta = 0$, we observe that the inequality $|m_{\beta\beta}^{\nu+N+S}| > |m_{\beta\beta}^{\nu}|$ is satisfied throughout the interval $1.4 \times 10^{-4} \text{ eV} \lesssim m_{1} \lesssim 4.7 \times 10^{-4}$~eV, indicating the dominance of non-standard contributions in this mass range. Within this interval, the effective mass satisfies $|m_{\beta\beta}^{\nu+N+S}| \gtrsim 4.7 \times 10^{-4}$~eV, corresponding to a $0\nu\beta\beta$ half-life of $T_{1/2}^{\nu+N+S} \gtrsim 10^{29}$~yrs. For the NO spectrum with $\alpha = \pi$, $\beta = 0$, we find $|m_{\beta\beta}^{\nu+N+S}| \gtrsim 9 \times 10^{-3}$~eV for $m_{1(3)} \lesssim 4.7 \times 10^{-5}$~eV, with the corresponding half-life satisfying $T_{1/2}^{\nu+N+S} \lesssim 10^{29}$~yrs. A pronounced minimum in the effective mass is observed at $m_{1(3)} \approx 2.40 \times 10^{-3}$~eV, where $|m_{\beta\beta}^{\nu+N+S}| \approx 7.41 \times 10^{-6}$~eV.

In contrast, for the IO scenario, no significant enhancement from non-standard contributions is observed within the charged lepton flavor violation (cLFV) allowed region. Across the full range of $m_{1(3)}$ and for all values of the Majorana phases $\alpha$ and $\beta$, the total effective mass remains comparable to the standard mechanism: $|m_{\beta\beta}^{\nu+N+S}| \simeq |m_{\beta\beta}^{\nu}|$. For this particular scenario, the half-life lies within the range $9.47\times 10^{26} \lesssim T_{1/2}^{\nu+N+S} \lesssim 2.35 \times 10^{28}$~yrs.

It is important to note that, in the present analysis, the effective contribution arising from interference scenarios has not been taken into account. This omission is justified by the fact that, within the cLFV-allowed parameter space, such contributions are negligible.

\section{Phenomenology with model-determined Dirac Mass Structure (Case II)}\label{sec:pehnoCaseII}

In this section, we study the phenomenology with Case II, where the Dirac mass matrices $M_D$ and $M_{RS}$ are not ad hoc but uniquely determined by the underlying double–seesaw dynamics and the assumed LR symmetry, as described in subsection~\ref{subsec:derivedMD} and Ref.~\cite{Patel:2023voj}. Unlike the scenario in which $M_D$ is taken proportional to the identity for calculational convenience, here the Dirac mass terms acquire a fully explicit structure:
\begin{equation}
	M_{RS} = i k\, U_\nu \sqrt{\hat{m}_\nu \hat{m}_N}\, U_\nu^T, \qquad
	M_D = i\, U_\nu \sqrt{\hat{m}_\nu \hat{m}_N}\, U_\nu^T,
\end{equation}
with $m_{S_i} = k^2 m_i$ [see eqs.~(\ref{DiracMassesDmD}),(\ref{phyMnuMs})].
All flavor structures are thus dictated by the low–energy PMNS parameters and the mass spectra $m_i$, $m_{N_i}$, and $m_{S_i}$; no additional complex angles or phases beyond those already present in neutrino oscillations are introduced.

This construction significantly reduces the arbitrariness typically present in the heavy sector and leads to genuine predictivity: cLFV and LNV observables follow directly from the low-energy neutrino parameters ($\hat m_\nu$, $U_\nu$) together with assumptions on the RHN spectrum (see Table~\ref{tab:RHN_hierarchy_cases}). In particular, the mixing submatrices entering the one–loop amplitudes,
\[
V^{NN} \simeq U_\nu, \qquad V^{NS} = \frac{i}{k}\,U_\nu\,\sqrt{\hat{m}_\nu^{-1}\hat{m}_N},
\]
are entirely model-determined. This allows us to probe and correlate rare lepton-number- and lepton-flavor-violating processes with minimal flavor input, providing a direct handle on seesaw-scale physics from low-energy data.

It is this fully specified structure---where the Yukawa/Dirac parameters are fixed rather than scanned---that distinguishes our analysis from broadly explored parameter studies in the LRSM literature and sharpens the connection between the observed neutrino sector and new-physics signatures at upcoming cLFV and $0\nu\beta\beta$ experiments. In what follows, we present detailed predictions for $0\nu\beta\beta$ and $\mu \to e\gamma$ within this framework and highlight the correlations that emerge when the Dirac mass structure is fully specified by the PMNS matrix and neutrino masses.

\subsection{$0\nu\beta\beta$ predictions in favourable parameter space}\label{subsec:0nubb_caseII_fav}
Following from subsection~\ref{subsec:derivedMD}, we obtain the neutrino mixing matrices and the masses of both intermediate and lightest sterile neutrino states, as summarized in table~\ref{tab:case2}. These parameters define the effective Majorana mass (EMM) terms relevant to the new physics contribution in case II, whose general expressions are given in eq.~(\ref{meffRHN}) and eq.~(\ref{meffSterile}). While the standard channel EMM term remains the same, it is worth emphasizing that for the RHN channel, the EMM term in the present case is independent of active neutrino masses and hence also of their mass spectrum ordering. The expressions of EMM for the RHN and sterile neutrino channels in case II are:
\begin{equation}\label{meffCaseII}
	\begin{aligned}
		|m_{\beta\beta,R}^N|&= \Bigg|C_N\left[\frac{|U_{e1}|^2}{m_{N_1}}+\frac{|U_{e2}|^2 e^{i\alpha}}{m_{N_2}}+\frac{|U_{e3}|^2 e^{i(\beta-2\delta)}}{m_{N_3}}\right]\Bigg|, \\
		|m_{\beta\beta,R}^S|_{\textbf{NO}} &= \Bigg|-C_N\frac{m_1^2+\Delta m_{atm}^2}{m_{S_3}^2}\left[\frac{|U_{e1}|^2 m_{N_1}}{m_1^2}+\frac{|U_{e2}|^2 e^{i\alpha}m_{N_2}}{m_1^2+\Delta m_{sol}^2}+\frac{|U_{e3}|^2 e^{i(\beta-2\delta)}m_{N_3}}{m_1^2+\Delta m_{atm}^2}\right]\Bigg|, \\
		|m_{\beta\beta,R}^S|_{\textbf{IO}} &= \Bigg|-C_N\frac{m_3^2+\Delta m_{sol}^2+\Delta m_{atm}^2}{m_{S_2}^2}\\
		&\times\left[\frac{|U_{e1}|^2 m_{N_1}}{m_3^2+\Delta m_{atm}^2}+\frac{|U_{e2}|^2 e^{i\alpha}m_{N_2}}{m_3^2+\Delta m_{sol}^2+\Delta m_{atm}^2}+\frac{|U_{e3}|^2 e^{i(\beta-2\delta)}m_{N_3}}{m_3^2}\right]\Bigg|.
	\end{aligned}
\end{equation}
Here, $m_{N_i}\ (i=1,2,3)$ are the masses of three generations of RHNs, and they may have any of the mass hierarchies as presented in table~\ref{tab:RHN_hierarchy_cases}. Thus, in what follows, we explore the relevant mass hierarchies of RHN masses for their experimentally significant contribution to $0\nu\beta\beta$ decay in case II. The EMM for the RHN channel ($|m_{\beta\beta,R}^N|$) in this case, as expressed in eq.~(\ref{meffCaseII}), is highly suppressed for the considered parameter space ranges (1–1000 GeV), which will be discussed in upcoming discussions. Thus, we will consider the sterile neutrino channel contribution to EMM ($|m_{\beta\beta,R}^S|$) for both NO and IO mass spectra of active neutrinos to sort the relevant RHN mass hierarchies listed in table~\ref{tab:RHN_hierarchy_cases}.\\

\textbf{\underline{Relevant RHN mass hierarchies:}}\\

\textbf{\underline{NO case}}\\

\noindent From eq.~(\ref{meffCaseII}), the expression of the effective Majorana mass term associated with the sterile neutrino channel for the NO mass spectrum of active neutrinos can be rewritten as:
\begin{equation}\label{meffTtermsNO}
	\begin{aligned}
		{m_{\beta\beta,R}^S}\bigg|_{\textbf{NO}} = -C_N&\bigg[\underbrace{|U_{e1}|^2\frac{m_1^2+\Delta m_{atm}^2}{m_1^2}\frac{m_{N_1}}{m_{S_3}^2}}_{T_1}+\underbrace{|U_{e2}|^2\frac{m_1^2+\Delta m_{atm}^2}{m_1^2+\Delta m_{sol}^2}\frac{m_{N_2}}{m_{S_3}^2}e^{i\alpha}}_{T_2}\\
		&+\underbrace{|U_{e3}|^2\frac{m_{N_3}}{m_{S_3}^2}e^{i(\beta-2\delta)}}_{T_3}\bigg].
	\end{aligned}
\end{equation}
Here, the terms $T_1$, $T_2$, and $T_3$ reflect the contributions from each generation of right-handed neutrino mass $m_{N_i}\ (i=1,2,3)$, along with the dependence on the lightest active neutrino mass $m_1$, the heaviest sterile neutrino mass $m_{S_3}$, PMNS matrix elements $U_{ei}\ (i=1,2,3)$, the Dirac CP phase $\delta$, and Majorana phases $\alpha$ and $\beta$.

We perform a systematic analytic and numerical analysis of all possible RHN mass hierarchies (details in Appendix \ref{app:CaseII_fav_details}) and find that only term $T_1$ is dominant, thereby reducing the $m_N$ cases to effectively three: $m_{N_{A,F,D}}$, with the remaining scenarios yielding practically indistinguishable results in the parameter regions of interest. For maximal contribution to $m_{\beta\beta}^S$, we fix the hierarchy $m_{N_3} < m_{N_2} < m_{N_1}$ (case $m_{N_F}$) for the NO case.\\

\textbf{\underline{IO case}}\\

\noindent The expression for EMM associated with the sterile neutrino channel in the IO case can be rewritten from eq.~(\ref{meffCaseII}) as:
\begin{equation}\label{meffTtermsIO}
	\begin{aligned}
		{m_{\beta\beta,R}^S}\bigg|_{\textbf{IO}} = -C_N&\bigg[\underbrace{|U_{e1}|^2\frac{\Delta m_{sol}^2+\Delta m_{atm}^2}{\Delta m_{atm}^2}\frac{m_{N_1}}{m_{S_2}^2}}_{T_1}+\underbrace{|U_{e2}|^2\frac{m_{N_2}}{m_{S_2}^2}e^{i\alpha}}_{T_2}\\
		&+\underbrace{|U_{e3}|^2\frac{\Delta m_{sol}^2+\Delta m_{atm}^2}{m_3^2}\frac{m_{N_3}}{m_{S_2}^2}e^{i(\beta-2\delta)}}_{T_3}\bigg].
	\end{aligned}
\end{equation}
Upon detailed analysis (see Appendix \ref{app:CaseII_fav_details}), we find that only $T_3$ is dominant; thus, similar to the NO case, the RHN mass hierarchies reduce to effectively three cases: $m_{N_{A,F,D}}$. The maximal contribution to $m_{\beta\beta}^S$ arises for case $m_{N_A}$ ($m_{N_1}<m_{N_2}<m_{N_3}$); hence we choose this RHN mass hierarchy for $0\nu\beta\beta$ predictions in Case II for the IO case.

We summarize the considered RHN mass hierarchies along with the corresponding dominant terms for both NO and IO mass spectra in table~\ref{tab:RelevantmNcases}. We also briefly discuss the individual non-standard contributions to the effective Majorana mass (EMM) in Case II. The standard contribution remains the same in both Case I and Case II and can be referred to from \cite{Patra:2023ltl}, along with the non-standard contributions in Case I.\\

\begin{table}
	\centering
	\footnotesize
	\renewcommand{\arraystretch}{1.14}
	\begin{tabular}{|c|c|c|}
		\hline
		\multicolumn{3}{|c|}{\textbf{Relevant RHN mass hierarchy for $0\nu\beta\beta$ predictions in Case II}} \\
		\hline
		\multirow{2}{*}{\textbf{NO/IO}}& \textbf{Preferred}&\textbf{Leading Non-Standard}\\
		&\textbf{$m_N$ Case}&\textbf{Contributing Expression}\\
		\hline
		\multirow{2}{*}{NO}&Case $m_{N_F}$&	\multirow{2}{*}{$\frac{1}{m_{N_1}}\cdot \frac{\Delta m^{2}_{\mbox{atm}}}{m^{2}_{1}}\cdot 10^{-2}\cdot |U_{e1}|^{2}$}\\
		&$(m_{N_3} < m_{N_2} < m_{N_1})$&\\
		\hline
		\multirow{2}{*}{IO}&Case $m_{N_A}$&	\multirow{2}{*}{$\frac{e^{i(\beta-2\delta)}}{m_{N_3}}\cdot \frac{\Delta m^{2}_{\mbox{sol}}+\Delta m^{2}_{\mbox{atm}}}{m^{2}_{3}}\cdot 10^{-2}\cdot |U_{e3}|^{2}$}\\
		&$(m_{N_1} < m_{N_2} < m_{N_3})$&\\
		\hline
	\end{tabular}
	\caption{Summary of the preferred RHN mass hierarchies and corresponding dominant non-standard contributions to the effective Majorana mass term in Case II, for both normal (NO) and inverted (IO) light neutrino mass spectra.}
	\label{tab:RelevantmNcases}
\end{table}

\textbf{\underline{EMM contribution due to exchange of RH neutrinos $N_{1,2,3}$}}\\

In general, one may expect the EMM contribution from the RHN channel to dominate over that from the sterile neutrino channel in the model under consideration, as the self-mixing of RHNs ($V^{NN}$) is much stronger than the mixing between RH and sterile neutrinos ($V^{NS}$). Moreover, the double seesaw requirement implies $1/m_N > 1/m_S$. Together, these imply the relation
\begin{equation*}
	\frac{|V^{NS}|^2}{M_{S_j}} \ll \frac{|V^{NN}|^2}{M_{N_j}},
\end{equation*}
indicating that the dominant non-standard contribution to the effective Majorana mass $m_{\beta\beta,R}^{N,S}$ originates from the right-handed neutrino exchange term, $m_{\beta\beta,R}^{N}$. However, it is important to emphasize that for Case~II in our framework, the right-handed neutrino contribution $m_{\beta\beta,R}^N$ is independent of the light neutrino mass spectrum and is highly suppressed by the RHN mass (see eq.~(\ref{meffCaseII})).\\

\textbf{\underline{EMM contribution due to exchange of sterile neutrinos $S_{1,2,3}$}}\\

The effective Majorana mass due to exchange of sterile neutrinos constitutes the dominant non-standard contribution to GEMM in Case II. Table~\ref{tab:RelevantmNcases} summarizes the non-standard contributions arising from sterile neutrino exchange. From the leading term expressions in the third column, for both NO and IO cases in table~\ref{tab:RelevantmNcases}, we observe that 
\begin{equation*}
	m_{\beta\beta,R}^S \propto  \frac{1}{m^{2}_{1(3)}}. 
\end{equation*}
This dependence of $m_{\beta\beta,R}^S$ on the lightest active neutrino mass $m_{1(3)}$ introduces a mass-enhancement feature into the sterile neutrino contribution, particularly relevant at small $m_{1(3)}$. As a result, even though $m_{\beta\beta,R}^{S}$ is suppressed by neutrino mixing, its scaling with $m_{1(3)}$ partially compensates, leading to a non-negligible effect in the regime where the lightest neutrino mass is sufficiently small. While there is no phase dependence in $m_{\beta\beta,R}^{S}$ for the NO case, the IO case exhibits a dependence on one Majorana phase $\beta$ and the Dirac phase $\delta$ as $e^{i(\beta-2\delta)}$.\\

\textbf{\underline{Favourable Parameter Space for $0\nu\beta\beta$ in Case II}}\\

\begin{figure}
	\begin{subfigure}[b]{0.49\textwidth}
		\includegraphics[width=\linewidth]{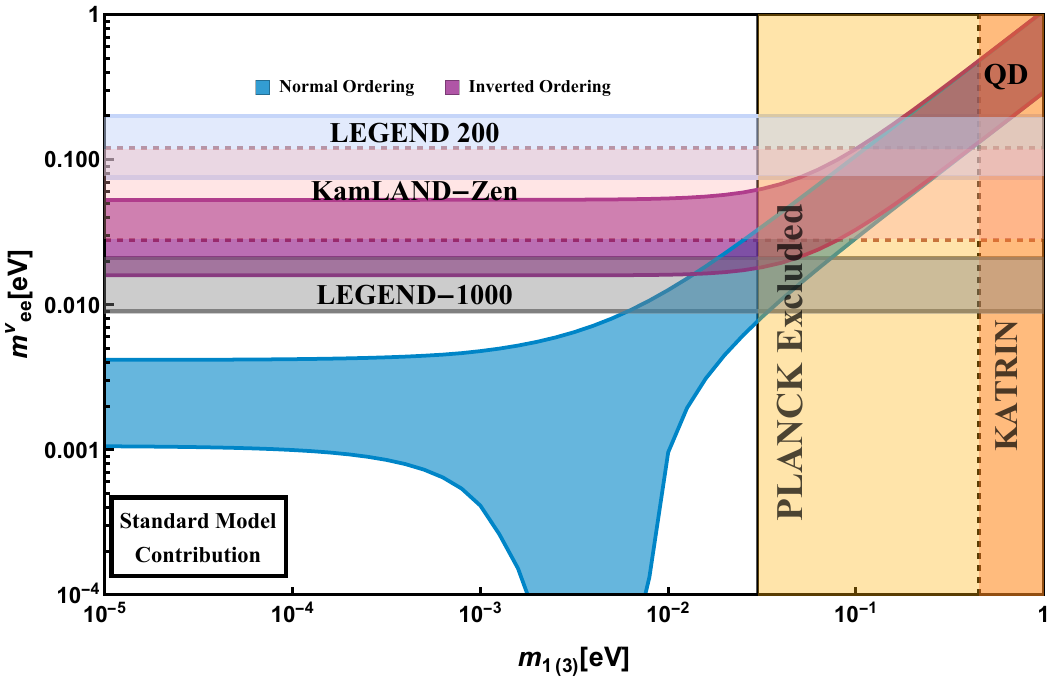}
		\caption{}
		\label{fig:NDMDlobSMee}
	\end{subfigure}
	\hspace{0.1cm}
	\begin{subfigure}[b]{0.49\textwidth}
		\includegraphics[width=\linewidth]{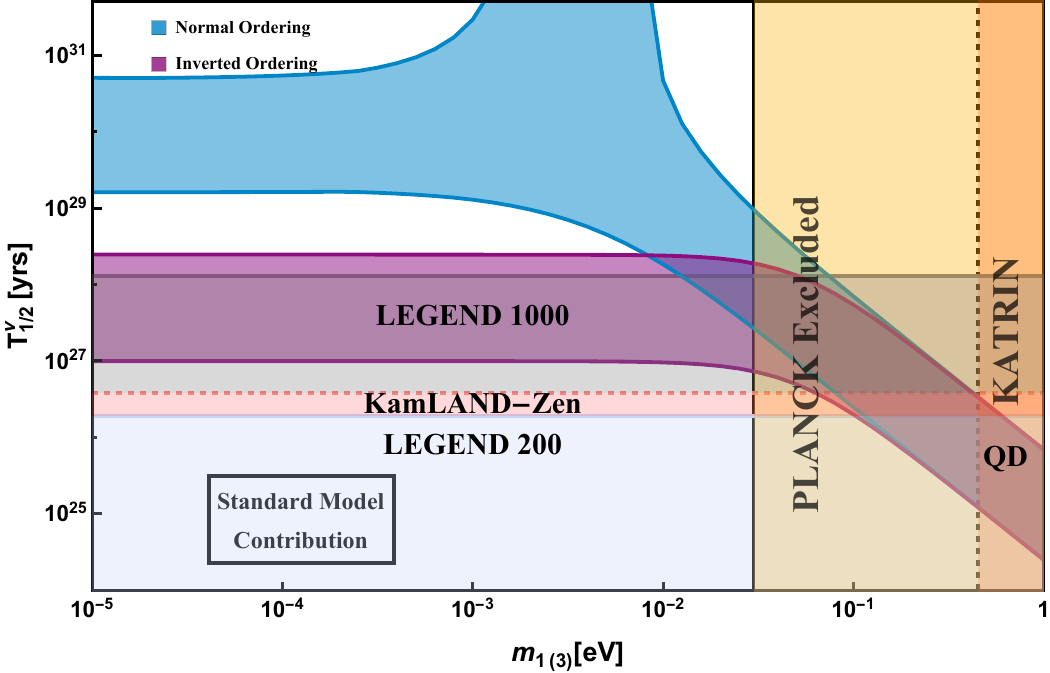}
		\caption{}
		\label{fig:NDMDlobSMhl}
	\end{subfigure}
	
	\vspace{0.3cm}
	
	\begin{subfigure}[b]{0.49\textwidth}
		\includegraphics[width=\linewidth]{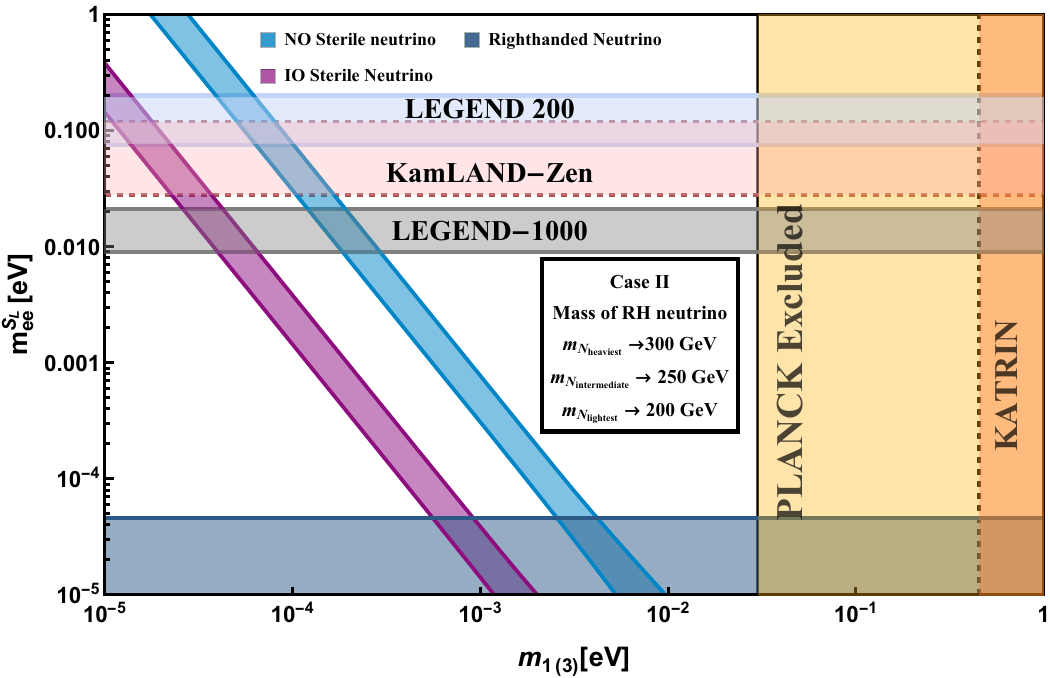}
		\caption{}
		\label{fig:NDMDBSMee}
	\end{subfigure}
	\hspace{0.1cm}
	\begin{subfigure}[b]{0.49\textwidth}
		\includegraphics[width=\linewidth]{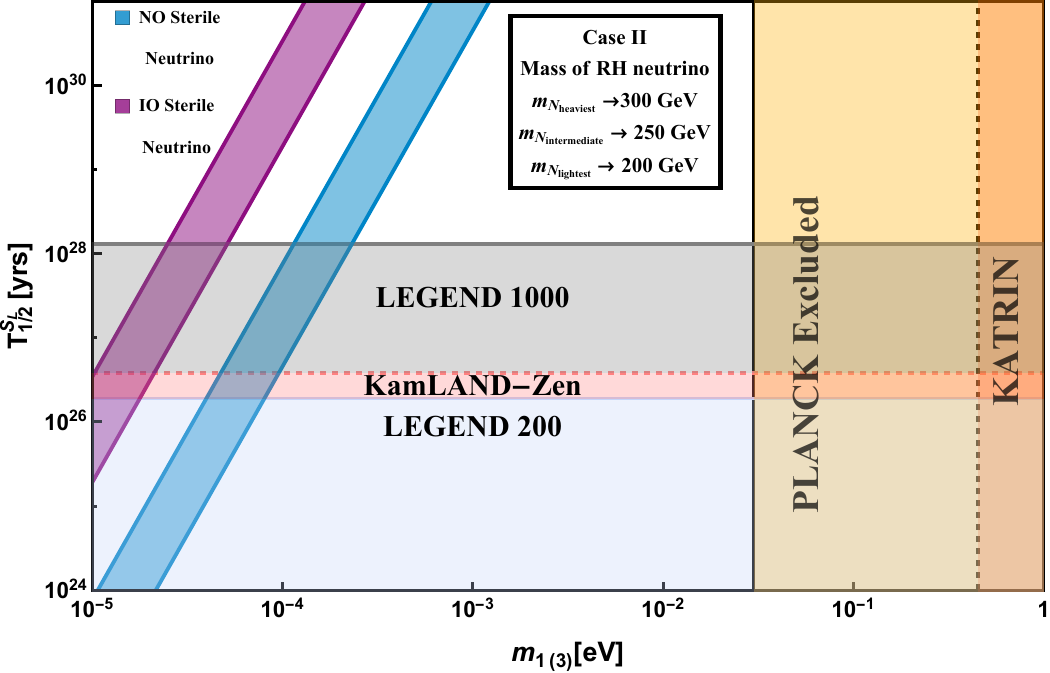}
		\caption{}
		\label{fig:NDMDBSMhl}
	\end{subfigure}
	
	\vspace{0.3cm}
	
	\begin{subfigure}[b]{0.49\textwidth}
		\includegraphics[width=\linewidth]{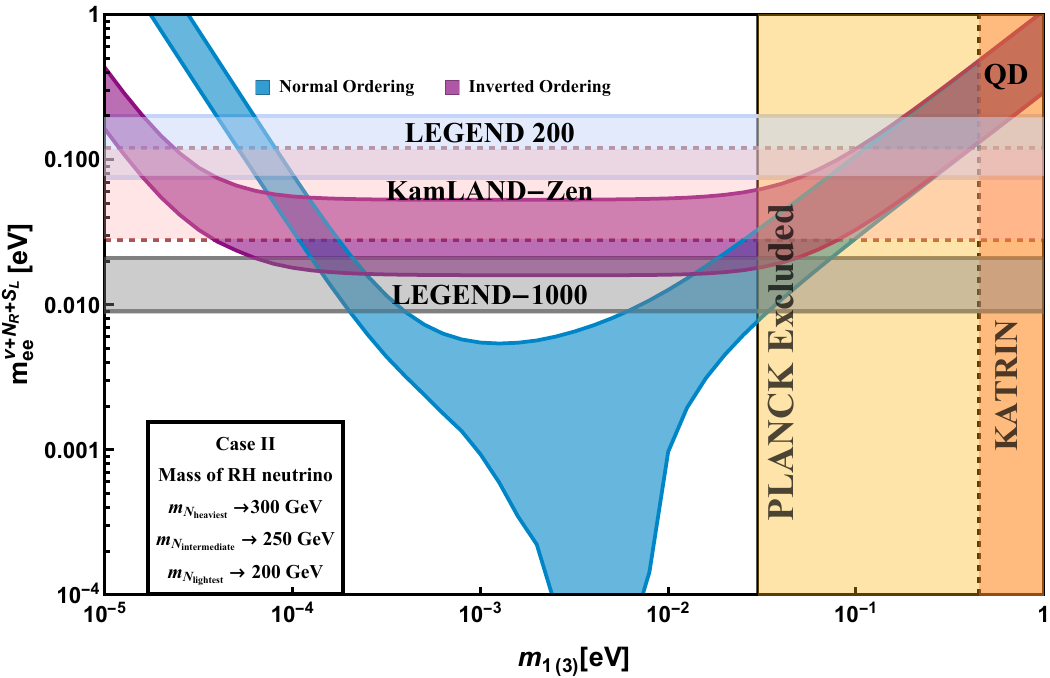}
		\caption{}
		\label{fig:NDMDlobBSMee}
	\end{subfigure}
	\hspace{0.1cm}
	\begin{subfigure}[b]{0.49\textwidth}
		\includegraphics[width=\linewidth]{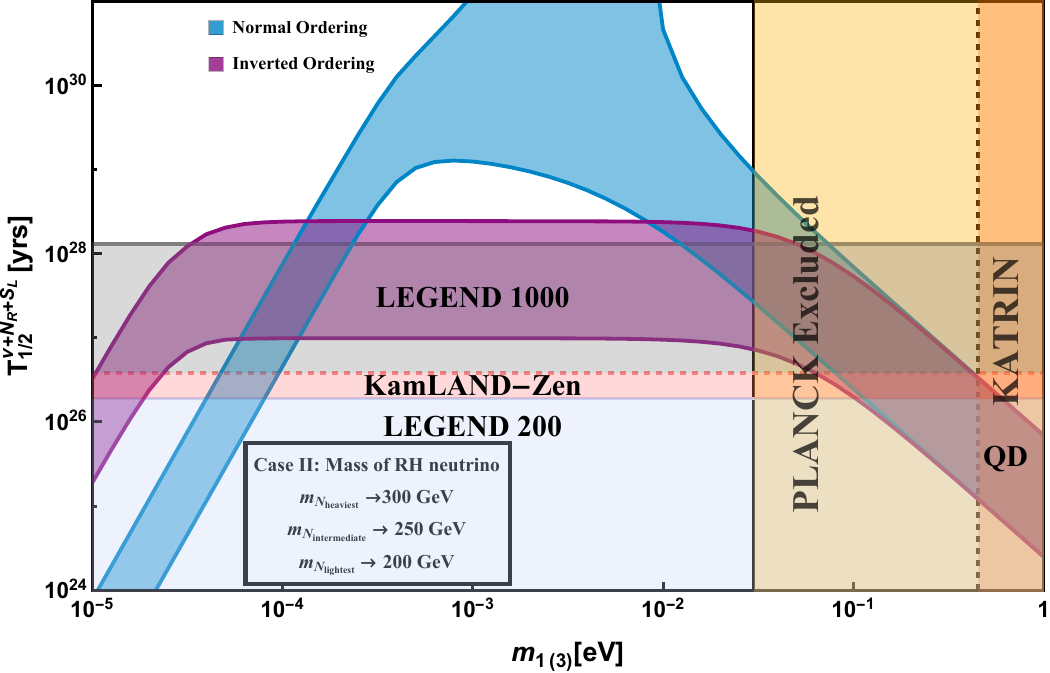}
		\caption{}
		\label{fig:NDMDlobBSMhl}
	\end{subfigure}
	\caption{Effective Majorana mass (left) and half-life (right) predictions for $0\nu\beta\beta$ decay in Case II for the favourable parameter space. The top row shows the standard mechanism, the middle row shows individual non-standard contributions from RH and sterile neutrino exchange, and the bottom row shows the combined contributions. The RHN mass hierarchies $m_{N_F}$ (for NO) and $m_{N_A}$ (for IO), as identified in Table~\ref{tab:RelevantmNcases}, are chosen to represent the favourable parameter space. Overlapping regions between NO and IO bands indicate the QD spectrum. Possible interference between RH and sterile neutrino contributions is not considered in this analysis. All experimental bands are the same as those given in figure~\ref{fig:onubbCaseI}.}
	\label{fig:LobsterPlotsCaseII_6panel}
\end{figure}

To study $0\nu\beta\beta$ decay within a favourable parameter space---i.e., one that yields ideal enhancement of decay rates in the framework under consideration for Case II---we adopt the following setup: the right-handed gauge boson mass is fixed at $M_{W_R}=7$ TeV, and the heaviest RHN mass is taken as 300 GeV. Consequently, the heaviest sterile neutrino mass $m_{S_{3(2)}}$ is set at 3 TeV for the NO and IO cases, respectively. The intermediate and lightest RHN masses are fixed at 250 and 200 GeV, respectively. With these parameters and the mass hierarchies $m_{N_F}$ $(m_{N_3} < m_{N_2} < m_{N_1})$ for NO and $m_{N_A}$ $(m_{N_1} < m_{N_2} < m_{N_3})$ for IO---taken from the preferred RHN mass hierarchies summarized in table~\ref{tab:RelevantmNcases}---we plot GEMM (left panel) and half-life (right panel) of $0\nu\beta\beta$ decay as a function of the lightest neutrino mass in figure~\ref{fig:LobsterPlotsCaseII_6panel}.

All the plots in figure~\ref{fig:LobsterPlotsCaseII_6panel} are generated by varying the neutrino oscillation parameters within their respective $3\sigma$ allowed ranges. The Majorana and Dirac phases $\alpha, \beta$, and $\delta$ are uniformly varied in the interval $[0,2\pi]$. The NME ratio is varied as: $22.2 \lesssim \mathcal{M}^{0\nu}_N/\mathcal{M}^{0\nu}_\nu \lesssim 76.3$.
Plots \ref{fig:NDMDlobSMee} and \ref{fig:NDMDlobSMhl} correspond to the standard mass mechanism. Plots \ref{fig:NDMDBSMee} and \ref{fig:NDMDBSMhl} show the individual non-standard contributions to GEMM and half-life from the RH and sterile neutrino channels in NO and IO cases. Finally, plots \ref{fig:NDMDlobBSMee} and \ref{fig:NDMDlobBSMhl} show the combined effect of standard and non-standard contributions. In all plots, cyan and magenta shaded bands represent the normal and inverted mass orderings, respectively; the overlap denotes the quasi-degenerate regime. Light-blue and light-pink horizontal bands represent current experimental limits on EMM from \textsf{LEGEND-200} and \textsf{KamLAND-Zen}, respectively, while the grey band shows the projected bound from \textsf{LEGEND-1000}. Vertical orange and yellow bands indicate constraints on absolute neutrino mass from \textsf{KATRIN} and \textsf{PLANCK}, respectively.

From the analysis, we find that in the case of NO mass spectrum, the EMM contribution from sterile neutrinos $S_{1,2,3}$ can dominate over that from active neutrinos $\nu_i$, while the contribution from RH neutrinos $N_{1,2,3}$ remains essentially independent of the lightest neutrino mass $m_{1(3)}$. Moreover, $\log m^{S}_{\beta\beta}$ varies nearly linearly with $\log m_{1(3)}$, with a slope governed by $-C_N$, indicating strong inverse dependence. For $m_{1(3)} \gtrsim 10^{-2}$ eV, $m^{S}_{\beta\beta}$ falls below $10^{-5}$ eV, far beneath the reach of next-generation experiments. However, the combined contribution $m_{\beta\beta}^{\nu+N+S} = \sqrt{|m^{\nu}_{\beta\beta,L}|^2 + |m^{N}_{\beta\beta,R}|^2 + |m^{S}_{\beta\beta,R}|^2}$ enhances the total GEMM in NO, pushing it within experimental sensitivity.

For $m_{\beta\beta}^{\nu+N+S}$ and $T_{1/2}^{\nu+N+S}$ in \ref{fig:NDMDlobBSMee} and \ref{fig:NDMDlobBSMhl}, we do not account for interference, as it is negligible. At small $m_{1(3)}$, non-standard contributions are dominant enough that $m_{1(3)} \lesssim 3.96\times 10^{-5}$ eV in NO is ruled out by existing bounds from \textsf{KamLAND-Zen} and \textsf{LEGEND-200}. In IO, total contributions are ruled out for $\alpha=\beta=0$, but allowed for $\alpha=\pi$, $\beta=0$.

For NO and $\alpha=\beta=0$, $|m^{\nu+N+S}_{\beta\beta}| \gg |m^{\nu}_{\beta\beta}|$ holds for $m_1 \cong (3.86\times 10^{-5} - 1.05\times 10^{-3})$ eV, with $|m^{\nu+N+S}_{\beta\beta}| \gtrsim 0.005$ eV. In this range, $T_{1/2} \lesssim 1.21\times 10^{29}$ yrs. For $\alpha=\pi$, $\beta=0$, $|m^{\nu+N+S}_{\beta\beta}| \gtrsim 0.010$ eV and $T_{1/2} \lesssim 10^{28}$ yrs for $m_{1} \lesssim 2.0 \times 10^{-3}$ eV. The minimum value is $\mbox{min}(m^{\nu+N+S}_{\beta\beta}) \cong 9\times 10^{-4}$ eV at $m_1 \cong 9\times 10^{-3}$ eV. Notably, $|m^{\nu}_{\beta\beta}|$ vanishes at $m_1 \cong 2.26\times 10^{-3}$ eV, where $|m^{\nu+N+S}_{\beta\beta}| \cong 18.3\times 10^{-3}$ eV.

For IO, $|m^{\nu+N+S}_{\beta\beta}| \gtrsim 0.015$ eV for $m_{3} < 1.43\times 10^{-4}$ eV, with $|m^{\nu+N+S}_{\beta\beta}| > |m^{\nu}_{\beta\beta}|$ for all considered values of $\alpha$ and $\beta$. The approximate equality $|m^{\nu+N+S}_{\beta\beta}| = |m^{\nu}_{\beta\beta}|$ holds at $m_3 \gtrsim 10^{-3}$ eV. For $m_{3}> 3.38\times 10^{-5}$ eV, the predicted value of $T^{\nu+N+S}_{1/2}$ ranges from $9.83\times 10^{26} ~\text{yrs}<T^{\nu+N+S}_{1/2}< 2.47\times 10^{28}~\text{yrs}$.

\subsection{Charged lepton flavor violation: $\mu\to e \gamma$ predictions}\label{subsec:cLFVcase2}
Following our discussion on cLFV predictions in Case I in subsection~\ref{subsec:cLFVcase1}, we now turn to the computation of BR($\mu \to e \gamma$) in Case II, where the neutrino mixing matrices $V^{NN}$ and $V^{NS}$ are determined by fully derived Dirac mass terms, as summarized in Table~\ref{tab:case2}. Unlike Case I, the RHN masses in this case are treated as input parameters, allowing for a variety of possible mass hierarchies, as outlined in Table~\ref{tab:RHN_hierarchy_cases}.

Given that the active neutrino mass spectrum may follow either normal or inverted ordering, and considering the variation of the Dirac CP phase for each RHN mass pattern, a full scan over all possibilities for BR($\mu \to e \gamma$) in Case II would be extensive and is beyond the scope of this work.

Since we have also investigated neutrinoless double beta decay in Case II, we restrict our discussion of cLFV predictions to the common parameter space relevant for both $0\nu\beta\beta$ and $\mu\to e\gamma$. The $0\nu\beta\beta$ analysis in Case II (see summary table~\ref{tab:RelevantmNcases}) favors the RHN mass hierarchies $m_{N_F}$ for NO and $m_{N_A}$ for IO, as defined in Table~\ref{tab:RHN_hierarchy_cases}. Accordingly, we limit our analysis of $\mathrm{BR}(\mu\to e\gamma)$ in Case II to these specific RHN patterns along with the degenerate RHN mass scenario, case $m_{N_G}$. For completeness, we complement this by presenting both NO and IO scenarios for each RHN mass hierarchy in Appendix~\ref{app:clfv_caseII}, including the corresponding analytical expressions and plots.

As RHN masses are input parameters in this framework, the stringent \textsf{ATLAS} bound of $M_{W_R} = 4.8$~TeV can now be consistently realized. However, we begin by analyzing the predictions for a relaxed value of $M_{W_R}$ before commenting on the more constrained scenario. Throughout our discussion, we fix the $W_L$-$W_R$ mixing angle at $\xi = 10^{-6}$ and initially consider $M_{W_R} = 7$~TeV, which remains compatible with \textsf{ATLAS} constraints.

The approximate expression for the branching ratio in Case II is given by:
\begin{equation}
	\text{BR}(\mu \to e \gamma) = \underbrace{\boxed{\frac{3\alpha_{em}}{2\pi}}}_{\textcolor{blue}{\approx 3 \times 10^{-3}}} \times \left| G_N + G_S \right|^2,
	\label{BRdefOOM}
\end{equation}
where the form factors $G_N$ and $G_S$ are defined as:
\begin{equation}\label{GNcase2}
	G_N = \underbrace{\boxed{\frac{m^2_{W_L}}{m^2_{W_R}}}}_{\textcolor{blue}{\approx 1.32 \times 10^{-4}}}
	\underbrace{\sum_{i=1}^{3} V_{\mu i}^{NN} \, {V_{e i}^{NN}}^* \, G^\gamma(y_i)}_{\text{$G_N$ Sum Term}},
\end{equation}
\begin{equation}\label{GScase2}
	G_S = \underbrace{\boxed{\frac{m^2_{W_L}}{m^2_{W_R}}}}_{\textcolor{blue}{\approx 1.32 \times 10^{-4}}}
	\underbrace{\sum_{i=1}^{3} V_{\mu i}^{NS} \, {V_{e i}^{NS}}^* \, G^\gamma(y_i)}_{\text{$G_S$ Sum Term}}.
\end{equation}
Here, the mixing matrices $V^{NN}$ and $V^{NS}$ are specific to Case II and are detailed in Table~\ref{tab:case2}, while the loop function $G^\gamma(y_i)$ is defined in eq.~(\ref{loopFunc}).

For the range of BR($\mu \to e \gamma$) most relevant to the \textsf{MEG II} experiment \cite{MEGII:2023ltw}:
\begin{equation*}
	6 \times 10^{-14} \leq \text{BR}(\mu \to e \gamma) < 3.1 \times 10^{-13},
\end{equation*}
the required absolute value of the sum of the $G_N$ and $G_S$ Sum Terms can be estimated as:
\begin{equation}\label{SumTermRange}
	3.39 \times 10^{-2} \leq \left| G_N \text{ Sum Term} + G_S \text{ Sum Term} \right| < 7.70 \times 10^{-2}.
\end{equation}

In the following subsections, we analyze whether these bounds can be achieved within the framework, for the RHN mass hierarchies motivated by the $0\nu\beta\beta$ study.

\subsubsection{Case $m_{N_F}$ ($m_{N_3} < m_{N_2} < m_{N_1}$)}

Before delving deeper into how $\left| G_N \text{ Sum Term} + G_S \text{ Sum Term} \right|$ behaves with respect to the required range stated in eq.~(\ref{SumTermRange}), it is worthwhile to highlight that the $G_N$ Sum Term, as expressed in eq.~(\ref{GNcase2}), is independent of the active neutrino masses and hence unaffected by the choice of mass ordering. However, it does depend on the Dirac CP phase $\delta$. 

The approximate analytical expression for the $G_N$ Sum Term, evaluated at the best-fit values of the low-energy oscillation parameters and expressed in terms of $\delta$ and the corresponding loop functions, is given by:
\begin{equation}\label{SymbolicGNsumTerm}
	\begin{aligned}
		G_N \text{ Sum Term}&\approx \left(-0.30-0.08\,e^{i\delta}\right)G^{\gamma}\left(\frac{m^2_{N_1}}{m^2_{W_R}}\right)+\left(0.30-0.03\,e^{i\delta}\right)G^{\gamma}\left(\frac{m^2_{N_2}}{m^2_{W_R}}\right)\\
		&+\left(0.11\,e^{i\delta}\right)G^{\gamma}\left(\frac{m^2_{N_3}}{m^2_{W_R}}\right)
	\end{aligned}
\end{equation}

This expression serves as the most general form of the $G_N$ Sum Term in Case II, independent of the active neutrino mass ordering (NO or IO). However, we do not include the explicit analytical expressions for the loop functions here due to their highly involved dependence on the RHN masses. The behavior of the loop function with respect to RHN mass is illustrated in figure \ref{fig:gGammamN} of appendix~\ref{app:loopfunc_caseII}, which shows that $G^{\gamma}\left(\frac{m^2_{N_i}}{m^2_{W_R}}\right)$ increases approximately linearly with increasing RHN mass. Therefore, for the $m_{N_F}$ case under consideration, the loop function $G^{\gamma}\left(\frac{m^2_{N_1}}{m^2_{W_R}}\right)$ contributes the most.

From the expression in eq.~(\ref{SymbolicGNsumTerm}), it is also evident that the Dirac CP phase $\delta$ has only a minor effect. It mildly enhances or suppresses the overall magnitude of the $G_N$ Sum Term for $\delta = 0$ and $\delta = \pi$, respectively. However, this variation is not significant, and the overall sign of the $G_N$ Sum Term remains negative in the $m_{N_F}$ case.

We now discuss the analytical expression for the $G_S$ Sum Term, which depends on the active neutrino mass ordering. For the normal ordering (NO) scenario, the approximate expression of the $G_S$ Sum Term at best-fit values of the low-energy oscillation parameters is:
\begin{equation}\label{SymbolicGSSumTermNOmNF}
	\begin{aligned}
		G_S \text{ Sum Term}\Big|^{\text{case}\,m_{N_F}}_{\text{NO}} &\approx \frac{(-1.51-0.39\,e^{i\delta})\times10^{-3}}{m_1\,(\text{eV})}G^{\gamma}\left(\frac{m^2_{S_1}}{m^2_{W_R}}\right)\\
		&+\frac{(1.51-0.18\,e^{i\delta})\times10^{-3}}{\sqrt{7.5\times10^{-5}+[m_1\,(\text{eV})]^2}}\left[\frac{m_{N_2}}{m_{N_1}}\right]G^{\gamma}\left(\frac{m^2_{S_2}}{m^2_{W_R}}\right)\\
		&+1.12\,e^{i\delta}\times10^{-2}\left[\frac{m_{N_3}}{m_{N_1}}\right]G^{\gamma}\left(\frac{m^2_{S_3}}{m^2_{W_R}}\right).
	\end{aligned}
\end{equation}

The $G_S$ Sum Term in eq.~(\ref{SymbolicGSSumTermNOmNF}) varies significantly depending on the loop functions, the Dirac CP phase $\delta$, and the lightest active neutrino mass $m_1$. It is worth noting that the loop function for the $\mu \to e \gamma$ process mediated by the heaviest sterile neutrino ($m_{S_3}$ in NO and $m_{S_2}$ in IO) depends solely on the heaviest RHN mass and is independent of the active neutrino mass ordering. The behaviour of this loop function, $G^{\gamma}(m^2_{S_{3(2)}}/m^2_{W_R})$, is illustrated in figure~\ref{fig:gGammaHeaviestSL} of appendix~\ref{app:loopfunc_caseII}.

The other loop functions, $G^{\gamma}(m^2_{S_{2,1}}/m^2_{W_R})$, appearing in eq.~(\ref{SymbolicGSSumTermNOmNF}), depend on the active neutrino mass ordering and vary with both the lightest neutrino mass $m_1$ and the heaviest RHN mass $M_{N_1}$ for the case under consideration. Their behaviour is depicted in the contour plots in figures~\ref{fig:gGammaNOIntermtdSL} and \ref{fig:gGammaNOLightestSL} of appendix~\ref{app:loopfunc_caseII}.

From the loop-function dynamics, we find that the $G_S$ Sum Term in eq.~(\ref{SymbolicGSSumTermNOmNF}) achieves its largest (negative) magnitude for $\delta = 0$, large heaviest RHN mass, and a strong hierarchy among RHN masses. This behaviour supports the requirement in eq.~(\ref{SumTermRange}), particularly given the result from eq.~(\ref{SymbolicGNsumTerm}) that the $G_N$ Sum Term remains negative in the $m_{N_F}$ scenario. On the other hand, the case $\delta = \pi$ disfavors hierarchical RHN masses, as it leads to cancellations in $|G_N + G_S|$.

The takeaways for case $m_{N_F}$ with hierarchical RHN masses are: Dirac CP phase $\delta = 0$, large RHN masses, and a strong mass hierarchy among them. We therefore fix the RHN mass hierarchy as follows:
\begin{equation}
	m_{N_2} = 0.5\,m_{N_1}; \,\, m_{N_3} = 0.4\,m_{N_1} \hspace{1cm}(\text{case } m_{N_F})
\end{equation}
We now move forward to explore the second scenario: case $m_{N_A}$ for the RHN mass hierarchy, which is favoured by the $0\nu\beta\beta$ study in the IO mass spectrum of active neutrinos.

\subsubsection{Case $m_{N_A}$ ($m_{N_1} < m_{N_2} < m_{N_3}$)}

The $G_N$ Sum Term, defined in eq.~(\ref{GNcase2}), retains the approximate analytical expression given in eq.~(\ref{SymbolicGNsumTerm}). However, unlike case $m_{N_F}$ discussed in the previous subsection, the $G_N$ Sum Term in the present case $m_{N_A}$ can be positive or negative depending on the Dirac CP phase $\delta$ and the strength of the RHN mass hierarchy. From eq.~(\ref{SymbolicGNsumTerm}), together with the loop-function behaviour shown in figure~\ref{fig:gGammamN} of appendix~\ref{app:loopfunc_caseII}, it follows that for the hierarchy $m_{N_1} < m_{N_2} < m_{N_3}$ a large positive value of the $G_N$ Sum Term is obtained when $\delta = 0$, the hierarchy between $m_{N_3}$ and $m_{N_2}$ is mild, and the hierarchy between $m_{N_3}$ and $m_{N_1}$ is strong.

The $G_S$ Sum Term depends on the active neutrino mass spectrum. Its approximate analytical form for the IO scenario, with the low-energy oscillation parameters fixed at their best-fit values, is:
\begin{equation}\label{SymbolicGSSumTermIOmNA}
	\begin{aligned}
		G_S \text{ Sum Term}\Big|^{\text{case}\,m_{N_A}}_{\text{IO}} &\approx (-2.98-0.77\,e^{i\delta})\times10^{-2}\left[\frac{m_{N_1}}{m_{N_3}}\right] G^{\gamma}\left(\frac{m^2_{S_1}}{m^2_{W_R}}\right)\\
		&+(2.98-0.35\,e^{i\delta})\times10^{-2}\left[\frac{m_{N_2}}{m_{N_3}}\right]G^{\gamma}\left(\frac{m^2_{S_2}}{m^2_{W_R}}\right)\\
		&+\frac{5.66\,e^{i\delta}\times10^{-6}}{m_3\text{ (eV)}}\,G^{\gamma}\left(\frac{m^2_{S_3}}{m^2_{W_R}}\right).
	\end{aligned}.
\end{equation}

It varies significantly with the loop functions, the Dirac CP phase $\delta$, and the lightest active neutrino mass $m_{3}$, and attains its maximum positive value for $\delta = 0$, a mild hierarchy between $m_{N_3}$ and $m_{N_2}$, and a strong hierarchy between $m_{N_3}$ and $m_{N_1}$.

The key conclusions for case $m_{N_A}$ with hierarchical RHN masses are: $\delta = 0$, large RHN masses, a mild hierarchy between $m_{N_3}$ and $m_{N_2}$, and a strong hierarchy between $m_{N_3}$ and $m_{N_1}$. We therefore fix the RHN mass hierarchy as follows:
\begin{equation}
	m_{N_2} = 0.9\,m_{N_3}; \,\, m_{N_1} = 0.4\,m_{N_3} \hspace{1cm}(\text{case } m_{N_A})
\end{equation}

\noindent\textit{Benchmark rationale:} The RHN mass ratios in both hierarchies are chosen to realize the analytic tendencies of eqs.~(\ref{SymbolicGNsumTerm}), (\ref{SymbolicGSSumTermNOmNF}), and (\ref{SymbolicGSSumTermIOmNA}), together with the loop–function behaviour shown in appendix~\ref{app:loopfunc_caseII}, so as to maximize $|G_N+G_S|$ within the target range of eq.~(\ref{SumTermRange}). For case $m_{N_F}$ (NO), $m_{N_2}=0.5\,m_{N_1}$ and $m_{N_3}=0.4\,m_{N_1}$ implement a \emph{strong} hierarchy that keeps the $G_N$ sum negative while enhancing the (negative) $G_S$ term at $\delta=0$, thereby avoiding cancellations and boosting the net magnitude. For case $m_{N_A}$ (IO), $m_{N_2}=0.9\,m_{N_3}$ and $m_{N_1}=0.4\,m_{N_3}$ realize the required mix of a \emph{mild} $m_{N_3}$-$m_{N_2}$ splitting and a \emph{strong} $m_{N_3}$-$m_{N_1}$ hierarchy, which drives a positive $G_N$ sum and a constructive $G_S$ contribution at $\delta=0$. In both patterns, the choices keep $y_i=m_{N_i}^2/M_{W_R}^2$ in a perturbative off–threshold regime and are representative rather than unique: nearby ratios (e.g.\ $m_{N_2}/m_{N_1}\!\sim\!0.45$-$0.55$, $m_{N_3}/m_{N_1}\!\sim\!0.35$-$0.45$ for $m_{N_F}$; $m_{N_2}/m_{N_3}\!\sim\!0.85$-$0.95$, $m_{N_1}/m_{N_3}\!\sim\!0.3$-$0.5$ for $m_{N_A}$) yield the same qualitative enhancement and are chosen here for clarity in the subsequent phenomenology.

We now proceed to the final scenario, namely case $m_{N_G}$, which assumes degenerate RHN masses.

\subsubsection{Case $m_{N_G}$ ($m_{N_1} = m_{N_2} = m_{N_3}$)}

The case of degenerate (or nearly degenerate) RHN masses is arguably the most intriguing scenario within the framework under consideration. As can be seen from the analytical expression for the $G_N$ Sum Term in eq.~(\ref{SymbolicGNsumTerm}), all loop-functions become identical in this limit. Consequently, the $G_N$ Sum Term vanishes, leading to a null contribution to BR($\mu \to e \gamma$). This phenomenon is a manifestation of GIM suppression, as the neutrino mixing matrix $V^{NN}$ in Case II is unitary and corresponds to the PMNS matrix, as summarized in Table~\ref{tab:case2}.

In this scenario, the BR($\mu \to e \gamma$) receives contributions solely from sterile neutrinos. Since the sterile neutrino mass spectrum is governed by the active neutrino mass spectrum, we now present the approximate analytical forms of the $G_S$ Sum Term for both normal (NO) and inverted (IO) ordering of active neutrino masses, with low-energy oscillation parameters fixed at their best-fit values:

\begin{equation}\label{SymbolicGSSumTermNOmNG}
	\begin{aligned}
		G_S \text{ Sum Term}\Big|^{\text{case}\,m_{N_G}}_{\text{NO}} &\approx \frac{(-1.51-0.39\,e^{i\delta})\times10^{-3}}{m_1\,(\text{eV})}G^{\gamma}\left(\frac{m^2_{S_1}}{m^2_{W_R}}\right)\\
		&+\frac{(1.51-0.18\,e^{i\delta})\times10^{-3}}{\sqrt{7.5\times10^{-5}+[m_1\,(\text{eV})]^2}}G^{\gamma}\left(\frac{m^2_{S_2}}{m^2_{W_R}}\right)\\
		&+1.12\,e^{i\delta}\times10^{-2}\,G^{\gamma}\left(\frac{m^2_{S_3}}{m^2_{W_R}}\right),
	\end{aligned}
\end{equation}

\begin{equation}\label{SymbolicGSSumTermIOmNG}
	\begin{aligned}
		G_S \text{ Sum Term}\Big|^{\text{case}\,m_{N_G}}_{\text{IO}} &\approx (-2.98-0.77\,e^{i\delta})\times10^{-2} \,G^{\gamma}\left(\frac{m^2_{S_1}}{m^2_{W_R}}\right)\\
		&+(2.98-0.35\,e^{i\delta})\times10^{-2}\,G^{\gamma}\left(\frac{m^2_{S_2}}{m^2_{W_R}}\right)\\
		&+\frac{5.66\,e^{i\delta}\times10^{-6}}{m_3\text{ (eV)}}\,G^{\gamma}\left(\frac{m^2_{S_3}}{m^2_{W_R}}\right).
	\end{aligned}
\end{equation}
From these expressions, we find that only for the NO mass spectrum of active neutrinos does the $G_S$ Sum Term attain the required magnitude when the Dirac CP phase $\delta = 0$ and RHN masses are large. In contrast, for the IO case, although the $G_S$ Sum Term also reaches its maximum for $\delta = 0$ and large RHN mass, it remains highly suppressed and does not fall within the required region specified in eq.~(\ref{SumTermRange}). 

\begin{figure}
	
	\begin{subfigure}[b]{0.50\textwidth}
		\includegraphics[width=\linewidth]{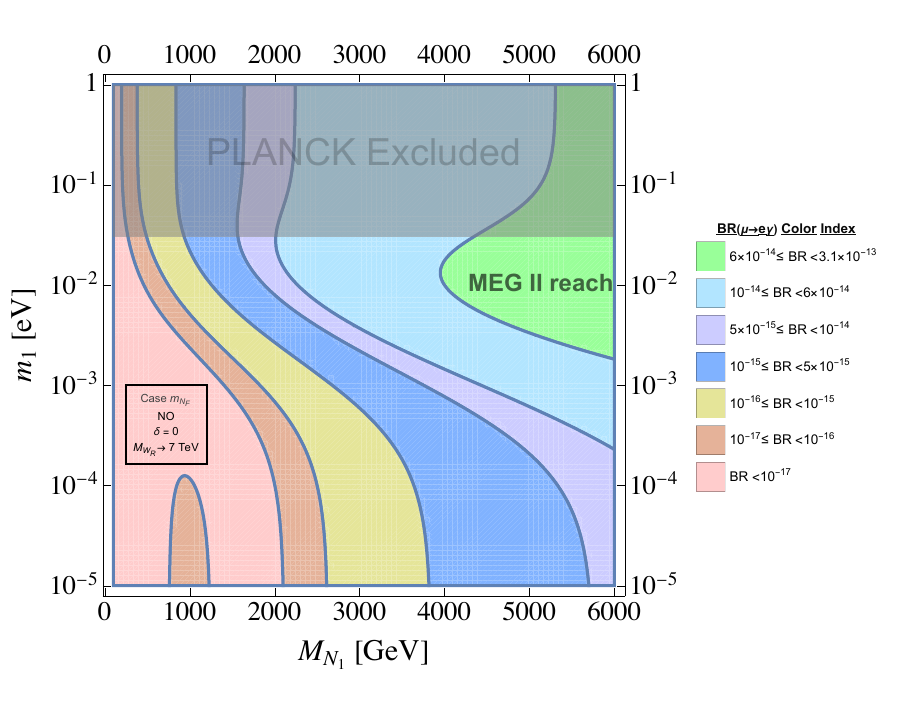}
		\caption{}
		\label{fig:RP_NDMD_mNFNO}
	\end{subfigure}
	\hspace{0.0001cm}
	\begin{subfigure}[b]{0.50\textwidth}
		\includegraphics[width=\linewidth]{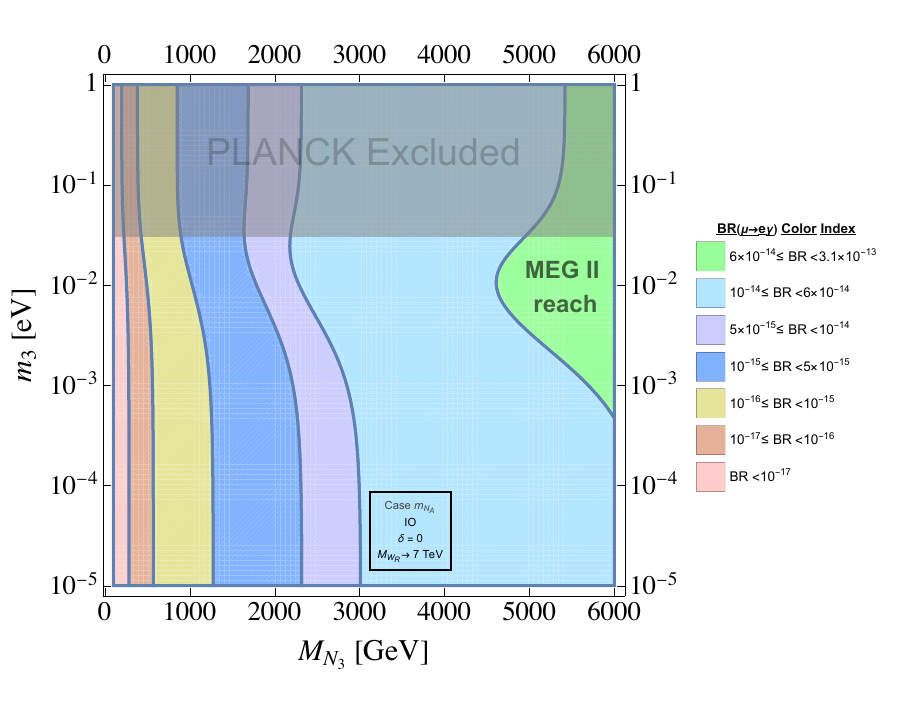}
		\caption{}
		\label{fig:RP_NDMD_mNAIO}
	\end{subfigure}
	
	\vspace{0.1cm}
	
	\begin{subfigure}[b]{0.50\textwidth}
		\includegraphics[width=\linewidth]{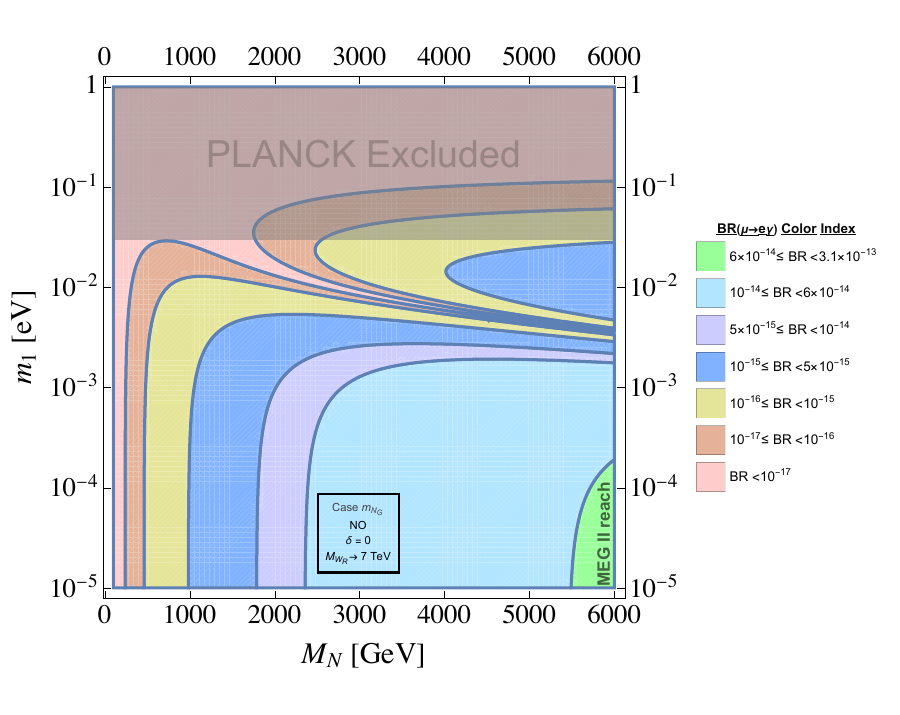}
		\caption{}
		\label{fig:RP_NDMD_mNGNO}
	\end{subfigure}
	\hspace{0.0001cm}
	\begin{subfigure}[b]{0.50\textwidth}
		\includegraphics[width=\linewidth]{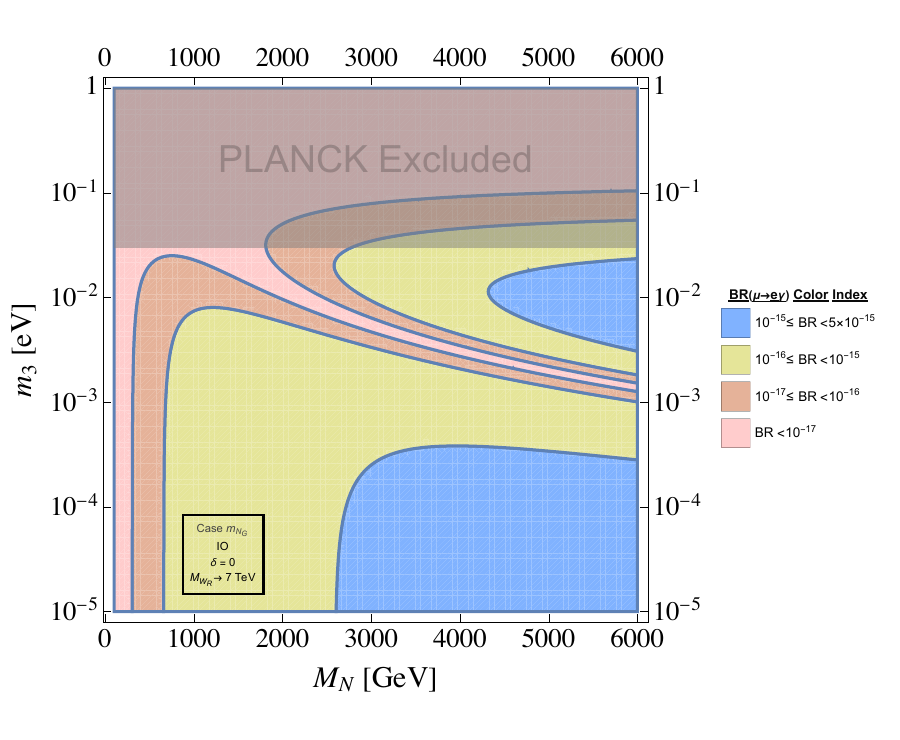}
		\caption{}
		\label{fig:RP_NDMD_mNGIO}
	\end{subfigure}
	\caption{Region plots of BR\((\mu \to e\gamma)\) in the plane of the heaviest RHN mass and the lightest active neutrino mass for Case II. Figures~\ref{fig:RP_NDMD_mNFNO} and \ref{fig:RP_NDMD_mNAIO} correspond to the RHN mass hierarchies \(m_{N_F}\) and \(m_{N_A}\) in NO and IO, respectively, as favoured by the $0\nu\beta\beta$ study in Case II. Figures~\ref{fig:RP_NDMD_mNGNO} and \ref{fig:RP_NDMD_mNGIO} correspond to the degenerate RHN mass hierarchy (case $m_{N_G}$) in the NO and IO mass spectra of active neutrinos. The green shaded region indicates the projected sensitivity of \textsf{MEG II}, while the translucent grey band marks the area excluded by the \textsf{PLANCK} bound on the sum of active neutrino masses. The input parameters follow the benchmark choices summarized in Table~\ref{tab:benchmarks_caseII}.}
	\label{fig:RegionPlotsCaseII_6panel}
\end{figure}

The key conclusions in the degenerate RHN mass scenario are the preference for a Dirac CP phase $\delta = 0$ and a large RHN mass. We conclude this discussion by summarizing the Key parameters fixed or varied in different RHN mass hierarchy cases $m_{N_{F,A,G}}$ in table~\ref{tab:benchmarks_caseII}.
\begin{table}
	\centering
	\scriptsize
	\renewcommand{\arraystretch}{1.3}
	\begin{tabular}{|c|c|c|c|}
		\hline
		\multicolumn{4}{|c|}{\textbf{cLFV Benchmark Parameters for Case II}} \\
		\hline
		\textbf{Parameter} & \multicolumn{3}{c|}{\textbf{Value or Range}} \\
		\hline
		\(M_{W_R}\) & \multicolumn{3}{c|}{7 TeV (fixed)} \\
		\hline
		\(\delta\) (Dirac CP phase) & \multicolumn{3}{c|}{0 (fixed)} \\
		\hline
		\(\xi\) ($W_L-W_R$ mixing)& \multicolumn{3}{c|}{\(10^{-6}\) (fixed)} \\
		\hline
		$m_N$ Cases&\textbf{Case }$m_{N_F}$&\textbf{Case }$m_{N_A}$&\textbf{Case }$m_{N_G}$\\
		\hline
		RHN mass hierarchy&$m_{N_3} < m_{N_2} < m_{N_1}$&$m_{N_1} < m_{N_2} < m_{N_3}$&$m_{N_1} = m_{N_2} = m_{N_3}=m_N$\\
		\hline
		Heaviest RHN mass & \(m_{N_1} = 100\text{--}6000\) GeV & \(m_{N_3} = 100\text{--}6000\) GeV &\(m_{N} = 100\text{--}6000\) GeV\\
		\hline
		\multirow{2}{*}{RHN mass ratio}& $m_{N_2}=0.5m_{N_1}$;& $m_{N_2}=0.9m_{N_3}$;& \multirow{2}{*}{(All Equal)}\\
		&$m_{N_3}=0.4m_{N_1}$&$m_{N_1}=0.4m_{N_3}$&\\
		\hline
		Heaviest sterile neutrino mass & \multirow{2}{*}{\(m_{S_{3(2)}} = 10\, m_{N_1}\)}&\multirow{2}{*}{ \(m_{S_{3(2)}} = 10\, m_{N_3}\)}&\multirow{2}{*}{\(m_{S_{3(2)}} = 10\, m_{N}\) }\\
		NO(IO)&&&\\
		\hline
			Lightest active neutrino mass & \multicolumn{3}{c|}{
				\begin{tabular}{c}
					$m_{1(3)} = (10^{-5} \text{--} 1)$ eV \\
					NO(IO)
				\end{tabular}
			} \\
			\hline
			\end{tabular}
\caption{Benchmark parameter choices used in the analysis of \(\mu \to e\gamma\) in Case II, for both Normal and Inverted mass ordering of active neutrinos. The RH gauge boson mass \(M_{W_R}\), Dirac CP phase \(\delta\), and left-right mixing angle \(\xi\) are fixed throughout the study. For each RHN mass hierarchy case ($m_{N_F}$, $m_{N_A}$, and $m_{N_G}$), the heaviest RHN mass is varied in a broad range, while specific ratios among RHN masses are chosen to capture representative features. The heaviest sterile neutrino mass is fixed to be ten times the heaviest RHN mass. The lightest active neutrino mass is varied across a wide range with consideration of experimental constraints.}
\label{tab:benchmarks_caseII}
\end{table}
Before proceeding to comment on the Case II study with the more stringent bound of \(M_{W_R} = 4.8\) TeV, we conclude our discussion of Case II with \(M_{W_R} = 7\) TeV by presenting region plots of BR(\(\mu \to e \gamma\)) as a function of the heaviest RHN mass and the lightest active neutrino mass, as shown in figure~\ref{fig:RegionPlotsCaseII_6panel}. These plots are generated using the benchmark parameters summarized in table~\ref{tab:benchmarks_caseII}.

Figure~\ref{fig:RP_NDMD_mNFNO}, corresponding to the NO scenario of case \(m_{N_F}\), shows that BR(\(\mu \to e\gamma\)) enters the \textsf{MEG II} sensitivity region only for large values of the heaviest RHN mass (\(m_{N_1} \gtrsim 4000\) GeV) and relatively large values of the lightest active neutrino mass. The IO scenario of case \(m_{N_A}\), shown in Figure~\ref{fig:RP_NDMD_mNAIO}, requires both larger values of the heaviest RHN mass (\(m_{N_3} \gtrsim 4610\) GeV) and higher lightest active neutrino masses to enter the \textsf{MEG II} sensitivity region.

Finally, Figures~\ref{fig:RP_NDMD_mNGNO} and~\ref{fig:RP_NDMD_mNGIO} show the region plots for BR(\(\mu \to e\gamma\)) in the NO and IO scenarios, respectively, for the degenerate RHN mass case \(m_{N_G}\). While the NO case exhibits a small \textsf{MEG II} sensitivity region for very large RHN masses (\(m_N \gtrsim 5490\) GeV) and very small lightest active neutrino masses, the IO case yields highly suppressed BR(\(\mu \to e\gamma\)) and does not enter the \textsf{MEG II} sensitivity region for any value in the explored parameter range.

The favourable ranges of the heaviest RHN mass and the lightest active neutrino mass, corresponding to each RHN mass hierarchy and active neutrino mass spectrum for which BR(\(\mu \to e\gamma\)) falls within the \textsf{MEG II} sensitivity region---as inferred from the region plots in Figure~\ref{fig:RegionPlotsCaseII_6panel}---are summarised in Table~\ref{tab:FavPara_caseII}.

\begin{table}
	\centering
	\scriptsize
	\renewcommand{\arraystretch}{1.3}
	\begin{tabular}{|c|c|c|c|c|}
		\hline
		\multicolumn{5}{|c|}{\textbf{cLFV Favourable Parameter Space for Case II}} \\
		\hline
		$m_N$ \textbf{Cases}&\textbf{Case }$m_{N_F}$&\textbf{Case }$m_{N_A}$&\multicolumn{2}{c|}{\textbf{Case }$m_{N_G}$}\\
		\hline
		$m_\nu$ \textbf{Ordering} &\textbf{NO}&\textbf{IO}&\textbf{NO}&\textbf{IO}\\
		\hline
		\textbf{Heaviest RHN}&\multirow{2}{*}{$m_{N_1}\gtrsim4000$}&\multirow{2}{*}{$m_{N_3}\gtrsim4610$}&\multirow{2}{*}{$m_{N}\gtrsim5490$}&\multirow{2}{*}{None}\\
		\textbf{mass (GeV)}&&&&\\
		\hline
		\textbf{Lightest active}&\multirow{2}{*}{$m_{1}\gtrsim2\times10^{-3}$}&\multirow{2}{*}{$m_{3}\gtrsim5\times10^{-4}$}&{$10^{-5}\lesssim m_{1}$}&\multirow{2}{*}{None}\\
		\textbf{$\nu$ mass (eV)}&&&$\lesssim1.58\times10^{-4}$&\\
		\hline
	\end{tabular}
	\caption{Favourable parameter space for BR\((\mu \to e\gamma)\) to enter the \textsf{MEG II} sensitivity zone in Case II. The table summarizes the required ranges of the heaviest RHN mass and the lightest active neutrino mass for different RHN mass hierarchies and for both normal (NO) and inverted (IO) ordering of active neutrinos. The lightest neutrino mass has a full range from \(10^{-5}\)~eV up to the \textsf{PLANCK}-excluded limit. The parameters \(M_{W_R} = 7\)~TeV, Dirac CP phase \(\delta = 0\), and left-right mixing angle \(\xi = 10^{-6}\) are fixed throughout.}
	\label{tab:FavPara_caseII}
\end{table}

After discussing cLFV predictions in Case II with \(M_{W_R} = 7\) TeV, we are now in a position to comment on the cLFV study in Case II with the most stringent bound of \(M_{W_R} = 4.8\) TeV.\\

\textbf{\underline{Comment on cLFV predictions in Case II with \(M_{W_R}=4.8\) TeV}}\\

In the analysis of BR($\mu\to e\gamma$) with the most stringent bound of \(M_{W_R} = 4.8\) TeV, the required absolute value of the sum of the $G_N$ and $G_S$ Sum Terms, as specified in eq.~(\ref{SumTermRange}), modifies to:
\begin{equation}\label{SumTermRange48}
	1.59 \times 10^{-2} \leq \left| G_N \text{ Sum Term} + G_S \text{ Sum Term} \right| < 3.62 \times 10^{-2}.
\end{equation}
The analytical expressions of the $G_N$ and $G_S$ Sum Terms, discussed earlier for the BR($\mu\to e\gamma$) analysis with \(M_{W_R} = 7\) TeV in Case II, remain unchanged, except that the loop functions are slightly enhanced. Thus, it is expected that the requirement in eq.~(\ref{SumTermRange48}) is satisfied within the allowed parameter space of this scenario.

However, it is worth noting that RHN masses in this case are constrained to lie within a narrow range: $3600 < m_N < 4800$ GeV. Consequently, the smallest achievable hierarchy among RHNs is theoretically limited to \(m_{N_{\text{lightest}}}/m_{N_{\text{heaviest}}} \approx 0.75\). We omit a detailed analysis of this stringent case and instead present the BR($\mu\to e\gamma$) behaviour through plots for the NO and IO spectra of active neutrinos in appendix \ref{48TeVcLFV}.

With this, we conclude our discussion on cLFV in case II and proceed to explore the implications of the cLFV-relevant parameter space for $0\nu\beta\beta$ decay in case II.

\subsection{$0\nu\beta\beta$ predictions within the cLFV-relevant region}\label{subsec:0nubb_caseII_common}
Following subsection~\ref{subsec:0nubb_caseII_fav}, we have the analytical expressions of EMM for the RHN and sterile neutrino channels in Case II (see eq.~(\ref{meffCaseII})), along with the preferred RHN mass hierarchies summarised in Table~\ref{tab:RelevantmNcases}. These results are general for Case II, and we now employ them in the context of the cLFV-preferred parameter space to study the corresponding $0\nu\beta\beta$ decay predictions. 

From Table~\ref{tab:FavPara_caseII}, we have RHN mass bounds for BR($\mu\to e \gamma$) to enter the \textsf{MEG II} sensitivity region in Case II for the RHN mass hierarchies: case $m_{N_F}$ (NO), case $m_{N_A}$ (IO), and case $m_{N_G}$ (NO). 

Taking these bounds into consideration, we fix the heaviest RHN mass at 5500~GeV. Accordingly, the representative values of the heaviest, intermediate, and lightest RHN masses are fixed as follows, as per Table~\ref{tab:FavPara_caseII}:
\begin{equation}\label{RHNmassesForCase2cLFV}
	\begin{aligned}
		m_{N_1}= 5500\text{ GeV };\hspace{1pt}m_{N_2}= 2750\text{ GeV};\hspace{1pt}m_{N_3}=2200\text{ GeV}\hspace{0.8cm}&\textbf{ case }\mathbf{m_{N_F}}\hspace{1pt}\textbf{(NO)}\\
		m_{N_3}= 5500\text{ GeV };\hspace{1pt}m_{N_2}= 4950\text{ GeV};\hspace{1pt}m_{N_1}=2200\text{ GeV}\hspace{0.8cm}&\textbf{ case }\mathbf{m_{N_A}}\hspace{1pt}\textbf{(IO)}\\
		m_{N_1}=m_{N_2}=m_{N_3}=5500\text{ GeV}\hspace{4.66cm}&\textbf{ case }\mathbf{m_{N_G}}\hspace{1pt}\textbf{(NO/IO)}
	\end{aligned}
\end{equation}

The RHN masses in eq.~(\ref{RHNmassesForCase2cLFV}) define a viable parameter space for $0\nu\beta\beta$ decay within the cLFV ($\mu\to e\gamma$)–relevant regime in Case II. With these RHN mass values, we find that the predictions of $0\nu\beta\beta$ decay rates in the cLFV-preferred region of Case~II are almost identical to those obtained in Case~I. For clarity, we omit separate plots for Case II in this regime and refer the reader to the representative results already provided for Case I (see subsection~\ref{subsec:0nubb_caseI_common}), since they numerically and qualitatively exhaust the relevant predictions for both scenarios. This similarity arises because in both cases the heaviest RHN mass is fixed at a large value (5500~GeV in the NO case), which strongly suppresses the role of neutrino mixing dynamics \cite{Fang:2021jfv,Goswami:2020loc}. As a consequence, the overall behaviour of the generalized effective Majorana mass (GEMM) and the corresponding half-life is dominantly governed by the contribution of the heaviest RHN, while the effects from neutrino mixing, and thus from the Dirac mass matrix texture, become negligible. For the NO spectrum, the results overlap almost exactly between Case~I and Case~II. A slight deviation is, however, observed in the IO scenario: in Case~I the heaviest RHN mass was taken to be 3000~GeV, whereas in Case~II it is fixed at 5500~GeV. This difference leads to a marginal shift in the predicted behaviour of GEMM at very small lightest neutrino masses, $m_3 \sim 10^{-5}$~eV. Overall, this reflects the generic feature that in both Case~I and Case~II the non-standard contributions scale inversely with the heaviest RHN mass, thereby making the heavy state the decisive parameter in shaping the $0\nu\beta\beta$ phenomenology within the cLFV-relevant regime.

It is worth emphasising that, although Case~II is theoretically distinct, in the phenomenologically relevant regime its predictions effectively converge with those of Case~I. This convergence highlights the robustness of the allowed parameter space, rather than signalling any redundancy of Case~II.
\\

\textbf{\underline{Comment on $0\nu\beta\beta$ in Case II with degenerate RHN masses}}\\

For the representative value $m_N = 5500$~GeV, the degenerate case $m_{N_G}$ yields $0\nu\beta\beta$ predictions (GEMM and half-life) essentially identical to those obtained for Case~II in the cLFV-relevant parameter space. This behaviour follows from the structure of the leading non-standard terms summarised in Table~\ref{tab:RelevantmNcases}: for both NO and IO, these terms are inversely proportional to the heaviest RHN mass and are independent of the intermediate and lightest RHN masses. Consequently, the predicted GEMM and half-life effectively depend only on the heaviest RHN mass.\\

\textbf{\underline{Comment on variation of $|m_{\beta\beta}^{\nu+N_R+S_L}|$ with respect to $\sum m_\nu$}}\\

\begin{figure}
	\centering
	\begin{minipage}[b]{0.56\textwidth}
		\centering
		\includegraphics[width=\textwidth]{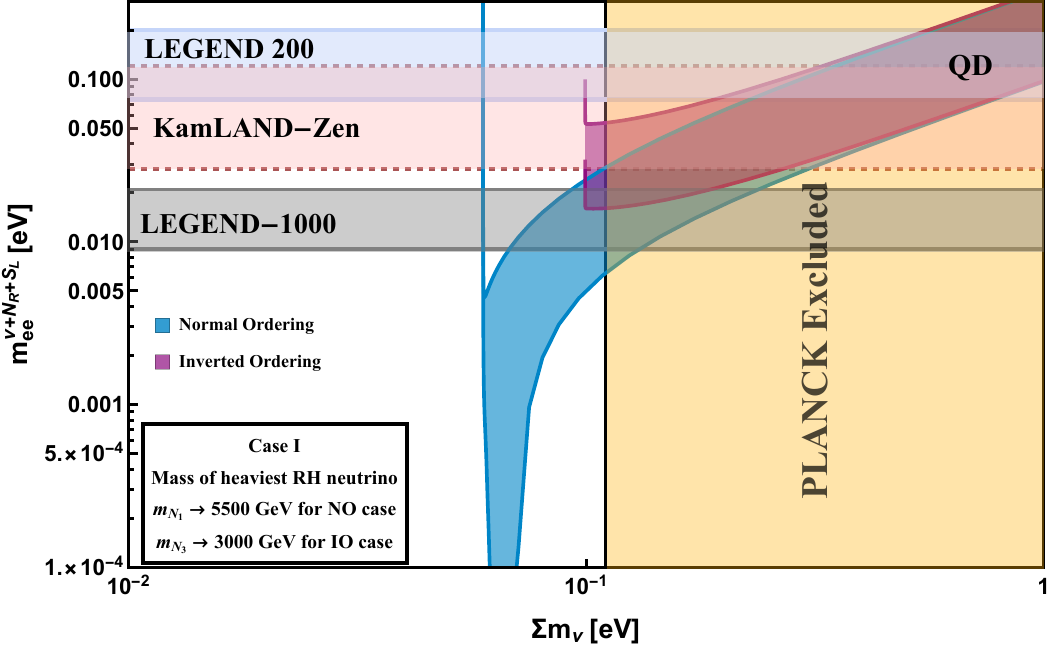}
	\end{minipage}
	\caption{Allowed region of the generalized effective Majorana mass parameter $|m_{\beta\beta}^{\nu+N_R+S_L}|$ as a function of the sum of light neutrino masses $\sum m_\nu$ for Case~I within the cLFV-relevant parameter space.}
	\label{fig:nusumMassPlot}
\end{figure}

Using the \textsf{PLANCK} bound on the sum of neutrino masses, $\sum m_\nu < 0.12\,\text{eV}$ at 95\% C.L.~\cite{Planck:2018vyg}, we show in figure~\ref{fig:nusumMassPlot} the generalized effective Majorana mass parameter as a function of $\sum m_\nu$ for Case~I within the cLFV-relevant parameter space. The $1\sigma$ bound on $\sum m_\nu$ ($\sum m_\nu < 84\,\text{meV}$ at $1\sigma$ C.L.) quoted in~\cite{Palanque-Delabrouille:2014jca,DellOro:2015kys} was inferred from earlier cosmological limits at 95\% C.L. With the more stringent recent \textsf{PLANCK} constraint, the corresponding $1\sigma$ limit is expected to be tighter. However, even for $\sum m_\nu < 84\,\text{meV}$, the inverted ordering (IO) band is disfavoured by cosmology.

For completeness, we note that Case~II yields a qualitatively similar plot in this cLFV-relevant (heavy RHN mass) regime, since mixing effects are suppressed and the predictions effectively converge with those of Case~I.

\textbf{\underline{Dependence on $M_{W_R}$ and LHC sensitivity}}\\

Besides the overall $M_{W_R}^{-4}$ scaling already noted,
lowering $M_{W_R}$ simultaneously (i) enhances $\mathrm{BR}(\mu\to e\gamma)$ and (ii) increases the production rate for $pp\to W_R\to \ell N$ at the LHC (and HL-LHC). For illustration, $(7/8)^4\simeq0.59$ and $(7/10)^4\simeq0.24$, whereas a modest decrease to $6.5$~TeV gives $(7/6.5)^4\simeq1.3$. Thus, a slightly smaller $M_{W_R}$ can move collider benchmarks closer to Run-3/HL-LHC sensitivity (see, e.g., \cite{Keung:1983uu,Nemevsek:2011hz,Deppisch:2015qwa}) while keeping the same heavy-fermion spectrum. In this work we keep $M_{W_R}=7$~TeV for a conservative, intensity-frontier-oriented baseline and to cleanly compare with our $0\nu\beta\beta$ analysis, but a collider-oriented benchmark with $M_{W_R}\!\sim\!6$–$6.5$~TeV could be included for completeness.



\section{Conclusion}
\label{sec:concl}
In this work, we have carried out a comprehensive analysis of new physics contributions to neutrinoless double beta decay and the charged lepton flavor-violating process $\mu \to e \gamma$ in the context of a left-right symmetric model with a double seesaw mechanism. The double seesaw setup naturally generates large Majorana masses for right-handed neutrinos (RHNs), leading to dominant new physics contributions to $\mu \to e \gamma$ and $0\nu\beta\beta$ decay rates through RHN self-mixing and RHN–sterile neutrino mixing, while the heavy–light neutrino mixing remains highly suppressed. We considered two distinct case studies to parametrize the heavy-neutrino mixing in terms of Dirac mass terms: in Case I, the Dirac mass matrix is symmetry-motivated and taken proportional to the identity, whereas in Case II it is fully determined by the model framework. We include Case~I because the choice $M_D \propto \mathbbm{I}$ is a commonly adopted LRSM texture for the Dirac neutrino mass matrix; it provides a clean reference in which flavor-structure effects are minimized and enter only through low-energy oscillation parameters, enabling straightforward comparison with known literature and a clear contrast with the model-determined flavor structure in Case~II. While both cases rely on low-energy oscillation inputs, Case~II in particular stands out for its predictive power. The flavor structure is entirely fixed by low-energy neutrino data and the discrete LR symmetry $\mathcal{C}$ (with no additional arbitrary phases), whereas the overall heavy-sector scales are treated as inputs constrained by phenomenology. This enables direct experimental tests and thus allows falsifiability of the framework. Furthermore, we explored $0\nu\beta\beta$ decay both in its favorable parameter space (chosen for optimal enhancement of the decay rate) and in the cLFV-preferred parameter space determined by $\mu \to e \gamma$, with the aim of identifying a common region that can simultaneously accommodate both processes.

Throughout our analysis, we have used experimentally allowed fixed benchmark values for the right-handed gauge boson mass ($M_{W_R} = 7$~TeV) and the $W_L$–$W_R$ mixing angle ($\xi = 10^{-6}$). In addition, the heaviest sterile neutrino mass is fixed to be ten times the heaviest RHN mass in order to realize the double seesaw approximation. For varying parameters, we considered the following reference ranges in our calculations and plots: $10^{-5}$--$1$~eV (for the lightest neutrino mass), $100$--$6000$~GeV (for the heaviest RHN mass), and $22.2$--$76.3$ (for the ratio of nuclear matrix elements $\mathcal{M}^{0\nu}_N/\mathcal{M}^{0\nu}_\nu$). The CP-violating phases (Dirac phase $\delta$ and Majorana phases $\alpha,\beta$) are varied within the range $[0,2\pi]$. However, for BR($\mu\to e\gamma$) calculations, we kept the Dirac CP phase fixed at $\delta=0$ to achieve the required enhancement for experimental relevance. In the Case~I scenario (i.e., Dirac mass matrix proportional to the identity matrix), the heaviest RHN mass together with the lightest neutrino mass completely determines the heavy-neutrino mixing, enabling straightforward phenomenological discussions. In contrast, for Case~II (i.e., Dirac mass matrix fully determined by the model framework), all three generations of RHN masses are treated as input parameters, leading to seven possible mass configurations.

The new physics contributions to BR($\mu\to e\gamma$) in Case~I arise from both right-handed and sterile neutrino–mediated one-loop processes. For the heaviest RHN mass $m_{N_1} \gtrsim 4210$~GeV, the combined contributions from heavy neutrinos enhance the BR($\mu\to e\gamma$) predictions over a wide range of $m_1$, bringing them within the sensitivity of the upcoming \textsf{MEG~II} experiment (figure~\ref{fig:DMD_RP_NO}). In contrast, in the IO scenario of Case~I, the sterile neutrinos provide the dominant contribution across the entire allowed range of the lightest neutrino mass $m_3$. In particular, for the heaviest RHN mass in the range $2400 \lesssim m_{N_3} \lesssim 3530$~GeV, the predicted BR($\mu\to e\gamma$) lies within the reach of \textsf{MEG~II} for a wide range of $m_3$, whereas for $m_{N_3} > 3530$~GeV, the predicted branching ratio is mostly excluded by current \textsf{MEG} limits at 90\%~C.L. (figure~\ref{fig:DMD_RP_IO}).

We next estimated the $0\nu\beta\beta$ decay rates in Case~I (figure~\ref{fig:onubbCaseI}) using representative values $m_{N_1} = 5500$~GeV (NO) and $m_{N_3} = 3000$~GeV (IO). For $m_{1(3)} \lesssim 3.85 \times 10^{-4}$~eV, non-standard contributions dominate over the standard light-neutrino exchange, while for $m_{1(3)} \gtrsim 5.75 \times 10^{-3}$~eV they are strongly suppressed, giving $|m_{\beta\beta}^{\nu+N+S}| \simeq |m_{\beta\beta}^{\nu}|$. In the NO case with $\alpha=\beta=0$, the inequality $|m_{\beta\beta}^{\nu+N+S}| > |m_{\beta\beta}^{\nu}|$ holds for $1.4 \times 10^{-4} \lesssim m_1 \lesssim 4.7 \times 10^{-4}$~eV. For very small masses, $m_{1(3)} \lesssim 1.05 \times 10^{-5}$~eV, the generalized effective Majorana mass (GEMM) $|m_{\beta\beta}^{\nu+N+S}|$ enters regions already excluded by \textsf{LEGEND-200} and \textsf{KamLAND-Zen}, with the IO case reaching exclusion at slightly lower $m_3$. Across a wide range of $m_3$ and for all $\alpha,\beta$, no significant enhancement appears in the IO scenario, leaving $|m_{\beta\beta}^{\nu+N+S}| \simeq |m_{\beta\beta}^{\nu}|$.

The Case~II scenario comprises six hierarchical and one degenerate mass-spectrum configurations for RHNs. Interestingly, in this case the leading non-standard contribution to $0\nu\beta\beta$ decay depends solely on the heaviest RHN mass, effectively reducing the six hierarchical scenarios to three independent ones. Among these three, we selected case $m_{N_F}(m_{N_3}<m_{N_2}<m_{N_1})$ for NO and case $m_{N_A}(m_{N_1}<m_{N_2}<m_{N_3})$ for IO, as they maximize the GEMM in the corresponding mass ordering. In $m_{N_F}$ and $m_{N_A}$, the new-physics contributions to BR($\mu\to e\gamma$) arise from both RH and sterile neutrinos, whereas in the degenerate case ($m_{N_G}$) the RHN contributions vanish and BR($\mu\to e\gamma$) receives input solely from sterile neutrinos.

For case $m_{N_F}$, an enhanced BR($\mu\to e\gamma$) requires a strong hierarchy among RHN masses. Specifically, with $m_{N_2}=0.5 \,m_{N_1}$ and $m_{N_3}=0.4 \,m_{N_1}$, we find that for the NO scenario and $m_{N_1}\gtrsim 4000$~GeV, the branching ratio lies within the \textsf{MEG II} sensitivity for $m_1 \gtrsim 2\times 10^{-3}$~eV (figure~\ref{fig:RP_NDMD_mNFNO}). For case $m_{N_A}$, adopting $m_{N_2}=0.9\,m_{N_3}$ and $m_{N_1}=0.4\,m_{N_3}$, the IO scenario predicts BR($\mu\to e\gamma$) within the \textsf{MEG II} reach when $m_{N_3}\gtrsim 4610$~GeV and $m_3 \gtrsim 5\times 10^{-4}$~eV (figure~\ref{fig:RP_NDMD_mNAIO}). The complementary cases $m_{N_F}$ (IO) and $m_{N_A}$ (NO) are discussed in appendix~\ref{app:clfv_caseII}. Finally, in the degenerate case $m_{N_G}$, the required enhancement is realized only in the NO scenario, with $m_N \gtrsim 5490$~GeV and $10^{-5}$~eV $\lesssim m_1 \lesssim 1.58\times 10^{-4}$~eV (figure~\ref{fig:RP_NDMD_mNGNO}) and no such enhancement is seen in the IO scenario (figure~\ref{fig:RP_NDMD_mNGIO}).

Within the cLFV-relevant parameter space, we analyzed the $0\nu\beta\beta$ decay rates in Case~II by fixing the heaviest RHN mass at $m_{N_{1(3)}} = 5500$~GeV for the $m_{N_F}$ (NO) and $m_{N_A}$ (IO) hierarchies, respectively. In this regime the predictions effectively coincide with those of Case~I: the GEMM and half-life are dominated by the heaviest RHN and become insensitive to the details of the Dirac texture, so mixing-induced differences are highly suppressed. Apart from a mild IO shift due to the different benchmark choice of the heaviest RHN mass, both cases collapse to the same effective description, underscoring the robustness of our bounds and justifying the use of the Case~I plots as representative for the cLFV-relevant region. We do not discuss the degenerate RHN case separately, since for \(m_N=5.5~\text{TeV}\) its \(0\nu\beta\beta\) and cLFV predictions effectively coincide with the hierarchical benchmarks—the rates are dominated by the heaviest RHN, rendering the lighter states irrelevant. Furthermore, in the cLFV-preferred parameter space, we find that the predicted GEMM in the IO case is largely excluded for both Case~I and Case~II when confronted with the sum of neutrino masses. In fact, the recent \textsf{PLANCK} bound on $\sum m_{\nu}$ already disfavors the IO case once the $1\sigma$ limit is applied.

In addition to the cLFV-preferred region, we also examined a favorable parameter space with smaller RHN masses, where the $0\nu\beta\beta$ rates receive optimal enhancement in Case~II (subsection~\ref{subsec:0nubb_caseII_fav}). Specifically, we considered representative values for the heaviest, intermediate, and lightest RHN masses as 300, 250, and 200~GeV, respectively. For NO with $\alpha=\beta=0$, we found that $|m^{\nu+N+S}_{\beta\beta}| \gg |m^{\nu}_{\beta\beta}|$ holds for $m_1 \cong (3.86\times 10^{-5} - 1.05\times 10^{-3})$~eV, predicting rates accessible in current and upcoming experiments. However, for very small lightest neutrino mass, $m_{1(3)}\lesssim 3.96\times 10^{-5}$~eV, the non-standard contributions in the NO case become excessively large and are already excluded by \textsf{KamLAND-Zen} and \textsf{LEGEND-200}. In the IO scenario, the total contributions are excluded for $\alpha=\beta=0$ but remain experimentally viable for $\alpha=\pi,\,\beta=0$. Moreover, for IO, the inequality $|m^{\nu+N+S}_{\beta\beta}| \gg |m^{\nu}_{\beta\beta}|$ holds for $m_3 < 1.43\times 10^{-4}$~eV across all $\alpha,\, \beta$, while the approximate equality $|m^{\nu+N+S}_{\beta\beta}| \approx |m^{\nu}_{\beta\beta}|$ is restored for $m_3 \gtrsim 10^{-3}$~eV. These findings for $0\nu\beta\beta$ decay in Case~II differ significantly from those of Case~I with smaller RHN masses (see \cite{Patra:2023ltl}), thereby establishing a clear phenomenological distinction between the two cases.\\

\noindent\textbf{Impact of $M_{W_R}$ and collider reach:}
Throughout, we adopt $M_{W_R}=7$~TeV as a collider–safe baseline above current LHC exclusions for sub-TeV $N_R$. This choice mildly suppresses the non-standard contributions to $0\nu\beta\beta$ and $\mu\to e\gamma$, making our intensity-frontier projections conservative. Our qualitative conclusions do not hinge on this value: a modest reduction of $M_{W_R}$ would simultaneously strengthen both signals and move the same benchmarks closer to direct production sensitivity at the \textsf{LHC}, without altering the pattern of allowed and excluded regions identified in this study. Looking ahead, the \textsf{HL-LHC} is expected to probe a substantial portion of the parameter space highlighted here, providing a complementary test of the double-seesaw LRSM.

We end our discussion by pointing that in the doublet LRSM considered here, $\mu\to 3e$ and $\mu$--$e$ conversion arise only at one loop with rates controlled by the same heavy-neutrino mixings that enter $\mu\to e\gamma$. This correlation yields dipole dominance, with subleading box contributions from right-handed currents. A dedicated analysis of $\mu$--$e$ conversion (including nuclear inputs) and $\mu\to 3e$ is therefore well motivated and complementary to our study. Our emphasis on the radiative muon mode is driven by experimental reach (\textsf{MEG II}) and its clean sensitivity to the dipole amplitude; forthcoming searches such as \textsf{Mu3e}, \textsf{Mu2e}, and \textsf{COMET} will probe complementary directions of parameter space and provide robust cross-checks. Taken together, improved cosmological constraints on $\sum m_\nu$, next-generation $0\nu\beta\beta$ searches, and the broader cLFV program will sharpen tests of the parameter space identified here. Recently the first \textsf{JUNO} reactor-oscillation results have appeared \cite{JUNO:2025gmd}, reporting sub-percent-level precision on \(\sin^2\theta_{12}\) and \(\Delta m^2_{ee}\); these improved oscillation inputs can be incorporated in our study without modifying the qualitative conclusions.

\section*{Acknowledgements}
PA would like to acknowledge the financial support obtained from the Ministry of Education, Government of India. RB would like to acknowledge the Ministry of Electronics and IT for the financial support through the
Visvesvaraya fellowship scheme for carrying out this research work. PS would like to acknowledge the Ministry of Education, Government of India for financial support through the Institute of Eminence funding to IIT Bombay. UP would like to acknowledge SINP, Kolkata as his research funding source. SP would like to acknowledge the funding support from SERB, Government of India, under MATRICS project with grant no. MTR/2023/000687.

\newpage
\appendix	

\section{Benchmark Viability of $M_{W_R}=4.8$~TeV in Case~I}\label{app:WRbenchmark}
Considering the NO and IO mass splittings of RH neutrinos as defined in eqs.~(\ref{mimNi_NO}) and~(\ref{mimNi_IO}), respectively, and using best-fit oscillation parameters from Table~\ref{tab:exp_nu}, we fix the lightest RHN mass at \(m_{N_{3(2)}} = 3.6\)~TeV. The intermediate and heaviest RHN masses can then be expressed in terms of the lightest active neutrino mass \(m_{1(3)}\) as follows:
\begin{itemize}
	\item[] \textbf{ \underline{For NO}}
	\begin{equation}
		\begin{aligned}
			m_{N_2}&\approx (3.6~\text{TeV})\sqrt{\frac{2.55\times10^{-3}~\text{eV}^2}{7.5\times10^{-5}~\text{eV}^2 + [m_{1}~(\text{eV})]^2}},\\
			m_{N_1}&\approx (3.6~\text{TeV})\sqrt{\frac{2.55\times10^{-3}~\text{eV}^2}{[m_{1}~(\text{eV})]^2}+1}.
		\end{aligned}
	\end{equation}
	\item[] \textbf{ \underline{For IO}}
	\begin{equation}
		\begin{aligned}
			m_{N_1}&\approx (3.6~\text{TeV})\sqrt{\frac{2.53\times10^{-3}~\text{eV}^2 + [m_{3}~(\text{eV})]^2}{2.45\times10^{-3}~\text{eV}^2 + [m_{3}~(\text{eV})]^2}},\\
			m_{N_3}&\approx (3.6~\text{TeV})\sqrt{\frac{2.53\times10^{-3}~\text{eV}^2}{[m_{3}~(\text{eV})]^2}+1}.
		\end{aligned}
	\end{equation}
\end{itemize}

For both NO and IO cases, the heaviest RHN mass scales inversely with the lightest active neutrino mass, i.e., \(m_{N_{1(3)}} \propto 1/m_{1(3)}\). Using the \textsf{PLANCK} upper bound on the lightest active neutrino mass, \(m_{1(3)} = 0.03\)~eV, we obtain the lower bounds on the intermediate and heaviest RHN masses for the lightest RHN mass, \(m_{N_{3(2)}} = 3.6\)~TeV:
\begin{equation}
	\begin{aligned}
		&m_{N_2} = 6.80~\text{TeV},\hspace{0.5cm} m_{N_1} = 7.05~\text{TeV} \hspace{0.5cm}\textbf{(for NO)},\\
		&m_{N_1} = 3.64~\text{TeV},\hspace{0.5cm} m_{N_3} = 7.03~\text{TeV} \hspace{0.5cm}\textbf{(for IO)}.
	\end{aligned}
\end{equation}

\begin{figure}
	\centering
	\begin{minipage}[b]{0.48\textwidth}
		\centering
		\includegraphics[width=\textwidth]{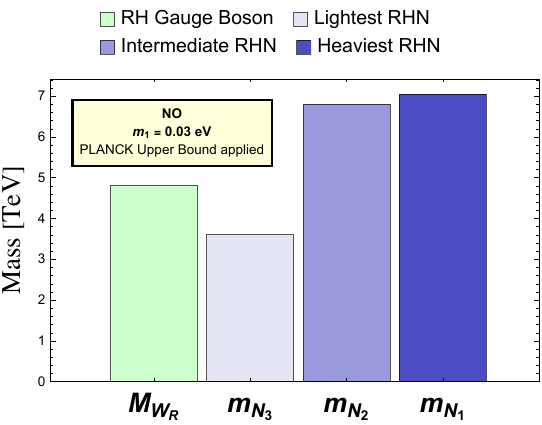}
	\end{minipage}
	\hfill
	\begin{minipage}[b]{0.48\textwidth}
		\centering
		\includegraphics[width=\textwidth]{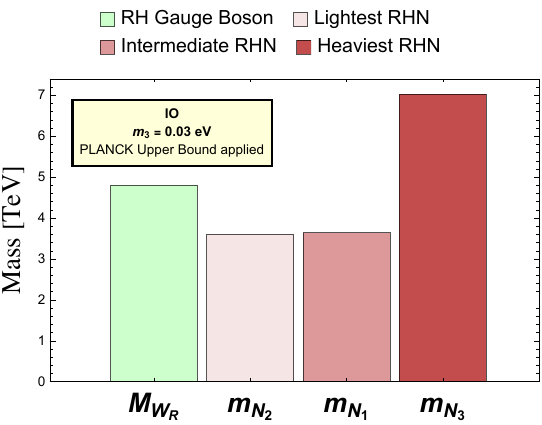}
	\end{minipage}
	\caption{Bar chart illustration of the mass spectrum of right-handed neutrinos in Case~I for both Normal Ordering (left) and Inverted Ordering (right) of active neutrino masses. The lightest RHN mass is fixed at \(3.6\)~TeV, and the remaining RHN masses are computed as functions of the lightest active neutrino mass using oscillation data. The values shown correspond to the maximal allowed lightest active neutrino mass, \(m_{1(3)} = 0.03\)~eV, as constrained by \textsf{PLANCK} data. The resulting RHN mass hierarchy demonstrates that in both cases, the heaviest RHN mass exceeds the benchmark RH gauge boson mass \(M_{W_R} = 4.8\)~TeV, thereby invalidating this benchmark under the assumption \(M_{W_R} > m_{N_i}\) used in collider bounds \cite{ATLAS:2023cjo}.}
	\label{fig:RHNsplitting}
\end{figure}

Since the heaviest RHN masses \(m_{N_{1(3)}}\) exceed the RH gauge boson mass \(M_{W_R} = 4.8\)~TeV in the considered case, this benchmark point is excluded based on the \textsf{ATLAS} result, which assumes \(M_{W_R} > m_{N_i}\). The above discussion is illustrated in figure~\ref{fig:RHNsplitting} via bar charts for both NO and IO scenarios of the active neutrino mass spectrum.

\section{Individual RHN and Sterile neutrino Contributions to the BR($\mu\to e\gamma$) (Case~I)}\label{app:tables_caseI}

For completeness, we collect here the detailed numerical breakdown of the form-factor sum terms corresponding to individual RHN and sterile neutrino contributions to the BR($\mu\to e\gamma$) in Case I. The tables \ref{tab:RHNOoM_caseI} and \ref{tab:SLOoM_caseI} correspond to the benchmarks summarized in Table~\ref{tab:benchmarks_caseI} and underpin the qualitative trends emphasized in subsection~5.1.

\begin{table}[h!]
	\centering
	\begin{tabular}{|c|c|c|c|c|c|}
		\hline
		\multicolumn{6}{|c|}{\textbf{BR} \boldmath\((\mu \to e\gamma)\) \textbf{RHN \& Sterile Neutrinos Contribution (Case I)}} \\
		\hline
		\multicolumn{6}{|c|}{\textbf{Normal Ordering (NO)}} \\
		\hline
		\multicolumn{3}{|c|}{\textbf{RHN Contribution}}&\multicolumn{3}{c|}{\textbf{Sterile Neutrino Contribution}}\\
		\hline
		\(\mathbf{m_{N_1}}\) & \(\mathbf{m_1}\) & \(\mathbf{G_N}\) \textbf{Sum Term}  & \(\mathbf{m_{N_1}}\) & \(\mathbf{m_1}\) & \(\mathbf{G_S}\) \textbf{Sum Term} \\
		\textbf{(GeV)} & \textbf{(eV)} & \textbf{Value} & \textbf{(GeV)} & \textbf{(eV)} & \textbf{Value} \\
		\hline
		\multirow{2}{*}{\textbf{100}} 
		& \(10^{-2}\) & \(-1.13\times10^{-5}\) & \multirow{2}{*}{\textbf{100}} & \(10^{-5}\) & \(-4.28\times10^{-10}\) \\
		\cline{2-3} \cline{5-6}
		& \(10^{-5}\) & \(-1.92\times10^{-5}\) & & \(10^{-2}\) & \(-4.02\times10^{-7}\) \\
		\hline
		\multirow{2}{*}{\textbf{2000}} 
		& \(10^{-2}\) & \(-3.86\times10^{-3}\) & \multirow{2}{*}{\textbf{2000}} & \(10^{-5}\) & \(-6.39\times10^{-6}\) \\
		\cline{2-3} \cline{5-6}
		& \(10^{-5}\) & \(-6.80\times10^{-3}\) & & \(10^{-2}\) & \(-3.57\times10^{-3}\) \\
		\hline
		\multirow{2}{*}{\textbf{4250}} 
		& \(10^{-2}\) & \(-1.27\times10^{-2}\) & \multirow{2}{*}{\textbf{4250}} & \(10^{-5}\) & \(-4.43\times10^{-5}\) \\
		\cline{2-3} \cline{5-6}
		& \(10^{-5}\) & \(-2.39\times10^{-2}\) & & \(6.90\times10^{-3}\) & \(-1.36\times10^{-2}\) \\
		\hline
		\multirow{2}{*}{\textbf{6000}} 
		& \(10^{-2}\) & \(-1.95\times10^{-2}\) & \multirow{2}{*}{\textbf{6000}} & \(10^{-5}\) & \(-1.02\times10^{-4}\) \\
		\cline{2-3} \cline{5-6}
		& \(10^{-5}\) & \(-3.87\times10^{-2}\) & & \(5.20\times10^{-3}\) & \(-2.37\times10^{-2}\) \\
		\hline
	\end{tabular}
	\caption{For selected benchmark values of the heaviest right-handed neutrino (RHN) mass \(m_{N_1}\), the table shows the corresponding maximum and minimum values of the form factor ($G_{N,S}$) sum terms, defined in Eqs.~(\ref{GNsumTerm},~\ref{GSsumTerm}), as the lightest active neutrino mass \(m_1\) is varied. Since the branching ratio \(\text{BR}(\mu \to e\gamma)\) depends on the square of the combined form factor, i.e., \(|G_N + G_S|^2\), and the signs of \(G_N\) and \(G_S\) remain the same across the parameter space, the relative magnitudes of these terms determine the minimum and maximum contributions to the branching ratio in Case I under Normal Ordering (NO) mass spectrum of active neutrinos.}
	\label{tab:RHNOoM_caseI}
\end{table}

From Table~\ref{tab:RHNOoM_caseI}, we observe that for smaller RHN masses (e.g., \(m_{N_1} = 100~\text{GeV}\)), the sterile neutrino contribution \(G_S\) \textit{sum term} is significantly smaller than the RHN contribution \(G_N\) \textit{sum term} across the entire range of \(m_1\). However, for larger RHN masses (e.g., \(m_{N_1} = 6000~\text{GeV}\)), the magnitude of \(G_S\) \textit{sum term} becomes comparable to \(G_N\) \textit{sum term} at higher values of \(m_1\), while it remains much smaller than \(G_N\) \textit{sum term} for the smallest values of \(m_1\) (e.g., \(10^{-5}~\text{eV}\)). The approximate equality $G_N$ \textit{sum term} $\approx$ $G_S$ \textit{sum term} occurs for \(m_1 \approx 10^{-2}~\text{eV}\) and \(m_{N_1} = 6000~\text{GeV}\), where $G_N\text{ \textit{sum term}} = G_S\text{ \textit{sum term}} \approx -1.95 \times 10^{-2}$.

Since the branching ratio depends on the square of the modulus of the total form factor, i.e., \(|G_N + G_S|^2\), it attains a maximum and minimum for specific parameter combinations. The maximum value occurs for:
\[
|G_N + G_S|_{\text{max}}^{(m_{N_1}=6000~\text{GeV},\, m_1=3.21\times 10^{-3}~\text{eV})} = \underbrace{\boxed{\frac{m^2_{W_L}}{m^2_{W_R}}}}_{\textcolor{blue}{\approx 1.32\times10^{-4}}}\times 5.5 \times 10^{-2}\approx 7.26\times10^{-6},
\]
while the minimum value is found for:
\[
|G_N + G_S|_{\text{min}}^{(m_{N_1}=100~\text{GeV},\, m_1=10^{-2}~\text{eV})} =  \underbrace{\boxed{\frac{m^2_{W_L}}{m^2_{W_R}}}}_{\textcolor{blue}{\approx 1.32\times10^{-4}}}\times1.17 \times 10^{-5}\approx 1.54\times10^{-9}.
\]
Thus, the maximum branching ratio in this setup is estimated as:
\begin{equation}
	\text{BR}^{\text{max}}(\mu \to e\gamma) = \underbrace{\boxed{\frac{3\alpha_{\text{em}}}{2\pi}}}_{\textcolor{blue}{\approx 3 \times 10^{-3}}} \left( |G_N + G_S|^2 \right) \approx 3 \times 10^{-3} \times (7.26 \times 10^{-6})^2 \approx 1.58 \times 10^{-13}.
\end{equation}

\begin{table}[h!]
	\centering
	\begin{tabular}{|c|c|c|c|c|c|}
		\hline
		\multicolumn{6}{|c|}{\textbf{BR} \boldmath\((\mu \to e\gamma)\) \textbf{RHN \& Sterile Neutrinos Contribution (Case I)}} \\
		\hline
		\multicolumn{6}{|c|}{\textbf{Inverted Ordering (IO)}} \\
		\hline
		\multicolumn{3}{|c|}{\textbf{RHN Contribution}}&\multicolumn{3}{c|}{\textbf{Sterile Neutrino Contribution}}\\
		\hline
		\(\mathbf{m_{N_3}}\) & \(\mathbf{m_3}\) & \(\mathbf{G_N}\) \textbf{Sum Term}  & \(\mathbf{m_{N_3}}\) & \(\mathbf{m_3}\) & \(\mathbf{G_S}\) \textbf{Sum Term} \\
		\textbf{(GeV)} & \textbf{(eV)} & \textbf{Value} & \textbf{(GeV)} & \textbf{(eV)} & \textbf{Value} \\
		\hline
		\multirow{2}{*}{\textbf{100}} 
		& \(10^{-5}\) & \(5.71\times10^{-6}\) & \multirow{2}{*}{\textbf{100}} & \(10^{-5}\) & \(1.82\times10^{-6}\) \\
		\cline{2-3} \cline{5-6}
		& \(10^{-2}\) & \(5.47\times10^{-6}\) & & \(10^{-2}\) & \(1.74\times10^{-6}\) \\
		\hline
		\multirow{2}{*}{\textbf{2400}} 
		& \(10^{-5}\) & \(2.81\times10^{-3}\) & \multirow{2}{*}{\textbf{2400}} & \(10^{-5}\) & \(2.88\times10^{-6}\) \\
		\cline{2-3} \cline{5-6}
		& \(10^{-2}\) & \(2.67\times10^{-3}\) & & \(10^{-2}\) & \(1.75\times10^{-3}\) \\
		\hline
		\multirow{2}{*}{\textbf{3000}} 
		& \(10^{-5}\) & \(4.11\times10^{-3}\) & \multirow{2}{*}{\textbf{3000}} & \(10^{-5}\) & \(4.69\times10^{-2}\) \\
		\cline{2-3} \cline{5-6}
		& \(10^{-2}\) & \(3.89\times10^{-3}\) & & \(10^{-2}\) & \(2.48\times10^{-2}\) \\
		\hline
		\multirow{2}{*}{\textbf{4000}} 
		& \(10^{-5}\) & \(6.50\times10^{-3}\) & \multirow{2}{*}{\textbf{4000}} & \(10^{-5}\) & \(8.64\times10^{-2}\) \\
		\cline{2-3} \cline{5-6}
		& \(10^{-2}\) & \(6.13\times10^{-3}\) & & \(10^{-2}\) & \(3.68\times10^{-2}\) \\
		\hline
	\end{tabular}
	\caption{For selected benchmark values of the heaviest right-handed neutrino (RHN) mass \(m_{N_3}\), the table shows the corresponding maximum and minimum values of the form factor ($G_{N,S}$) sum terms, defined in Eqs.~(\ref{GNsumTerm},~\ref{GSsumTerm}), as the lightest active neutrino mass \(m_1\) is varied in Case I under Inverted Ordering (IO) mass spectrum of active neutrinos.}
	\label{tab:SLOoM_caseI}
\end{table}

In contrast to the NO scenario, the $G_{N,S}$ \textit{sum terms} are positive in the IO case. Notably, the $G_S$ \textit{sum term} is approximately an order of magnitude larger than the $G_N$ \textit{sum term} for higher RHN masses—specifically, for \(m_{N_3} = 3000~\text{GeV}\) and \(4000~\text{GeV}\)—across the entire range of \(m_3\). The $G_S$ \textit{sum term} reaches its maximum value for \(m_{N_3} = 4000~\text{GeV}\) at \(m_3 = 10^{-5}~\text{eV}\). Accordingly, the maximum branching ratio for the considered benchmark ranges is estimated as:
\begin{equation}
	\text{BR}^{\text{max}}(\mu \to e\gamma) = \underbrace{\boxed{\frac{3\alpha_{\text{em}}}{2\pi}}}_{\textcolor{blue}{\approx 3 \times 10^{-3}}} \left( |G_N + G_S|^2 \right) \approx 3 \times 10^{-3} \times (1.23 \times 10^{-5})^2 \approx 4.54 \times 10^{-13},
\end{equation}
while values within the \textsf{MEG II} reach are obtained, for instance, for \(m_{N_3}=4000~\text{GeV}\) and \(m_3=10^{-2}~\text{eV}\), where
\[
|G_N + G_S|^{(m_{N_3}=4000~\text{GeV},\, m_3=10^{-2}~\text{eV})} \approx 5.66 \times 10^{-6}
\]
corresponds to $
\text{BR}(\mu \to e\gamma) \approx 9.61 \times 10^{-14}.
$

\section{Term-wise Decomposition and RHN-Hierarchy Selection for $0\nu\beta\beta$ in Case II}\label{app:CaseII_fav_details}

We present here a detailed analytical and numerical analysis to fix the RHN mass hierarchies for $0\nu\beta\beta$ predictions in Case II. For ready reference, we rewrite the EMM terms associated with the sterile-neutrino channel for the NO and IO cases:
\begin{equation}\label{meffTtermsNOApp}
	\begin{aligned}
		{m_{\beta\beta,R}^S}\bigg|_{\textbf{NO}} = -C_N&\bigg[\underbrace{|U_{e1}|^2\frac{m_1^2+\Delta m_{atm}^2}{m_1^2}\frac{m_{N_1}}{m_{S_3}^2}}_{T_1}+\underbrace{|U_{e2}|^2\frac{m_1^2+\Delta m_{atm}^2}{m_1^2+\Delta m_{sol}^2}\frac{m_{N_2}}{m_{S_3}^2}e^{i\alpha}}_{T_2}\\
		&+\underbrace{|U_{e3}|^2\frac{m_{N_3}}{m_{S_3}^2}e^{i(\beta-2\delta)}}_{T_3}\bigg],
	\end{aligned}
\end{equation}
\begin{equation}\label{meffTtermsIOApp}
	\begin{aligned}
		{m_{\beta\beta,R}^S}\bigg|_{\textbf{IO}} = -C_N&\bigg[\underbrace{|U_{e1}|^2\frac{\Delta m_{sol}^2+\Delta m_{atm}^2}{\Delta m_{atm}^2}\frac{m_{N_1}}{m_{S_2}^2}}_{T_1}+\underbrace{|U_{e2}|^2\frac{m_{N_2}}{m_{S_2}^2}e^{i\alpha}}_{T_2}\\
		&+\underbrace{|U_{e3}|^2\frac{\Delta m_{sol}^2+\Delta m_{atm}^2}{m_3^2}\frac{m_{N_3}}{m_{S_2}^2}e^{i(\beta-2\delta)}}_{T_3}\bigg].
	\end{aligned}
\end{equation}
In the NO case, the dependence of the EMM for the sterile-neutrino channel on the lightest active neutrino mass $m_1$ arises from the terms $T_1$ and $T_2$ in eq.~(\ref{meffTtermsNOApp}). However, the enhancement required for experimental relevance is provided mainly by the term \( T_1 \) for a hierarchical light-neutrino mass spectrum. In this regime, where the lightest neutrino mass \( m_1 \) is much smaller than the mass splittings, we can safely approximate $m_1^2+\Delta m_{atm}^2 \approx \Delta m_{atm}^2$ and $m_1^2+\Delta m_{sol}^2 \approx \Delta m_{sol}^2$. With these considerations, we have tabulated the expressions and values of the terms $T_1$, $T_2$, and $T_3$ in table~\ref{tab:terms} for different RHN mass hierarchies listed in table~\ref{tab:RHN_hierarchy_cases}. The lower panel of table~\ref{tab:terms} presents the numerical values of these terms for the heaviest RHN mass at 300~GeV, the intermediate at 250~GeV, and the lightest at 200~GeV in any RHN mass hierarchy. For the double-seesaw approximation, the heaviest sterile-neutrino mass is taken to be $m_{S_3} = 10 \times$ (heaviest RHN mass) in each case. The lightest neutrino mass is taken to be $10^{-5}$~eV and $10^{-2}$~eV for its minimum and maximum values, respectively, while the oscillation parameters are fixed at their best-fit values. Since only the first term ($T_1$) has an explicit dependence on the lightest neutrino mass ($m_1$), it exhibits a corresponding range of values. 
\begin{table}
	\centering
	\footnotesize
	\renewcommand{\arraystretch}{1.3}
	\begin{tabular}{|c|c|c|c|}
		\hline
		\multicolumn{4}{|c|}{\centering \textbf{Term-wise expression of $m^{S}_{\beta\beta}$ (Normal Ordering (NO))}} \\
		\hline
		\textbf{$m_N$ \textbf{Cases}} & \textbf{Term I} & \textbf{Term II} & \textbf{Term III} \\
		\hline
		\textbf{Case} $m_{N_A}$ & 
		$\frac{m_{N_{1}}}{m_{N_3}^{2}}\cdot \frac{\Delta m^{2}_{\mbox{atm}}}{m^{2}_{1}}\cdot 10^{-2}\cdot |U_{e1}|^{2}$ & 
		$e^{i\alpha}\frac{m_{N_{2}}}{m_{N_3}^{2}}\cdot \frac{\Delta m^{2}_{\mbox{atm}}}{\Delta m^{2}_{\mbox{sol}}}\cdot 10^{-2}\cdot |U_{e2}|^{2}$ & $\frac{e^{i(\beta-2\delta)}}{m_{N_{3}}} 10^{-2}\cdot |U_{e3}|^{2}$\\
		\hline
		\textbf{Case} $m_{N_F}$ & 
		$\frac{1}{m_{N_1}}\cdot \frac{\Delta m^{2}_{\mbox{atm}}}{m^{2}_{1}}\cdot 10^{-2}\cdot |U_{e1}|^{2}$  & 
		$e^{i\alpha}\frac{m_{N_{2}}}{m_{N_1}^{2}}\cdot \frac{\Delta m^{2}_{\mbox{atm}}}{\Delta m^{2}_{\mbox{sol}}}\cdot 10^{-2}\cdot |U_{e2}|^{2}$ & $\frac{e^{i(\beta-2\delta)}m_{N_{3}}}{m_{N_{1}}^2} 10^{-2}\cdot |U_{e3}|^{2}$  \\
		\hline
		\textbf{Case} $m_{N_D}$ & 
		$\frac{m_{N_1}}{m_{N_2}^{2}}\cdot\frac{\Delta m^{2}_{\mbox{atm}}}{m^{2}_{1}}\cdot 10^{-2}\cdot |U_{e1}|^{2}$  & 
		$e^{i\alpha}\frac{1}{m_{N_2}}\cdot \frac{\Delta m^{2}_{\mbox{atm}}}{\Delta m^{2}_{\mbox{sol}}}\cdot 10^{-2}\cdot |U_{e2}|^{2}$ & $\frac{e^{i(\beta-2\delta)}m_{N_{3}}}{m_{N_{2}}^2} 10^{-2}\cdot |U_{e3}|^{2}$  \\
		\hline
		\hline
		\multicolumn{4}{|c|}{\centering \textbf{Term-wise values of $m^{S}_{\beta\beta}$}} \\
		\hline
		\textbf{$m_N$ \textbf{Cases}} & \textbf{Term I} & \textbf{Term II} & \textbf{Term III} \\
		\hline
		\textbf{Case} $m_{N_A}$ & 
		$3.81\times 10^{2}- 3.81\times 10^{-2}$ & 
		$e^{i\alpha}\,2.84\times 10^{-4}$ & $e^{i(\beta-2\delta)}\,7.38\times 10^{-7}$\\
		\hline
		\textbf{Case} $m_{N_F}$ & 
		$4.76 \times 10^{2}-4.76\times 10^{-2}$  & 
		$e^{i\alpha}\,2.84\times 10^{-4}$ & $e^{i(\beta-2\delta)}\,4.92\times 10^{-7}$  \\
		\hline
		\textbf{Case} $m_{N_D}$& 
		$2.54\times 10^{2}-2.54\times 10^{-2}$  & 
		$e^{i\alpha}\,7.63\times 10^{-4}$ & $e^{i(\beta-2\delta)}\,4.92\times 10^{-7}$  \\
		\hline
	\end{tabular}
	\caption{Analytical expressions (top) and numerical estimates (bottom) of the sterile neutrino contribution to $m^{S}_{\beta\beta}$ for different RHN mass hierarchies in normal ordering. The heaviest RHN is fixed at 300~GeV with the intermediate and lightest RHN masses set to 250~GeV and 200~GeV, respectively, in all cases. The heaviest sterile-neutrino mass $m_{S_3}$ is taken to be $10\times$ (heaviest RHN mass) to ensure the double-seesaw approximation. Oscillation parameters are taken at their best-fit values, and variation in the lightest active neutrino mass ($10^{-5}$–$10^{-2}$~eV) affects only Term~I.}
	\label{tab:terms}
\end{table}
\begin{figure}
	\centering
	\begin{minipage}[b]{0.45\textwidth}
		\centering
		\includegraphics[width=\textwidth]{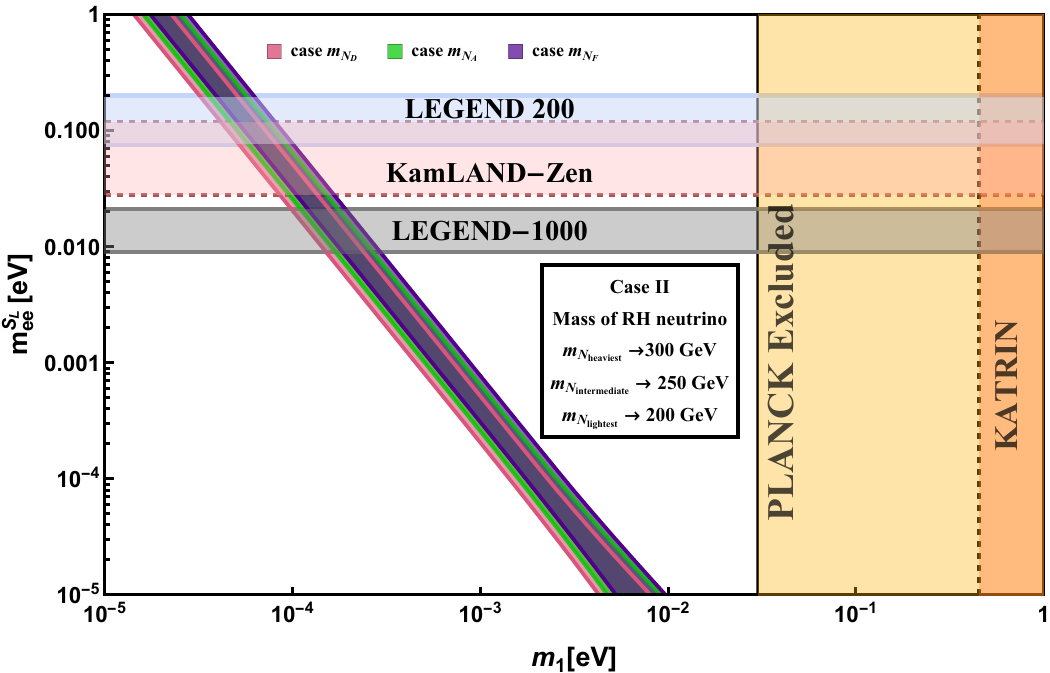}
	\end{minipage}
	\hfill
	\begin{minipage}[b]{0.45\textwidth}
		\centering
		\includegraphics[width=\textwidth]{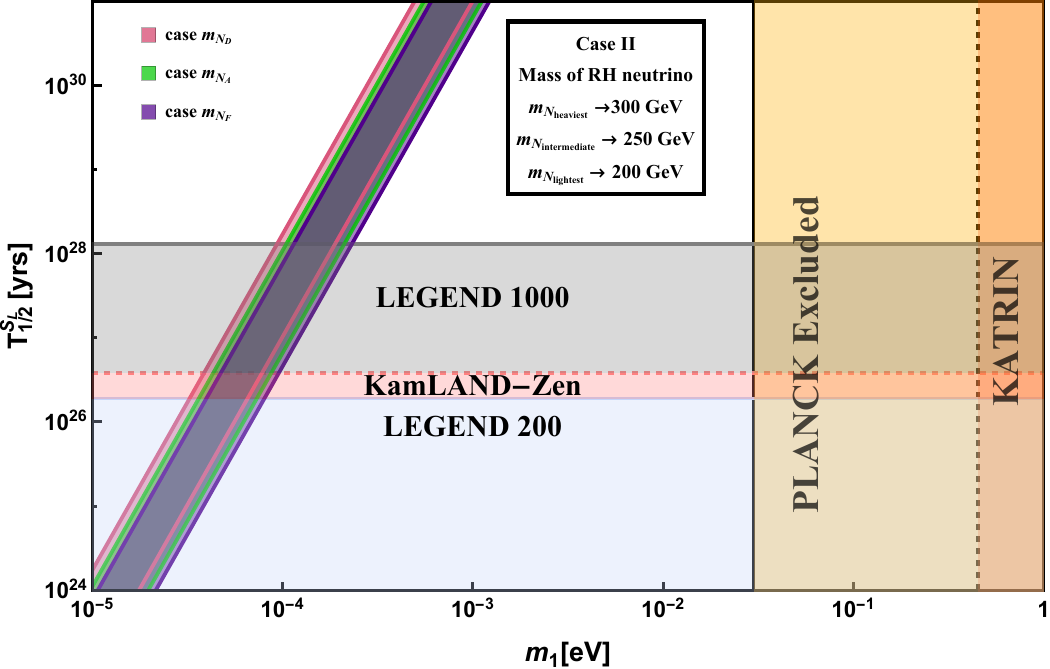}
	\end{minipage}
	\caption{Variation of the generalized effective Majorana mass associated with the sterile-neutrino channel, $m^{S_L}_{ee}$ (left), and the corresponding half-life $T^{S_L}_{1/2}$ (right), for the NO mass spectrum of active neutrinos. The plots are shown as a function of the lightest active neutrino mass $m_1$, for different RHN mass hierarchies $m_{N_{A,F,D}}$. The light-blue and light-pink horizontal bands represent current bounds on EMM from the \textsf{LEGEND-200} and \textsf{KamLAND-Zen} experiments, respectively. The grey horizontal band indicates the projected bound from the upcoming \textsf{LEGEND-1000} experiment. The vertical yellow and orange bands represent bounds corresponding to the sum of active neutrino masses from \textsf{PLANCK} and \textsf{KATRIN}, respectively.}
	\label{fig:terms}
\end{figure}
From the term-wise values of $m^{S}_{\beta\beta}$ presented in table~\ref{tab:terms}, it is evident that only term $T_1$ is dominant for the cases $m_{N_{A,F,D}}$. Thus, we restrict our discussion to these $m_N$ cases, as the others listed in table~\ref{tab:RHN_hierarchy_cases} are effectively equivalent due to the suppression of $T_2$ and $T_3$, making their total contribution indistinguishable from those already considered. To explore the dynamics of the generalized effective Majorana mass associated with the sterile-neutrino channel for the NO case with respect to the lightest neutrino mass $m_1$, and for different $m_N$ cases—namely, $m_{N_A}$, $m_{N_F}$, and $m_{N_D}$—we have plotted $m^{S_L}_{ee}$ and $T^{S_L}_{1/2}$ as functions of $m_1$ in figure~\ref{fig:terms}. It is evident from the plots of $m^{S_L}_{ee}$ (left panel of figure~\ref{fig:terms}) and $T^{S_L}_{1/2}$ (right panel of figure~\ref{fig:terms}) that the contributions are almost identical for the three $m_N$ cases under consideration: cases $m_{N_{A,F,D}}$. The bands in figure~\ref{fig:terms} are obtained by varying the neutrino oscillation parameters within their respective $3\sigma$ allowed ranges. The Dirac phase $\delta$ and Majorana phases $\alpha$ and $\beta$ are varied in the interval $[0,2\pi]$. The NME ratio is also varied as $22.2\lesssim\mathcal{M}^{0\nu}_N/\mathcal{M}^{0\nu}_\nu\lesssim76.3$, as discussed in table~\ref{tab:NMEratio}. The magenta, green, and purple bands in figure~\ref{fig:terms} correspond to the cases $m_{N_D}$, $m_{N_A}$, and $m_{N_F}$, respectively. 

Although the variation of GEMM with respect to $m_1$ appears very similar across different $m_N$ cases, a closer look at the numerical estimates of the first term (Term~I) in the lower panel of table~\ref{tab:terms} reveals that the contribution to $m^{S}_{\beta\beta}$ is maximal in the case $m_{N_F}$. Therefore, without loss of generality, we fix the mass hierarchy of right-handed neutrinos as case $m_{N_F}$ ($m_{N_3} < m_{N_2} < m_{N_1}$) in the NO mass spectrum of active neutrinos.

For the IO case, we consider a hierarchical light-neutrino mass spectrum, for which the following approximations are used: $m_3^2+\Delta m_{sol}^2+\Delta m_{atm}^2 \approx \Delta m_{sol}^2+\Delta m_{atm}^2$ and $m_3^2+\Delta m_{atm}^2 \approx \Delta m_{atm}^2$. From eq.~(\ref{meffTtermsIOApp}), it is evident that the dominant contribution arises from the third term ($T_3$). Similar to the NO scenario, the EMM for the sterile-neutrino channel in the IO case is nearly identical for the hierarchies $m_{N_{A,F,D}}$, as the other hierarchies listed in table~\ref{tab:RHN_hierarchy_cases} are effectively equivalent due to the suppression of subdominant terms. Since $T_3$ contributes maximally in the $m_{N_A}$ case ($m_{N_1}<m_{N_2}<m_{N_3}$), we fix the RHN mass hierarchy as $m_{N_A}$ for the IO spectrum, without loss of generality.

\section{Loop function behaviour for $\mu\to e\gamma$ in Case II ($M_{W_R}=7$~TeV)}\label{app:loopfunc_caseII}
\begin{figure}[H]
	
	\begin{subfigure}[b]{0.46\textwidth}
		\includegraphics[width=\linewidth]{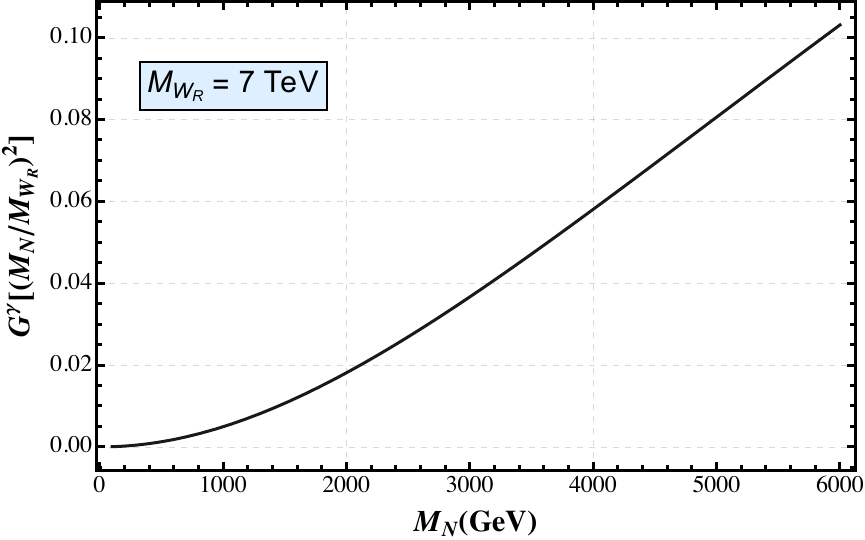}
		\caption{}
		\label{fig:gGammamN}
	\end{subfigure}
	\hfill
	\begin{subfigure}[b]{0.46\textwidth}
		\includegraphics[width=\linewidth]{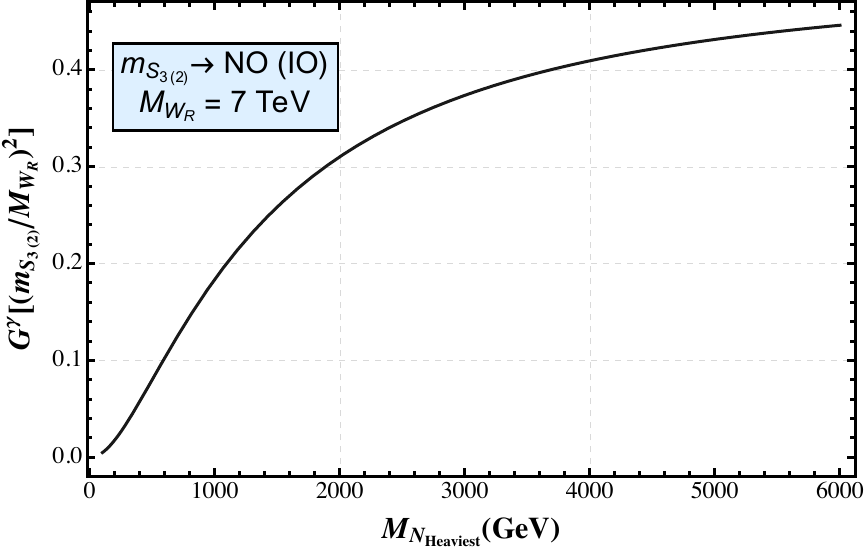}
		\caption{}
		\label{fig:gGammaHeaviestSL}
	\end{subfigure}
	
	\vspace{0.5cm}
	
	\begin{subfigure}[b]{0.50\textwidth}
		\includegraphics[width=\linewidth]{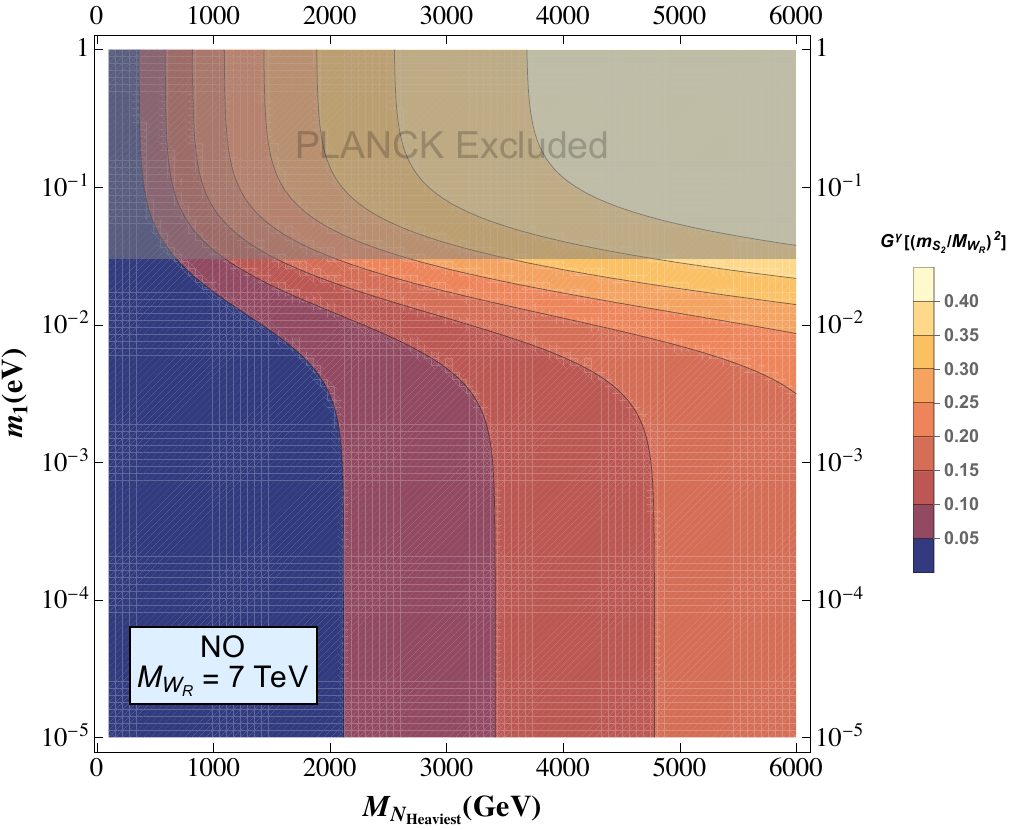}
		\caption{}
		\label{fig:gGammaNOIntermtdSL}
	\end{subfigure}
	\hfill
	\begin{subfigure}[b]{0.50\textwidth}
		\includegraphics[width=\linewidth]{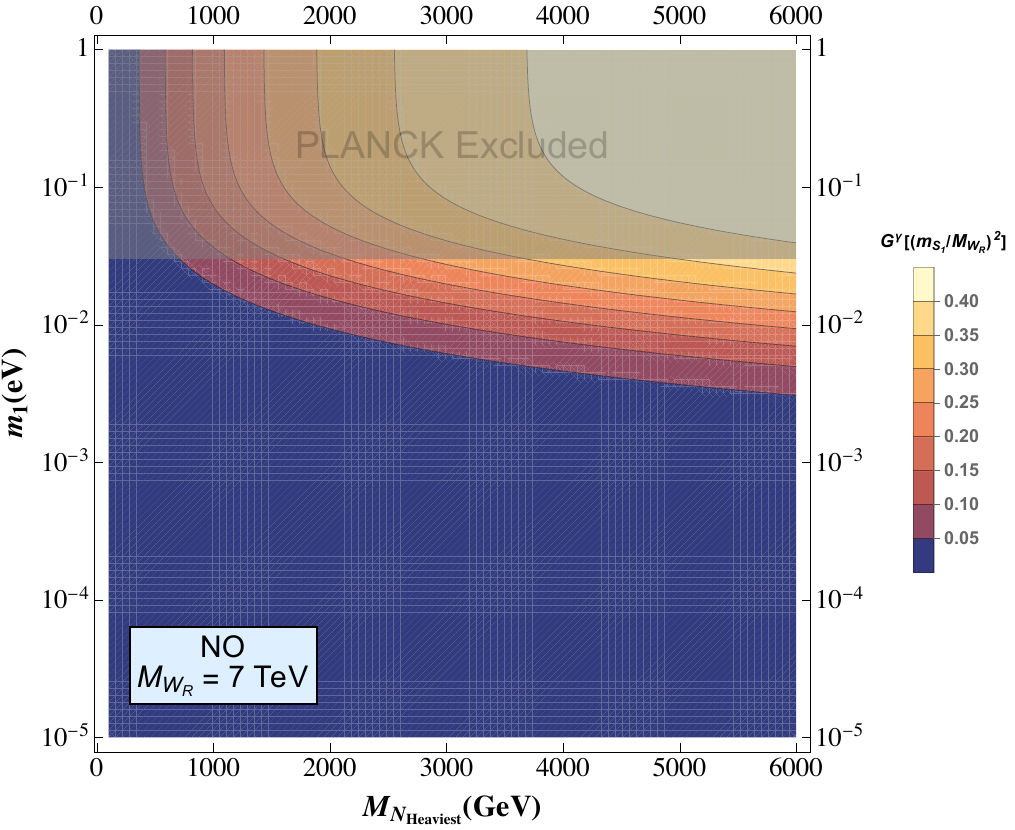}
		\caption{}
		\label{fig:gGammaNOLightestSL}
	\end{subfigure}
	
	\vspace{0.5cm}
	
	\begin{subfigure}[b]{0.50\textwidth}
		\includegraphics[width=\linewidth]{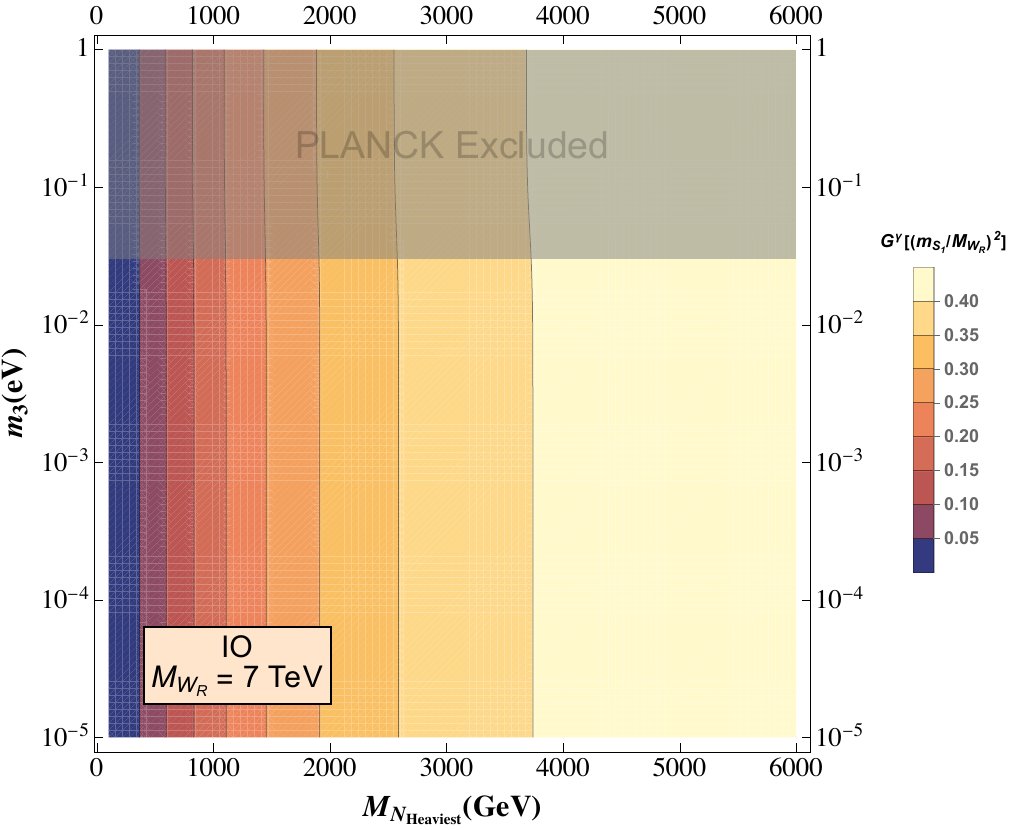}
		\caption{}
		\label{fig:gGammaIOIntermtdSL}
	\end{subfigure}
	\hfill
	\begin{subfigure}[b]{0.50\textwidth}
		\includegraphics[width=\linewidth]{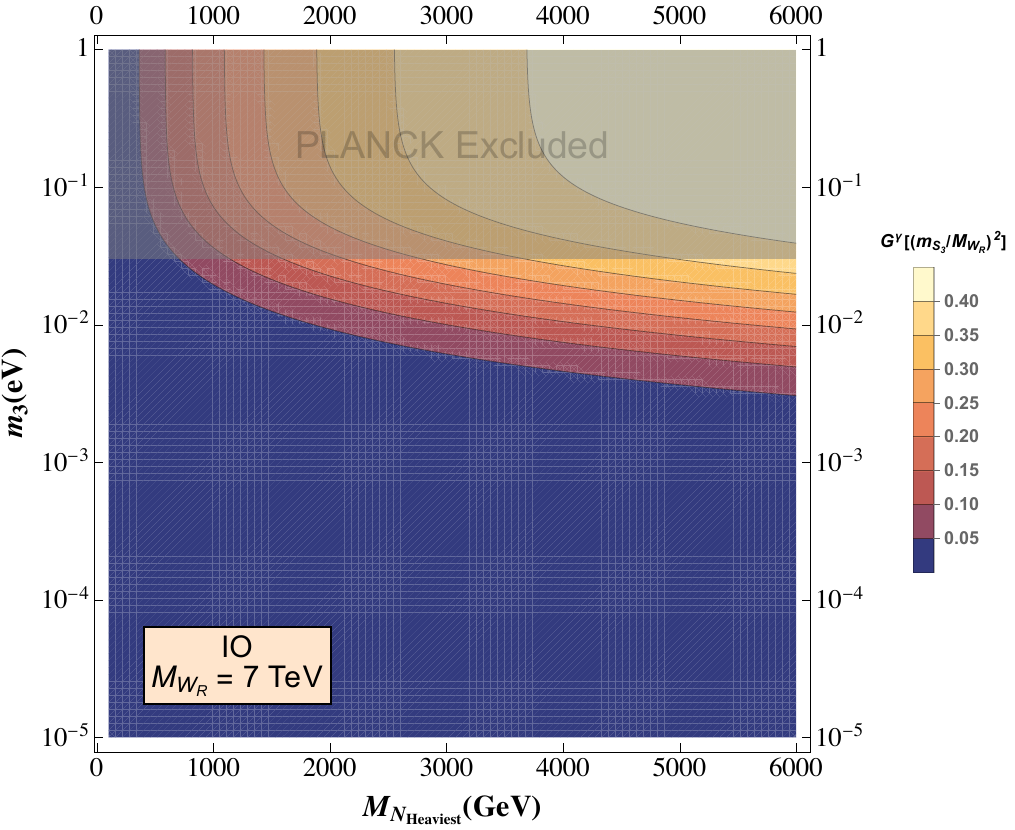}
		\caption{}
		\label{fig:gGammaIOLightestSL}
	\end{subfigure}
	\caption{Loop function variations relevant for the $\mu \to e \gamma$ process in case II with $M_{W_R} = 7$~TeV. Figure~\ref{fig:gGammamN} shows the RHN-mediated contribution versus RHN mass, while Figure~\ref{fig:gGammaHeaviestSL} displays the contribution from the heaviest sterile neutrino versus the heaviest RHN mass. Figures~\ref{fig:gGammaNOIntermtdSL}–\ref{fig:gGammaIOLightestSL} show contour plots of loop functions from intermediate and lightest sterile states for NO and IO, plotted against the heaviest RHN mass and lightest active neutrino mass ($m_1$ for NO, $m_3$ for IO). Grey bands denote \textsf{PLANCK}-excluded regions. In this framework, sterile masses follow $m_{S_1} < m_{S_2} < m_{S_3}$ (NO) and $m_{S_3} < m_{S_1} < m_{S_2}$ (IO).}
	\label{fig:LoopFunctions_6panel}
\end{figure}

For completeness, figure~\ref{fig:LoopFunctions_6panel} shows the loop functions used in the Case~II prediction of BR($\mu\to e\gamma$) at the benchmark $M_{W_R}=7$~TeV. The panels display how the loop functions change with the heavy masses and the lightest active neutrino mass, separate from mixing effects. These plots help estimate the sign and size of the $G_N$ and $G_S$ Sum Terms used in the main text, and they support the hierarchy choices made in subsection~\ref{subsec:cLFVcase2}.

\section{Supplementary $\mu\to e\gamma$ Results in Case II: RHN Mass Hierarchies under NO/IO}\label{app:clfv_caseII}
We discuss here BR($\mu\to e \gamma$) predictions for Case II with RHN mass hierarchies: case $m_{N_F}$ (in the IO mass spectrum of active neutrinos) and case $m_{N_F}$ (in the NO mass spectrum of active neutrinos). These case studies complement the cases already favoured by the $0\nu\beta\beta$ study in Case II discussed in subsection~\ref{subsec:0nubb_caseII_fav}.\\

\textbf{\underline{Case $m_{N_F}$ in IO}}\\

The $G_N$ Sum Term discussion and results remain valid for the RHN mass hierarchy case $m_{N_F}$ in the IO scenario, as the expression is independent of active neutrino masses. For inverted ordering, the approximate expression for the $G_S$ Sum Term at best-fit values of the low-energy oscillation parameters is:
\begin{equation}\label{SymbolicGSSumTermIOmNF}
	\begin{aligned}
		G_S \text{ Sum Term}\Big|^{\text{case}\,m_{N_F}}_{\text{IO}} &\approx (-2.98-0.77\,e^{i\delta})\times10^{-2}\times G^{\gamma}\left(\frac{m^2_{S_1}}{m^2_{W_R}}\right)\\
		&+(2.98-0.35\,e^{i\delta})\times10^{-2}\left[\frac{m_{N_2}}{m_{N_1}}\right]G^{\gamma}\left(\frac{m^2_{S_2}}{m^2_{W_R}}\right)\\
		&+\frac{5.66\,e^{i\delta}\times10^{-6}}{m_3\text{ (eV)}}\left[\frac{m_{N_3}}{m_{N_1}}\right]G^{\gamma}\left(\frac{m^2_{S_3}}{m^2_{W_R}}\right).
	\end{aligned}
\end{equation}
Similar to the NO case, the $G_S$ Sum Term in eq.~(\ref{SymbolicGSSumTermIOmNF}) varies significantly with the loop functions, the Dirac CP phase $\delta$, and the lightest active neutrino mass $m_3$. In IO, the sterile neutrino masses follow $m_{S_3} < m_{S_1} < m_{S_2}$; corresponding loop-function dynamics are illustrated in figures~\ref{fig:gGammaHeaviestSL}, \ref{fig:gGammaIOIntermtdSL}, and \ref{fig:gGammaIOLightestSL} of appendix~\ref{app:loopfunc_caseII}. The $G_S$ Sum Term prefers a strong RHN mass hierarchy and $\delta = 0$ to attain a large (negative) magnitude needed to enhance $|G_N + G_S|$ for the desired BR($\mu \to e \gamma$). Specifically, we choose the RHN mass hierarchy as:
\begin{equation}
	m_{N_2}=0.5\,m_{N_1};\quad m_{N_3}=0.4\,m_{N_1}.
\end{equation}

\textbf{\underline{Case $m_{N_A}$ in NO}}\\

\begin{figure}
	\centering
	\begin{minipage}[b]{0.48\textwidth}
		\centering
		\includegraphics[width=\textwidth]{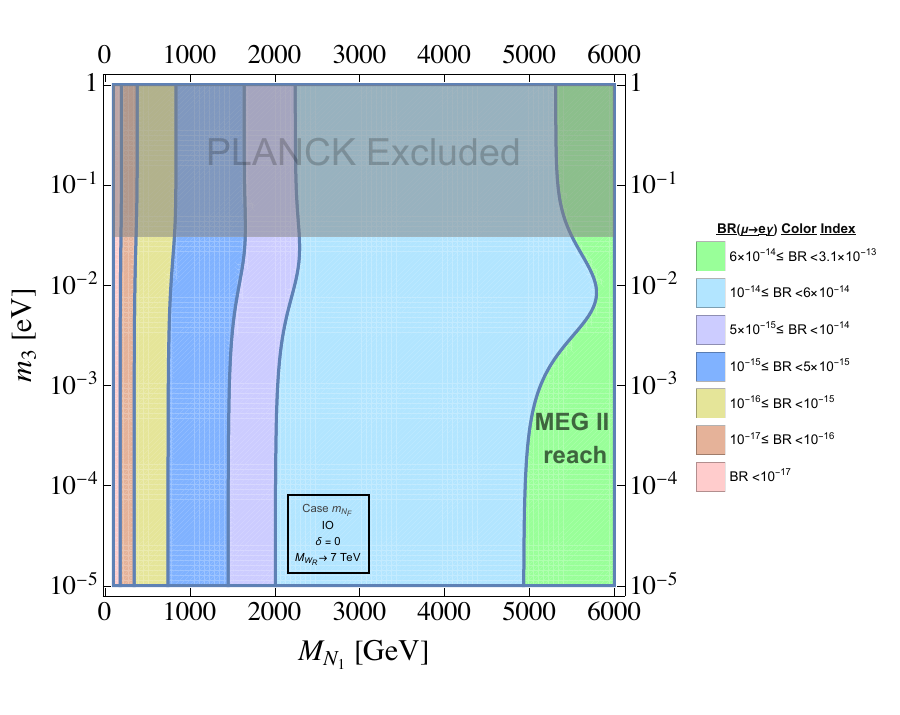}
	\end{minipage}
	\hfill
	\begin{minipage}[b]{0.48\textwidth}
		\centering
		\includegraphics[width=\textwidth]{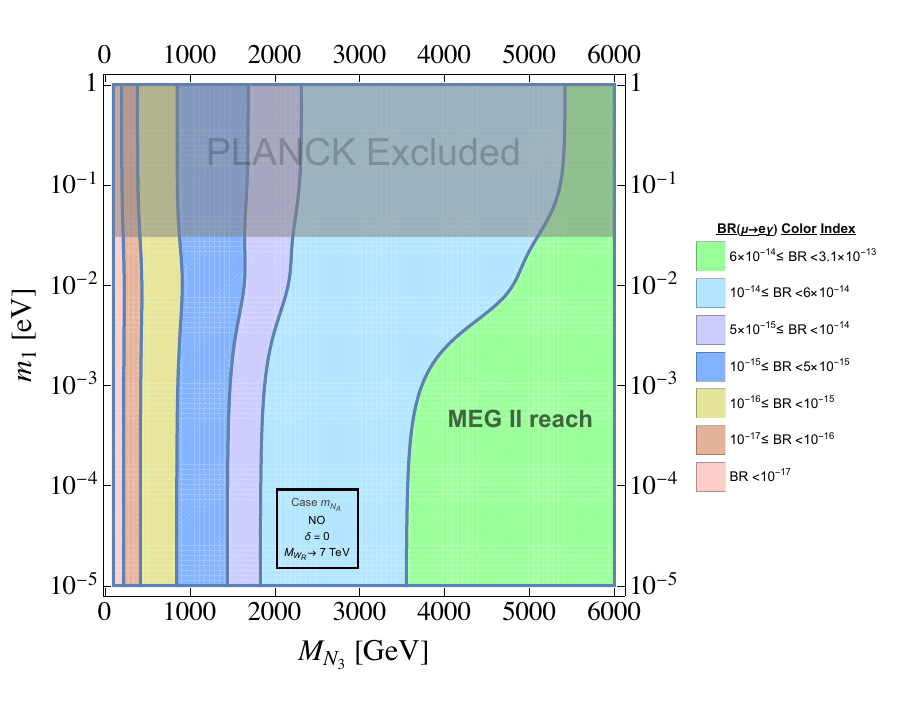}
	\end{minipage}
	\caption{Region plots of BR($\mu\to e\gamma$) in Case II with $M_{W_R}=7$~TeV, $\delta=0$, $\xi=10^{-6}$, the heaviest sterile neutrino mass set to ten times the heaviest RHN mass, and the lightest active neutrino mass varied from $1$ to $10^{-5}$~eV. Left: case $m_{N_F}$ (IO). Right: case $m_{N_A}$ (NO). The green shaded region indicates the projected sensitivity of \textsf{MEG II}, while the translucent grey band marks the area excluded by the \textsf{PLANCK} bound on the sum of active neutrino masses.}
	\label{fig:RegionPlotsCaseII_2panel}
\end{figure}

The $G_N$ Sum Term in case $m_{N_A}$ varies irrespective of the active neutrino mass ordering and attains a large positive value for $\delta = 0$, a mild hierarchy between $m_{N_3}$ and $m_{N_2}$, and a strong hierarchy between $m_{N_3}$ and $m_{N_1}$. For the NO case, the approximate analytic expression for the $G_S$ Sum Term at best-fit values of the neutrino oscillation parameters is:
\begin{equation}\label{SymbolicGSSumTermNOmNA}
	\begin{aligned}
		G_S \text{ Sum Term}\Big|^{\text{case}\,m_{N_A}}_{\text{NO}} &\approx \frac{(-1.51-0.39\,e^{i\delta})\times10^{-3}}{m_1\,(\text{eV})}\left[\frac{m_{N_1}}{m_{N_3}}\right]G^{\gamma}\left(\frac{m^2_{S_1}}{m^2_{W_R}}\right)\\
		&+\frac{(1.51-0.18\,e^{i\delta})\times10^{-3}}{\sqrt{7.5\times10^{-5}+[m_1\,(\text{eV})]^2}}\left[\frac{m_{N_2}}{m_{N_3}}\right]G^{\gamma}\left(\frac{m^2_{S_2}}{m^2_{W_R}}\right)\\
		&+1.12\,e^{i\delta}\times10^{-2}\,G^{\gamma}\left(\frac{m^2_{S_3}}{m^2_{W_R}}\right).
	\end{aligned}
\end{equation}
The expression reaches its maximum positive value for $\delta = 0$, requires a mild hierarchy between $m_{N_3}$ and $m_{N_2}$, and remains effectively independent of $m_{N_1}$. Therefore, for $|G_N + G_S|$ to attain a large value and enhance BR($\mu \to e \gamma$) for case $m_{N_A}$ in NO, we choose the RHN mass hierarchy as:
\begin{equation}
	m_{N_2}=0.9\,m_{N_3};\quad m_{N_1}=0.4\,m_{N_3}.
\end{equation}
With RHN mass hierarchies chosen for case $m_{N_F}$ in IO and case $m_{N_A}$ in NO, along with other fixed parameters such as $M_{W_R}=7$ TeV, $\delta=0$, $\xi=10^{-6}$, the heaviest sterile neutrino mass fixed to be ten times the heaviest RHN mass, and the lightest active neutrino mass varied from 1 to $10^{-5}$ eV, the region plots in figure~\ref{fig:RegionPlotsCaseII_2panel} show the variation of BR($\mu\to e\gamma$) as a function of the heaviest RHN mass and the lightest active neutrino mass.

The left panel of figure~\ref{fig:RegionPlotsCaseII_2panel}, for the IO scenario of case \(m_{N_F}\), indicates that over the entire range of the lightest active neutrino mass (from \(10^{-5}\) eV up to the \textsf{PLANCK}-excluded bound) and for \(m_{N_1} \gtrsim 4933\) GeV, BR(\(\mu \to e\gamma\)) lies within the \textsf{MEG II} reach.

The right panel of figure~\ref{fig:RegionPlotsCaseII_2panel}, corresponding to the NO scenario of case \(m_{N_A}\), represents the least constrained scenario, as BR(\(\mu \to e\gamma\)) enters the \textsf{MEG II} sensitivity region over a wide range of RHN masses (\(m_{N_3} \gtrsim 3560\) GeV) and across the entire range of the lightest active neutrino mass.

\section{cLFV predictions in Case II with $M_{W_R}=4.8$ TeV}\label{48TeVcLFV}

\begin{figure}
	\centering
	\begin{minipage}[b]{0.48\textwidth}
		\centering
		\includegraphics[width=\textwidth]{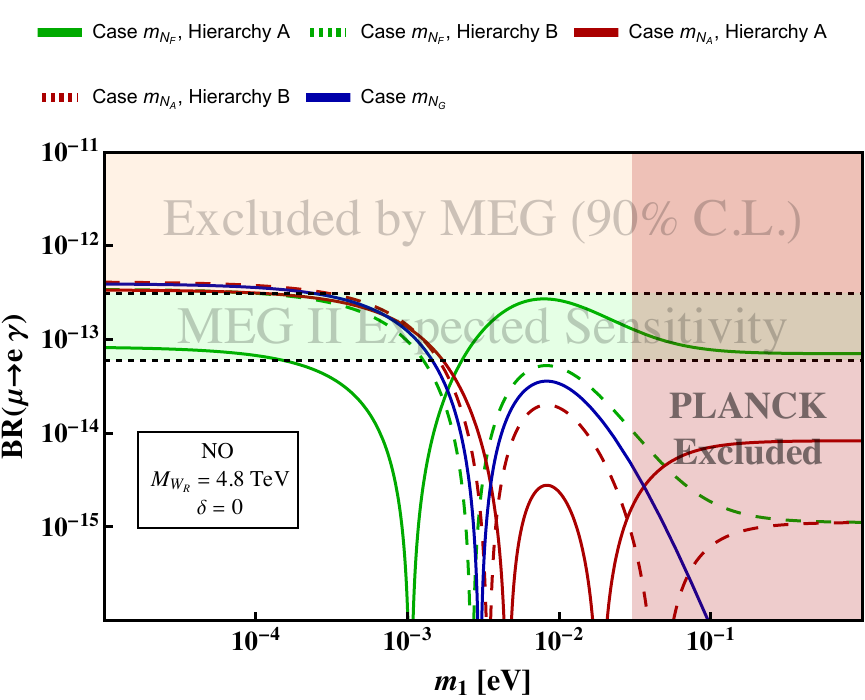}
	\end{minipage}
	\hfill
	\begin{minipage}[b]{0.48\textwidth}
		\centering
		\includegraphics[width=\textwidth]{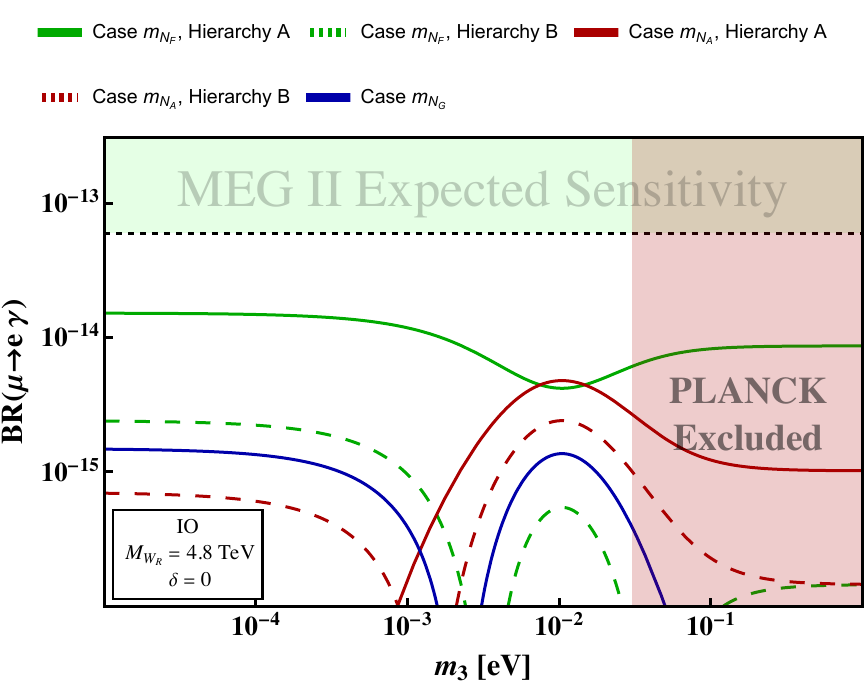}
	\end{minipage}
	\caption{Variation of BR\((\mu\to e\gamma)\) with the lightest active neutrino mass \(m_{1(3)}\) in Case II at \(M_{W_R} = 4.8\)~TeV and \(\delta=0\), for the normal ordering (NO, left panel) and inverted ordering (IO, right panel) mass spectra. Each curve corresponds to a different RHN mass hierarchy case \(m_{N_{F,A,G}}\), where the heaviest RHN mass is fixed at 4700~GeV and the other two are chosen to study overlap regions of parameter space for both \(0\nu\beta\beta\) and cLFV.}
	\label{fig:BRvsLightestMass48TeV}
\end{figure}

Figure~\ref{fig:BRvsLightestMass48TeV} shows the variation of BR\((\mu\to e\gamma)\) in Case II at \(M_{W_R} = 4.8\)~TeV and \(\delta = 0\), with respect to the lightest active neutrino mass \(m_{1(3)}\) for normal ordering (NO) (left panel) and inverted ordering (IO) (right panel) mass spectra of active neutrinos. The plots are generated for selected RHN mass hierarchies, as mentioned in table~\ref{tab:RHN_hierarchy_cases}, motivated by the search for a common parameter space relevant to both \(0\nu\beta\beta\) and cLFV (\(\mu \to e\gamma\)) processes. For each RHN mass hierarchy case (\(m_{N_{F,A,G}}\)), the heaviest RHN mass is fixed at 4700~GeV, and the intermediate and lightest RHN masses are chosen as follows:
\begin{equation*}
	\begin{aligned}
		&\textbf{(Hierarchy A)}\\
		&\text{Heaviest } m_N: 4700~\text{GeV},\quad \text{Intermediate } m_N: 3700~\text{GeV},\quad \text{Lightest } m_N: 3650~\text{GeV},\\
		&\textbf{(Hierarchy B)}\\
		&\text{Heaviest } m_N: 4700~\text{GeV},\quad \text{Intermediate } m_N: 4600~\text{GeV},\quad \text{Lightest } m_N: 4500~\text{GeV}.
	\end{aligned}
\end{equation*}

From the left panel of figure~\ref{fig:BRvsLightestMass48TeV}, which corresponds to the NO mass spectrum, it is evident that BR\((\mu\to e\gamma)\) enters the \textsf{MEG II} sensitivity region and even reaches the current \textsf{MEG} exclusion zone. In contrast, the right panel for the IO mass spectrum shows that BR\((\mu\to e\gamma)\) remains highly suppressed and does not reach the \textsf{MEG II} sensitivity zone for any of the parameter space considered in this study.



\clearpage
\bibliographystyle{utcaps_mod}
\bibliography{DSS_onubb_LR}
\end{document}